%% file: main.tex
\documentclass[reqno]{article}

\input{header}

\title{Bayesian and Frequentist Semantics for Common Variations of Differential Privacy: Applications to the 2020 Census}

\author[1]{Daniel Kifer}
\author[2]{John M. Abowd}
\author[2]{Robert Ashmead}
\author[2]{Ryan Cumings-Menon}
\author[2]{Philip Leclerc}
\author[3]{Ashwin Machanavajjhala}
\author[3]{William Sexton}
\author[2]{Pavel Zhuravlev}
\affil[1]{U.S. Census Bureau and Penn State University}
\affil[2]{U.S. Census Bureau}
\affil[3]{Tumult Labs}

\begin{document}

\maketitle

%\textcolor{red}{\Large Please do not add comments yet! But if you do, please use overleaf comment functionality}

\begin{abstract}
\input{abstract}

\end{abstract}

\section{Introduction}\label{sec:intro}
\input{intro}

\section{Informal Introduction to Privacy Semantics}\label{sec:informal}
\input{informal}

\section{Notation}\label{sec:notation}
\input{notation}

\section{The Privacy-Loss Random Variables and Technical Privacy Desiderata}\label{sec:plrv}
\input{plrv}

\section{Major Variants of Differential Privacy and Parameter Interpretability}\label{sec:privdefs}
\input{privdefs}

\section{Frequentist Hypothesis Testing Semantics}\label{sec:freq}

\input{freq}

\section{Bayesian Posterior-to-Posterior Semantics}\label{sec:bayes}
\input{bayes}

\section{Fine-grained, per-attribute Semantics}\label{sec:per}
\input{per}

\section{Related Work}\label{sec:related}
\input{related}

\section{Conclusions}\label{sec:conc}
\input{conc}

\section*{Acknowledgments and Disclaimer}
The views expressed in this technical paper are those of the authors and not those of the U.S. Census Bureau. We are grateful for the helpful comments provided by Mark Bun and Marco Gaboardi.

\bibliographystyle{plain}
\bibliography{refs}

\clearpage
\appendix
\input{appendix}
\end{document}

%% file: header.tex
\usepackage{amsmath}
\usepackage{amssymb}
\usepackage[utf8]{inputenc}
\usepackage[margin=1in]{geometry}
\usepackage{hyperref}
\hypersetup{
  colorlinks   = true, 
  urlcolor     = blue, 
  linkcolor    = blue, 
  citecolor   = blue
}
\usepackage{graphicx}
\usepackage{color}
\usepackage{amsthm}
\usepackage{xspace}
\usepackage{xargs}
\usepackage[conf={restate,no link to proof}]{proof-at-the-end}
\usepackage{todonotes}
\usepackage{comment}
\usepackage{authblk}
\usepackage{lineno}
\input{notation_defs}

\newtheorem{definition}{Definition}[section]
\newtheorem{example}[definition]{Example}
\newtheorem{theorem}[definition]{Theorem}

\newtheorem{remark}[definition]{Remark}

\allowdisplaybreaks

%% file: notation_defs.tex
\def\notationcolor{blue} % change to black in final version

\newcommand{\notation}[2]{\newcommand{#1}{{\textcolor{\notationcolor}{\ensuremath{#2}}}}}

\newcommand{\term}[2]{\newcommand{#1}{\textcolor{\notationcolor}{#2}\xspace}}

\notation{\mech}{M}
\notation{\data}{\mathcal{D}}
\notation{\datarv}{\mathcal{X}}
\notation{\dataspace}{\mathcal{U}}
\notation{\record}{r}
\notation{\outp}{\omega}
\notation{\randalg}{A}
\notation{\datb}{\data^\prime}
\notation{\query}{q}
\notation{\power}{1-\beta}
\notation{\typetwo}{\beta}
\notation{\level}{\ell}
\notation{\prior}{P_\mathfrak{A}}
\notation{\recordrv}{R}
\notation{\datasize}{n}
\notation{\psample}{\textup{psample}} % posterior sample
\newcommandx*\plrv[3][1=\ensuremath{\data_1}, 2=\ensuremath{\data_2}, 3=\mech]{\ensuremath{\mathcal{\textcolor{\notationcolor}{L}}_{#1,#2}^{#3}}\xspace} 

\newcommandx*\rem[1][1=\ensuremath{\data}]{\ensuremath{\textcolor{\notationcolor}{#1^{-}}}} % remove a record

\newcommand{\add}[2]{\ensuremath{#2\textcolor{blue}{\cup\{}#1\textcolor{blue}{\}}}}

\newcommandx*\repl[2][1=\ensuremath{\data},2=\record]{\add{#2}{\rem[#1]}}

\term{\pbdp}{pbdp}
\term{\pbdplong}{probabilistically-bounded differential privacy}
\term{\Pbdplong}{Probabilistically-bounded differential privacy}
\term{\target}{Jessie}

%% file: abstract.tex
The purpose of this paper is to guide interpretation of the semantic privacy guarantees for some of the major variations of differential privacy, which include pure, approximate, R\'enyi, zero-concentrated, and $f$ differential privacy. We interpret privacy-loss accounting parameters, frequentist semantics, and Bayesian semantics (including new results). The driving application is the interpretation of the  confidentiality protections for the 2020 Census Public Law 94-171  Redistricting Data Summary File released August 12, 2021, which, for the first time, were produced with formal privacy guarantees.

%% file: intro.tex
Differential privacy \cite{dmns06} is the formal privacy framework adopted for disclosure avoidance in the 2020 United States Decennial Census of Population and Housing. Officially invented in 2006, it allows for rigorous mathematical reasoning about confidentiality protections and utility of privacy-protected data.
Initially, the term \textit{differential privacy} referred to a single privacy definition \cite{dmns06} (now known as \emph{pure} differential privacy) but it has evolved into a family of related privacy definitions, each with its own parameters and interpretations.

The purpose of this paper is to introduce, to a statistically-oriented audience, the major variations of differential privacy that are particularly relevant to the disclosure avoidance system for the 2020 Census. These are pure differential privacy \cite{dmns06}, approximate differential privacy \cite{ourdata}, R\'enyi differential privacy \cite{renyidp},  zero-concentrated differential privacy \cite{zcdp}, and $f$-differential privacy \cite{gaussdp}.\footnote{For other variations of differential privacy, the interested reader can consult the survey article by Desfontaines and Pejó \cite{sokdps}.} We focus primarily on the privacy semantics of these definitions and consider several different viewpoints:
%The goal of this paper is to explain the privacy semantics of these privacy definitions and the roles played by their privacy-loss parameters because much of the discourse about privacy protections and privacy-loss parameter values is based on erroneous and incomplete understanding of advances in privacy technology since 2004. As part of this goal we show how the privacy semantics directly relate to the confidentiality protections assessed in the statistical disclosure limitation (SDL) literature.
%
%Thus we bring together results from the literature, filling in gaps with novel results as needed, to explain two major styles of semantics:
\begin{itemize}
\item \emph{Parameter interpretability}: Each privacy definition has its own set of privacy-loss parameters, such as the $\epsilon$ in pure differential privacy. Since privacy-loss parameters must be chosen by policymakers, it is important to gain a  comprehensible, minimally technical  understanding of the parameters that can be used in policy discussions. Thus, where possible, we discuss the interpretations of the different parameters. Not all parameters have minimally technical interpretations, however,  and, in particular, the $\delta$ parameter of the approximate differential privacy framework is often misinterpreted. We also propose a more interpretable version of approximate differential privacy, which we call \pbdplong.
\item \emph{Frequentist hypothesis testing}: All variations of differential privacy covered in this paper are based on a concept called the \emph{privacy-loss random variable} (PLRV). The PLRV has close connections to the likelihood-ratio test and, as a consequence, all these definitions provide protections that can be framed as frequentist hypothesis tests. If an attacker is using the privacy-protected data to perform a hypothesis test related to the confidential information supplied by an individual, the relationship between the significance level and power of \emph{any} such test is governed by the privacy-loss parameters of the variation's definitions. %semantics, which explain the information-theoretic limits of an attacker conducting a hypothesis test to determine confidential information about an individual.
\item \emph{Bayesian semantics:} There are fewer results related to Bayesian semantics owing to the mathematical complexity of analyzing Bayes' rule when differentially private randomization is combined with an attacker's prior information.
We explain the Bayesian semantics of pure differential privacy and introduce new results for R\'enyi and zero-concentrated differential privacy. These results have close connections to the \pbdplong definition we introduce in this paper. Traditional Bayesian disclosure avoidance methodology compares what can be learned about a target individual based on a data release to what would have been known without the data release (also known as a \emph{prior-to-posterior} comparison). This comparison is problematic because \emph{anything} learned from a data release (including generalizable knowledge such as smoking causes cancer) is treated as a privacy violation. The literature on differential privacy examines the more nuanced question of what can be learned about an individual based on a data release compared to what could have been learned from the data release had the individual's information been scrubbed beforehand. This comparison is known as a \emph{posterior-to-posterior} comparison and avoids the pitfalls of prior-to-posterior comparisons. Thus the Bayesian semantics we discuss are of the posterior-to-posterior variety.
\end{itemize}

When discussing these semantics, we also apply them to the production settings of the 2020 Census Disclosure Avoidance System (DAS). We show that different privacy semantics can be fine-grained and tailored to the types of inferences an attacker may try to make. This stands in contrast to the common mistaken belief that only one coarse-grained guarantee, summarized by a single number, is possible.

One might ask why this paper covers six variations of differential privacy. Isn't one enough? It turns out that if we want comprehensive privacy guarantees from the viewpoints  of parameter interpretability as well as frequentist and Bayesian statistics, then all six variations are needed. Specifically, pure $\epsilon$-differential privacy \cite{dmns06}, the ancestor of all these variations, is the starting point. It provides the strongest privacy protections and is easy to analyze from both the frequentist and Bayesian points of view. In certain cases, it may over-estimate the privacy risks that individuals face even against extremely powerful attackers. Hence, approximate $(\epsilon,\delta)$-differential privacy was introduced to better track those privacy risks. Approximate differential privacy represents a continuum of privacy definitions -- an  $(\epsilon,\delta)$-curve \cite{privacyprofiles,gaussdp} -- although many articles unfortunately consider just a single point on that curve (like trying to understand an elephant by studying a single hair on its tail). Approximate differential privacy is easy to study from the frequentist point of view \cite{kairouz15}, but Bayesian results are more limited \cite{kasiviswanathan2014semantics}. Furthermore, the $\delta$ parameter is not directly interpretable and is often misinterpreted. To address this issue, we propose  $(\epsilon,\delta)$-\pbdplong, which has more interpretable privacy-loss parameters. When it is treated as a continuum of privacy definitions (i.e., an $(\epsilon,\delta)$-curve), \pbdp is a reparametrization of $f$-differential privacy ($f$-DP) \cite{gaussdp} implying that $f$-
DP is a natural consequence of addressing the difficulties in interpreting the $\delta$ parameter of approximate differential privacy.
In addition to parameter interpretability, $f$-DP also has natural frequentist semantics because it was specifically developed to manage the trade-off between the level and power of re-identification inferences \cite{gaussdp}. Except in special cases, the privacy parameters of $f$-DP (and \pbdplong) are difficult to compute, but R\'enyi and zero-concentrated differential privacy can be used to compute upper bounds on those parameters and on the frequentist semantics of $f$-DP and \pbdplong.  Furthermore, we propose new Bayesian privacy semantics for R\'enyi and zero-concentrated differential privacy that, except in special cases, are difficult to obtain when working directly with $f$-DP and \pbdplong. Unfortunately, however, the privacy parameters of R\'enyi and zero-concentrated differential privacy are not easy to interpret. To summarize, parameter and frequentist interpretability is a strength of $f$-DP and \pbdp while Bayesian semantic interpretability and ease of parameter computation are strengths of R\'enyi and zero-concentrated DP.

We note that there is a natural trade-off between privacy and utility in a data release.  It is up to the data collector (e.g., statistical agency) to carefully balance the privacy and utility requirements of their applications. The utility side is dataset-specific, but privacy can be treated in a more general way. Thus, this paper discusses various ways in which privacy protections can be quantified in order for them to be used in privacy/utility discussions.

This paper is organized as follows.
In Section \ref{sec:informal}, we provide an accessible and informal introduction to what is and isn't considered a confidentiality breach in modern statistical disclosure limitation (SDL). Such a discussion is especially important given that many (peer-reviewed) articles about privacy and confidentiality are based on intuitive but faulty premises --  faulty in the sense that the natural conclusions of those premises are contradictory. In Section \ref{sec:notation}, we introduce the notation used in this paper. In Section \ref{sec:plrv}, we discuss the privacy-loss random variable (PLRV), which is the technical underpinning of all the privacy definitions we consider and has close connections to the likelihood ratio test. We also discuss the desirable technical properties that a ``good'' formal privacy definition should have. In Section \ref{sec:privdefs}, we explain all the privacy definitions, propose \pbdplong, and discuss the interpretability of each definition's parameters. 
Frequentist privacy semantics are discussed in Section \ref{sec:freq} and Bayesian semantics are discussed in Section \ref{sec:bayes}, including new semantics for R\'enyi and zero-concentrated differential privacy. Privacy analyses can be fine-grained, depending on the information an attacker may wish to infer about a person potentially represented in the confidential data. These fine-grained attacks, with applications to the 2020 Census, are discussed in Section \ref{sec:per}. Finally, we discuss additional related work in Section \ref{sec:related} and conclude in Section \ref{sec:conc}.

%One common thread that serves as the foundation of these privacy definitions is the concept of a \emph{privacy-loss random variable}. Thus, after an informal introduction to what a privacy breach is and isn't (a mistake routinely made in many peer-reviewed publications) in Section \ref{sec:informal}, and an explanation of our notation in Section \ref{sec:notation},  we explain in detail the concept of a privacy-loss random variable (Section \ref{sec:plrv}). Then, in Section \ref{sec:privdefs}, we present the mathematical definitions of pure, approximate, R\'enyi, and zero-concentrated differential privacy with explanations of how they related to the privacy-loss random variable ($f$-differential privacy is deferred to Section \ref{sec:freq}). We discuss the frequentist interpretations of these privacy definitions in Section \ref{sec:freq} and the Bayesian interpretations in Section \ref{sec:bayes}. In Section \ref{sec:per} we show how these ideas can be used to form more fine-grained analyses of information disclosure (we call this per-attribute semantics). We then review related work in Section \ref{sec:related} and present conclusions in Section \ref{sec:conc}.

%% file: informal.tex
Privacy and confidentiality are two different concepts. Informally, privacy relates to a person's ability to control access to information about themselves and confidentiality refers to a third-party's (e.g., a statistical agency's) ability to prevent disclosure of information it has collected. However, in the computer science literature, the word ``privacy'' is frequently used to mean ``confidentiality'' (as in the case of differential privacy) and that is also the way the term privacy will be used in this paper.

With this caveat about terminology, we note that there is a common belief that a privacy breach occurs whenever a public dataset (derived from confidential data) is used to make unwanted or potentially harmful inferences about an individual. This belief is quite wrong, as it encompasses situations that are not privacy breaches. One important case is when the unwanted inference  depends on generalizable (statistical) knowledge contained in the public dataset rather than on the specific datum that a target individual may or may not have contributed to the underlying confidential database. It is difficult to over-emphasize this distinction, and we take pains here to make it as clear as possible. 

Thus, consider the canonical example of
the first Cancer Prevention Study, also known as CPS-I \cite{hammond:horn:1954, cps1}, discussed by Dwork and Pottenger \cite{dworkpottenger}. This large-scale study, which followed a cohort of volunteers from 1952 to 1972 was the first study to conclusively establish the link between smoking and cancer.  Based on the scientific knowledge gained in this study, if you see a Millenial chain smoker, you can be virtually certain that they have a much higher chance of developing lung cancer than if they never smoked. Such an inference is definitely unwanted, as it can result in higher health and life insurance premiums. However, for people born after 1972, like our Millenial, this study cannot possibly be considered a privacy breach because their data could not have been used in the study. For such people, one would say that the inference is purely statistical in nature.

Therefore, an unwanted inference is only a privacy breach if it is specifically \emph{caused} by the inclusion of the individual's information in the dataset from which the inference was made. Consequently, an empirical privacy analysis needs to distinguish between the casual effect on  inference due to a particular individual's data being used and the general statistical information provided by the dataset on a group of other individuals. A privacy-protection methodology must work to limit the causal inference about a particular individual (to protect privacy) while leaving intact statistical inferences due to information in the data, which is basically the same as leaving the data fit for statistical purposes \cite{dpcausal}. 

To relate the causal inference interpretation of a privacy breech to concepts that have been used historically, consider the textbook definitions of identity and attribute disclosure.  In traditional SDL, an \textit{identity disclosure} occurs when ``a data subject is identified from released data'' \cite[p.~174]{duncan:etal:2011}, an \textit{attribute disclosure} occurs when ``information [is disclosed] about a population unit without (necessarily) the identification of the unit within the data set'' \cite[p.~172]{duncan:etal:2011}, and an \textit{inferential disclosure} occurs when either an identity or attribute disclosure can be inferred with high probability \cite{dalenius:1977}. The traditional SDL literature has not been careful to distinguish between identity and attribute inferences that depend upon the use of the respondent's data and those that are possible without using the individual's information. The confidentiality-breaching versions of identity, attribute and inferential disclosure all have the same mathematical formulation, which we now illustrate.

The differential privacy framework for SDL methods is designed precisely for the purpose of distinguishing between confidentiality breaches and valid scientific inferences. Typically, one reasons about causality by studying \emph{interventions} \cite{pearlcausality}, which are direct manipulations of a system. In the case of differential privacy, the intervention is to replace an individual's data record with something else and to reason about the effect of this intervention \cite{dpcausal}.

More concretely, suppose \target is a 52-year-old Hispanic Asian woman and let her survey response (e.g., to the decennial census) be denoted by $\record$. Let $\data$ be the collection of responses of everyone else, so that $\data\cup\{\record\}$ is the complete dataset (\target's record combined with everyone else's records). 
Let $\mech$ be an algorithm that applies SDL protections while producing an output $\outp$. Mathematically,  $\outp = \mech(\data\cup\{\record\})$. Both $\outp$ and the source code of the mechanism $\mech$ are then made public. One may wonder what does $\outp$ (and knowledge of $\mech$) reveal about \target, and how much of this is due to their record being part of the input to $\mech$. 

Following the causal view \cite{dpcausal}, one might ask what would have happened in an alternate world in which the data collector (e.g., the Census Bureau) replaced \target's submitted record $\record$ with some other fixed, pre-selected record $\record^*$ (e.g., a 60-year-old White Non-Hispanic male) before running $\mech$. In such a world, the output would be denoted as $\outp^*$. Mathematically, $\outp^* = \mech(\data\cup\{\record^*\})$. Clearly, the specific contents of the record $\record$ that \target submitted can have no causal effect on $\outp^*$ since $\record$ was never used. Therefore, $\outp^*$ has no causal effect on inferences about \target's recorded data. All inferences about \target based on $\outp^*$ are statistical inferences (e.g., based on the demographic composition of \target's community), not privacy-breaching inferences about how \target differs from their community.

One could then compare inference about \target in the actual world, where the output of  $\mech(\data\cup\{\record\})$ is observed, to inference about \target in her \emph{privacy-preserving counterfactual world}, in which $\mech(\data \cup \{\record^*\})$ is observed. The way this difference is measured is at the heart of the different commonly-accepted variations of differential privacy \cite{dmns06,ourdata,zcdp,renyidp,gaussdp}. 

We briefly note that the degree to which an inference is unwanted (i.e., potentially harmful to the particular individual or entity) is also an important consideration. For example, a healthy individual may have fewer concerns about revealing some medical data than an individual who is not. It is possible to add specifications about which types of inference are unwanted in formal privacy definitions (e.g, \cite{kifer14:tods,He14:sigmod,osdp,goodpdp}). However, for simplicity of the explanations, in this document we consider the case where all such inferences are considered potentially sensitive, which is consistent with the statutory framework regulating the Census Bureau and other statistical agencies in the United States  (13 U.S. Code §§ 8(b) \& 9, 44 U.S. Code §§ 3561(11) \& 3563). This also allows individuals to retroactively change their mind about which inferences they would consider harmful.

Now, comparing inferences about \target when her record is used (i.e., based on $\mech(\data\cup\{\record\})$ to when her record is not (i.e., $\mech(\data\cup\{\record^*\})$) is also tricky. To see why, suppose $\mech$ simply outputs the number of 52-year-old females in the input data and suppose, with \target's record, there would be 523 of them (i.e., $\mech(\data\cup\{\record\})=523$). In \target's privacy-preserving counterfactual world, the answer would be 522 (i.e., $\mech(\data\cup\{\record^*\})=522$). How does inference about \target's age compare to the inference in the counterfactual world? In the counterfactual world, nothing is revealed about \target's age. However, in the actual world, it is not clear what the answer 523 reveals -- it all depends on what an attacker knows, what additional information is available to the attacker, and how the attacker chooses to perform the statistical inference. There is no general agreement about reasonable choices here -- in any room of 10 economists, statisticians, social scientists, or demographers, there would be at least 23 different, and often mutually contradictory, suggestions.

Differential privacy provides a clever way of sidestepping the question of what does 523 reveal about \target compared to 522. The main idea is to force $\mech$ to be randomized: sometimes $\mech(\data\cup\{\record\})$ will produce 523, sometimes it will produce 522, and other times it will produce other numbers. Similarly, sometimes $\mech(\data\cup\{\record^*\})$ will produce 523, sometimes it will produce 522, etc. Thus instead of reasoning about 522 vs. 523, differential privacy reasons about how likely you are to see 523 if the input is $\data\cup\{\record\}$ compared to when the input is $\data\cup\{\record^*\}$. If the output distribution, when $\data\cup\{\record\}$ is the input, is exactly the same as when $\data\cup\{\record^*\}$ is the input, then clearly nothing is revealed as the result of using \target's record. If the output distributions are ``slightly'' different, then it is likely that nothing ``statistically meaningful'' is revealed as the result of using \target's record. This ``statistical meaningfulness'' is formalized as follows: given an output $\outp$, how well can a statistical procedure determine whether the input was  $\data\cup\{\record\}$ or $\data\cup\{\record^*\}$? The choice of how to measure differences between distributions (to determine whether they are ``slightly'' different) is what creates the different variations of differential privacy.

This change in perspective -- from reasoning about specific outputs to reasoning about output distributions -- allows differential privacy to have some appealing hypothesis testing and Bayesian inference interpretations that do not require any assumptions about the knowledge of an attacker. Explaining these guarantees and filling in missing pieces, such as the Bayesian semantics of concentrated differential privacy, are the goals of this paper. 

%% file: notation.tex
In this section, we summarize the notation used in this paper.

Let $\data$ be a collection of $\datasize$ records (e.g., census or sample survey responses) from individuals. For notational convenience, we assume that all the data an individual provides are collected in one record so that there is a one-to-one correspondence between records and individuals in the data. We let $\record$ denote a single record -- the data on one individual.

A function that can be computed over a dataset is called a query and denoted as $\query$. One example of a query is the total population in each county in the United States. Note that this is a vector-valued query with one component for each county. We refer to the different components of a vector-valued query using subscripts a follows: $\query(\data)_i$.

An algorithm or piece of software that applies SDL protections when producing an output is called a \emph{mechanism} and denoted by $\mech$. A mechanism can be randomized, so that the output may be different each time it is run.
We let $\outp=\mech(\data)$ denote the output of mechanism $\mech$, given $\data$ as input. Often we need to refer to a secondary analysis performed on the output of $\mech$. We use the notation $\randalg$ to denote a (possibly randomized) algorithm. Secondary analysis of the output of $\mech$ is denoted as $\randalg(\outp)$ or $\randalg(\mech(\data))$. The algorithm that applies $\mech$ to the data and then applies $\randalg$ to the output is denoted by $\randalg\circ\mech$; that is,  $(\randalg\circ\mech)(\data)\equiv \randalg(\mech(\data))$.

When an attacker is making inferences about a target individual, we refer to the target individual as \target. The dataset with \target's record removed is denoted by $\rem[\data]$.

The following symbols are reserved for the parameters of the various privacy definitions we discuss: $\epsilon,\delta,\rho,\alpha,\gamma$.
For this reason, when discussing hypothesis tests, we use the following symbols for Type I error probability/significance level ($\level$), Type II error probability ($\typetwo$), and power ($\power$).

For Bayesian analysis, we let $\prior$ denote an attacker's prior over datasets and let $\datarv$ denote the random variable corresponding to the confidential dataset, so that different values of $\datarv$ correspond to different datasets. $\dataspace$ is the set of possible datasets (universe or domain of $\datarv$) and  $\recordrv$ is the random variable corresponding to \target's record.

Finally, and importantly, all logarithms in this paper are \emph{natural} logarithms (base $e$).

%% file: plrv.tex
As explained in Section \ref{sec:informal}, the main idea underlying all major variants of differential privacy is the ability to distinguish between pairs of datasets that are formally called \emph{neighbors}. There are two commonly used versions of neighbors: \emph{bounded} and \emph{unbounded} 

\begin{definition}[Bounded Neighbors]\label{def:bneigh}
Two datasets $\data_1$ and $\data_2$ are bounded neighbors if $\data_1$ can be obtained from $\data_2$ by \emph{modifying} the record belonging to a single individual (recall that in this document, we assume all data about an individual is encapsulated in one record). In this setting, all datasets under consideration have the same size: $\datasize$ records, and $\datasize$ is public.
\end{definition}

\begin{definition}[Unbounded Neighbors]
Two datasets $\data_1$ and $\data_2$ are unbounded neighbors if $\data_1$ can be obtained from $\data_2$ by \emph{adding or removing} the record belonging to a single individual. In this case, the size of the  dataset is not public.
\end{definition}

For example, $\data_1=\data\cup\{\record\}$ and $\data_2=\data\cup\{\record^\prime\}$ are \emph{bounded} neighbors since they only differ on the record supplied by one person, say \target, and have the same number of respondents (hence the ``bounded'' terminology). On the other hand, $\data^\prime_1=\data\cup\{\record\}$ and $\data^\prime_2=\data$ are unbounded neighbors. Note that a record can contain missing values.

Unbounded neighbors represent the cleanest mathematical formulation of privacy comparisons to a counterfactual world in which an individual's data are not used (i.e., would it be difficult to figure out if the input to the privacy mechanism $\mech$ was $\data\cup\{\record\}$ or just $\data$). For this reason, most theoretical frameworks begin with unbounded neighbors. On the other hand, bounded neighbors model the situation where the \emph{existence} of a survey response can be presumed (as in a full enumeration census), but the \emph{contents} of the response must also be protected. Thus, when the population or sample size is revealed without using any randomization, one must use bounded neighbors. For the 2020 Census, the Census Bureau uses bounded neighbors to provide privacy semantics because the size of the U.S. population as of census day, 331,449,281, was published on April 26, 2021. This is the total number of person records in the confidential Census Edited File (CEF). Thus, in the remainder of this document, we focus on bounded neighbors and refer to them simply as neighbors.

The comparison in inference between pairs of neighbors is formalized by the \emph{privacy-loss random variable}, which  can be motivated as follows.
Suppose a mechanism $\mech$ was run with either $\data_1$ or $\data_2$ as the input. We observe the output $\outp$ from which we can compute $P(\mech(\data_1)=\outp)$ and $P(\mech(\data_2)=\outp)$.\footnote{For absolutely continuous distributions use the density function (Radon-Nikodym derivative) instead of the probability mass function.} Note that all of the randomness here comes from $\mech$ and not from the data.
%\footnote{For continuous distributions, these would be replaced with probability densities. In general, in this paper, we prefer to use discrete notation to avoid making the notation even more complex.} 
We now wish to use the information provided by $\outp$ to distinguish between whether $\data_1$ or $\data_2$ was the input.

This is a classical problem in statistics for which Neyman and Pearson provide the uniformly most powerful test \cite{nptest}. Treating, say, $\data_1$ as the null hypothesis and $\data_2$ as the alternative, we have a point-null, point-alternative testing framework. The choice of null or alternative is arbitrary because the definition of neighbor is symmetric. One forms the likelihood ratio test statistic\footnote{As noted in Section \ref{sec:notation}, $\log$ is the natural logarithm.} $\log\frac{P(\mech(\data_1)=\outp)}{P(\mech(\data_2)=\outp)}$   and defines a decision rule using a threshold $t$ and a ``tie-breaker'' probability $c$. These two numbers are used as follows:
\begin{itemize}
\item If $\log\frac{P(\mech(\data_1)=\omega)}{P(\mech(\data_2)=\omega)} < t$, then reject the null hypothesis.
\item If $\log\frac{P(\mech(\data_1)=\omega)}{P(\mech(\data_2)=\omega)} = t$, then reject the null hypothesis with probability $c$.
\item Otherwise, fail to reject the null hypothesis.
\end{itemize}

A privacy-loss random variable \cite{Dinur2003,DworkN04} is simply the distribution of the log-likelihood ratio test statistic  $\log\frac{P(\mech(\data_1)=\outp)}{P(\mech(\data_2)=\outp)}$ under the null hypothesis -- i.e., when $\outp$ is obtained from $\mech(\data_1)$. In other words, we can sample the privacy-loss random variable by first getting the  output $\outp=\mech(\data_1)$ and then computing $\log\frac{P(\mech(\data_1)=\outp)}{P(\mech(\data_2)=\outp)}$. 

Also note that a privacy-loss random variable can be defined for every pair of datasets $\data_1, \data_2$ that are neighbors of each other, so we use the notation $\plrv$ to refer to a specific privacy-loss random variable. To summarize:
\begin{align*}
\plrv[\data_1][\data_2][\mech] \text{ has the distribution of } \log\frac{P(\mech(\data_1)=\outp)}{P(\mech(\data_2)=\outp)} \text{ for } \outp \sim P(\mech(\data_1)).
\end{align*}
We emphasize, since it is easy to overlook, that for $\plrv[\data_1][\data_2][\mech]$, the dataset \emph{$\data_1$ is used to sample the output $\outp$ and also appears in the top of the log ratio.}

We next present a few examples of privacy-loss random variables and then show how they are used to create privacy definitions.
\subsection{Examples of Privacy-Loss Random Variables}

For simple mechanisms, the privacy-loss random variable can be computed algebraically. These include \emph{Randomized Response} \cite{warnerrr}, the Geometric Mechanism \cite{universal}, and the Gaussian Mechanism \cite{dpbook}. We also illustrate the computation of the privacy loss random variable when a mechanism releases a subset of its data.

\begin{example}[Randomized Response]\label{ex:plrv:rr}
Consider a dataset $\data=(\record_1,\dots,\record_n)$ of $n$ records (where $n$ is publicly known). Each record $\record_i$ corresponds to one individual and contains a unique identifier, a name, whether the person has cancer, and potentially other information. A target person \target may or may not be in the dataset. If \target is in the dataset and the corresponding record indicates that they have cancer, we say that \target has a cancer record in the dataset. If \target's record indicates that they do not have cancer or if \target's record is not in $\data$ then we say \target does not have a cancer record in the dataset. Alternatively, we can consider the case where \target is known to be in the dataset (either way, it does not affect what follows).
Consider the following \emph{query} function over a dataset:
$$\query(\data)=\begin{cases}1 & \text{ \target has a cancer record in $\data$}\\0 & \text{ \target does not have a cancer record in $\data$ }\end{cases}.$$
Let $\mech$ be a mechanism that implements randomized response \cite{warnerrr}. That is, given a parameter $\epsilon \geq 0$, $\mech$ flips the result of $q$  with probability $\frac{1}{1+e^{\epsilon}}$. Mathematically,
$$\mech(\data)=\begin{cases}\query(\data) & \text{ with probability }\frac{e^{\epsilon}}{1+e^\epsilon}\\1-\query(\data) & \text{ with probability }\frac{1}{1+e^\epsilon}\end{cases}.$$ 

There is a privacy-loss random variable $\plrv[\data_1][\data_2]$ for each pair\footnote{Note that \emph{all} pairs of neighboring datasets are considered, not just datasets that are neighboring to the specific dataset collected by the data curator.} of datasets $\data_1$ and $\data_2$ that are neighbors of each other -- one pair of neighbors can differ on \target, another pair can differ on Bob, and so on. Each of the privacy-loss random variables summarizes the protections available to different pieces of information, and each has one of the following forms: 
\begin{itemize}
\item If  $\data_1$ and $\data_2$ differ on $\target$ and if \target has a cancer record in either $\data_1$ or $\data_2$ (but not both) then: 
$$ \plrv[\data_1][\data_2] = \plrv[\data_2][\data_1] = \begin{cases}\epsilon & \text{ with probability }\frac{e^\epsilon}{1+e^\epsilon}\\-\epsilon & \text{ with probability }\frac{1}{1+e^\epsilon}\end{cases}.$$
This privacy-loss random variable is relevant to inferences about \target's cancer status.
\item If $\data_1$ and $\data_2$ differ on $\target$ and either (a) \target has no cancer records in both datasets or (b) has cancer records in both datasets, then: 
 $$\plrv[\data_1][\data_2]=\plrv[\data_2][\data_1]=0 \text{ with probability 1}.$$
 This privacy-loss random variable is relevant to inferences about whatever is different between $\data_1$ and $\data_2$. Since the cancer status of \target is the same in both $\data_1$ and $\data_2$, the privacy-loss random variable $\plrv[\data_1][\data_2]$ is providing information on how some other difference between $\data_1$ and $\data_2$ is being protected.
 \item Similarly, if $\data_1$ and $\data_2$ differ on a  person other than \target, then $$\plrv[\data_1][\data_2]=\plrv[\data_2][\data_1]=0 \text{ with probability 1},$$ because the differences between $\data_1$ and $\data_2$ would not affect the output distribution of $\mech$.
\end{itemize}
Taken together, the set of all privacy-loss random variables model how well different pieces of information are protected. 
As an example of the calculations used to derive this result, suppose $\data_1$ and $\data_2$ differ only on \target's record. Furthermore, suppose that $\query(\data_1)=1$ (i.e., \target's record in $\data_1$ is a cancer record) and $\query(\data_2)=0$ (\target does not have a cancer record in $\data_2$). Then for the output $\outp=1$, $P(\mech(\data_1)=1)= e^\epsilon/(1+e^\epsilon)$ and $P(\mech(\data_2)=1)=1/(1+e^\epsilon)$. The log of their ratio is $\epsilon$ and the output $\outp=1$ is produced from $\data_1$ with probability $e^\epsilon/(1+e^\epsilon)$. Thus $P(\plrv[\data_1][\data_2]=\epsilon)=e^\epsilon/(1+e^\epsilon)$. Similarly, for the output $\outp=0$, $P(\mech(\data_1)=0)= 1/(1+e^\epsilon)$ and $P(\mech(\data_2)=0)=e^\epsilon/(1+e^\epsilon)$. The log of their ratio is $-\epsilon$ and the output 0 is produced from $\data_1$ with probability $1/(1+e^\epsilon)$. Thus $P(\plrv[\data_1][\data_2]=-\epsilon)=1/(1+e^\epsilon)$.

%\begin{align*}
%\lefteqn{\text{ if Bob is in either $\data_1$ or $\data_2$ but not both:}}\\
% \plrv[\data_1][\data_2] &= \begin{cases}\epsilon & \text{ with probability }\frac{e^\epsilon}{1+e^\epsilon}\\-\epsilon & \text{ with probability }\frac{1}{1+e^\epsilon}\end{cases}\\
% \lefteqn{\text{ if Bob is in both  $\data_1$ and $\data_2$, or Bob is in neither:}}\\
% \plrv[\data_1][\data_2] &=0 \text{ with probability 1}
%\end{align*}
\end{example}

\begin{example}[Geometric Mechanism]\label{ex:plrv:geom}
The two-sided geometric random variable, with parameter $\epsilon$, is integer-valued and has the probability distribution $P(k)=\frac{1-e^{-\epsilon}}{1+e^{-\epsilon}}e^{-\epsilon|k|}$. Consider the mechanism $\mech$ that counts the number of cancer records in the dataset and adds two-sided geometric noise to the result.\footnote{Note that the noisy count may be negative and an end-user might wish to perform post-processing such as truncating negative counts or using proper statistical inference since the distribution of the added noise is known. None of these post-processing steps harm privacy, a property of differential privacy that is known as \emph{postprocessing invariance} and is discussed in Section \ref{sec:desiderata}.} This is known as the Geometric Mechanism \cite{universal}. Simple calculations show that the forms of the  privacy-loss random variables are exactly the same as in Example \ref{ex:plrv:rr}.
\end{example}

These two examples show that the privacy-loss random variables do not uniquely define the privacy mechanism $M$.

\begin{example}[Gaussian Mechanism]\label{ex:plrv:gauss}
Now consider a variation of Example \ref{ex:plrv:geom} that uses Gaussian noise instead of double geometric. That is, $\mech$ adds $N(0,\sigma^2)$ noise to the total number of people with cancer records in the data. This is called the \emph{Gaussian Mechanism} \cite{dpbook}.
There is a privacy-loss random variable $\plrv[\data_1][\data_2]$ for each pair of datasets $\data_1$ and $\data_2$ that are neighbors of each other and they have the following forms: 
\begin{itemize}
\item Without loss of generality, let Alice % NOT \target
be the person whose records differ between $\data_1$ and $\data_2$. 
If Alice % NOT \target
has a cancer record in either $\data_1$ or $\data_2$ (but not both) then 
$$ \plrv[\data_1][\data_2] \sim N\left(\frac{1}{2\sigma^2}, \frac{1}{\sigma^2}\right).$$
\item Let Alice % NOT \target
be the person whose records differ between $\data_1$ and $\data_2$. If Alice % NOT \target
has no cancer records in both datasets, then: 
 $$\plrv[\data_1][\data_2]=\plrv[\data_2][\data_1]=0 \text{ with probability 1}.$$
\end{itemize}

%\begin{align*}
% \plrv[\data_1][\data_2] &= N\left(\frac{1}{2\sigma^2}, \frac{1}{\sigma^2}\right)
%\end{align*}
\end{example}

We note that the privacy-loss random variables in Example \ref{ex:plrv:rr} and \ref{ex:plrv:geom} are bounded in absolute value by the parameter $\epsilon$ used by the mechanisms, while for Example \ref{ex:plrv:gauss}, it is only concentrated around the mean, with the mean and variance decreasing as $\sigma^2$ (the variance of the privacy noise) increases. 

\begin{example}[Composition]\label{ex:comp} Let $\mech_1$ and $\mech_2$ be two mechanisms whose sources of randomness are independent of each other. Let $\data_1$ and $\data_2$ be two neighboring datasets, and let $\mech^*$ be the mechanism that, on input $\data$, outputs both $\mech_1(\data)$ and $\mech_2(\data)$ (i.e., it releases the output of both mechanisms). It is easy to show that:
$$\plrv[\data_1][\data_2][\mech^*] = \plrv[\data_1][\data_2][\mech_1] + \plrv[\data_1][\data_2][\mech_2].$$
that is, the privacy-loss random variable of a combined data release is the sum of the individual privacy-loss random variables.
The reason is that the distribution of $\plrv[\data_1][\data_2][\mech^*]$ is the distribution of the random variable $\log\frac{P(\mech_1(\data_1)=\outp_1)P(\mech_2(\data_1)=\outp_2)}{P(\mech_1(\data_2)=\outp_1)P(\mech_2(\data_2)=\outp_2)}$ when $\outp_1$ and $\outp_2$ are sampled independently from the output distributions $P(\mech_1(\data_1))$ and $P(\mech_2(\data_2))$, respectively, which is the same as the distribution of $\log\frac{P(\mech_1(\data_1)=\outp_1)}{P(\mech_1(\data_2)=\outp_1)} + \log\frac{P(\mech_2(\data_1)=\outp_2)}{P(\mech_2(\data_2)=\outp_2)}$, when $\outp_1$ and $\outp_2$ are sampled independently from the output distributions $P(\mech_1(\data_1))$ and $P(\mech_2(\data_2))$, respectively.
%, by showing that the density function of $\plrv[\data_1][\data_2][\mech^*]$ is the convolution of the densities of $\plrv[\data_1][\data_2][\mech_1]$ and $\plrv[\data_1][\data_2][\mech_2].$ 
In general, the analysis of the combined privacy cost of several mechanisms is known as \emph{composition} \cite{gantacomposition}.
\end{example}

\begin{example}[Random Sampling]\label{ex:plrv:sample}
Consider again a dataset $\data_1=(\record_1,\dots,\record_n)$ of $n$ records, where $n$ is publicly known and where each record $\record_i$ corresponds to one individual. Suppose \target has a record in $\data_1$, say $\record_1$. Let $\data_2=(\record^\prime_1, \record_2\dots,\record_n)$ be a neighboring dataset in which there is some other record in place of \target's record, but everything else is the same. Consider a mechanism $\mech$ such that $\mech(\data)$ returns a set of $m$ records uniformly sampled (without replacement) from its input $\data$.

If $\outp$ is a set of $m$ records and $\outp\subseteq\{\record_2,\dots, \record_n\}$ then $P(\mech(\data_1)=\outp) = P(\mech(\data_2)=\outp)$. Note that the log of the ratio of the probabilities is 0 and that $\outp$ provides no information at all about \target. When $\data_1$ is the input, such an $\outp$ is produced with probability $\frac{{n-1 \choose m}}{{n \choose m}}=\frac{n-m}{n}$.

If $\record_1\in \outp$ and the rest of the records in $\outp$ are a subset of $\{\record_2,\dots, \record_n\}$ then 
$P(\mech(\data_1)=\outp)>0$ while $P(\mech(\data_2)=\outp)=0$, so the ratio of the probabilities is $\infty$. In this situation, anyone can clearly tell that the input to $\mech$ could not have been $\data_2$. When $\data_1$ is the input, such an even happens with probability $1-\frac{n-m}{n}=\frac{m}{n}$.

If $\record^\prime_1\in \outp$ and the rest of the records in $\outp$ are a subset of $\{\record_2,\dots, \record_n\}$ then 
$P(\mech(\data_1)=\outp)=0$ while $P(\mech(\data_2)=\outp)>0$. However, if $\data_1$ is the input, such an event happens with probability $0$. Putting all this together, the privacy loss random variable $\plrv$ has the distribution:
\begin{align*}
P(\plrv = 0) &= \frac{n-m}{n}\\
P(\plrv = \infty) &= \frac{m}{n}
\end{align*}
This privacy loss random variable says that the output $\outp$ either provides no information to an attacker trying to distinguish between $\data_1$ and $\data_2$ or allows the attacker to distinguish between them perfectly. The former situation happens with probability $\frac{n-m}{n}$ and the later with probability $\frac{m}{n}$.
\end{example}

\subsection{From Privacy-Loss Random Variables to Privacy Definitions and Privacy Accounting Frameworks}\label{sec:desiderata}

The set of privacy-loss random variables $\plrv[\data_1][\data_2][\mech]$ (one for each pair of neighboring datasets) associated with a mechanism $\mech$ capture the privacy properties of the mechanism. \emph{Privacy definitions} are statements of desirable properties that the mechanism $\mech$ and its privacy-loss random variables should have. 

For example, let us consider the case of \emph{pure differential privacy}, also known as $\epsilon$-differential privacy:

\begin{definition}[$\epsilon$-differential privacy/pure differential privacy]\label{def:dp} Given an $\epsilon\geq 0$, 
a mechanism $\mech$ satisfies $\epsilon$-differential privacy if for all pairs of neighbors $\data_1, \data_2$ and all (measurable) sets $S$, 
\begin{align}
P(\mech(\data_1)\in S) \leq e^\epsilon P(\mech(\data_2)\in S).\label{eq:puredp}
\end{align}
%(note that symmetry in the choice of neighbors also means that we require $P(\mech(\data_2)\in S) \leq e^\epsilon P(\mech(\data_1)\in S)$). 
Equivalently, a mechanism $\mech$ satisfies $\epsilon$-differential privacy if for any pair $(\data_1,\data_2)$ that are neighbors of each other, the corresponding privacy-loss random variable satisfies: $\plrv[\data_1][\data_2]\leq\epsilon$ with probability 1. Note that the probabilities in this definition are taken only with respect to $\mech$ and not with respect to any randomness in the data.
\end{definition}

First, we note the symmetry in the definition. If $(\data_1, \data_2)$ is a pair of neighbors, then so is $(\data_2,\data_1)$ and therefore Definition \ref{def:dp} also requires $P(\mech(\data_2)\in S)\leq e^\epsilon P(\mech(\data_1)\in S)$. This kind of symmetry holds for all the privacy definitions we discuss.

From Definition \ref{def:dp}, we see that pure differential privacy summarizes the properties of $\mech$ using a single number $\epsilon$, referred to as a \emph{privacy loss} (or privacy cost). When $\epsilon$ is close to 0, it guarantees that $P(\mech(\data_1)\in S) \approx P(\mech(\data_2)\in S)$ so that changing the content of any record barely affects the probability of any outcome. In terms of the privacy-loss random variable, it guarantees that the log-likelihood ratio test statistic is always bounded by $\epsilon$ (we cover these semantics in greater detail in Sections \ref{sec:freq} and \ref{sec:bayes}).

The general pattern in the formal privacy literature is that some property of the privacy-loss random variables is treated as a summary of the privacy loss  of a mechanism. For example, the maximum value achievable by the privacy-loss random variables is the privacy-loss $\epsilon$ in pure differential privacy. Such an assignment of cost is also known as a privacy accounting framework. However, not all properties of the privacy-loss random variables are useful measures of privacy cost. There are certain additional properties that a privacy accounting framework should also have. These are called  \emph{transparency}, \emph{post-processing invariance}, and \emph{composition}.

%In the formal privacy literature, an algorithm that operates on confidential data is called a \emph{mechanism} and is denoted by $\mech$. Mechanisms can (and often do) use random number generators, so their output can be modeled as a random variable. We let $\mech(\data)$ refer to a realized output when $\data$ is the input, and we let $\distr(\mech(\data))$ refer to the probability distribution that governs the output of $\mech$ when $\data$ is the input.

%The goal of a formal privacy accounting framework is to take a mechanism $\mech$ and assign one or more values to it that can be interpreted the amount of confidential information that it reveals -- this amount is referred to as the amount of privacy loss budget that was spent by $\mech$. A privacy loss of $0$ means that $\mech$ completely ignored its input. 

%In order to be useful, a privacy loss accounting framework must satisfy three important properties, known as \emph{transparency}, \emph{postprocessing invariance}, and \emph{composition}.

\textbf{Transparency} means that the accounting framework must assume that the attacker knows how $\mech$ works (i.e., may have access to its source code). This is a crucial property for data quality -- SDL methods that are \emph{not} transparent must hide their source code and other key details, meaning that a statistician would have no way of adjusting inferences to account for SDL perturbations. An example of a non-transparent system is data-swapping, which was used for the 2010 Census and for which exact details, such as the swap rate and likelihood of different households to be swapped, were kept confidential. On the other hand,  differential privacy and its variants support transparency because they can be translated into properties of the privacy-loss random variable which, by construction, makes use of knowledge of $\mech$ and the relations between its inputs and outputs.

Another reason transparency is important for data quality is that it lets the public audit the software used for the production of official statistics. There is strong evidence that disclosure avoidance implementation mistakes have happened in the past. For example, Alexander et al. \cite{davern} analyzed data from the American Community Survey (ACS), Current Population Survey (CPS), and the 2000 Census and found errors of up to 15\% in statistics for men and women over the age of 65. The errors were attributed by them to mistakes in the disclosure avoidance system used at the time. The errors were detectable for the 2000 Census and ACS data because the disclosure avoidance was applied using different methods for published tabular summaries and microdata \cite{abowd:schmutte:2015}. Had identical methods been used, as was the case for the CPS, significant data quality issue could have remained undetected.
More generally, if code and algorithmic details are not provided, coding mistakes can have a significant impact on data quality and be difficult, if not impossible, to catch.

\textbf{Post-processing invariance} means that secondary analysis of the output of $\mech$ is permissible (i.e., secondary analysis does not increase privacy risks). For example, if the input dataset is $\data$, the output $\mech(\data)$ could take the form of privacy-protected microdata or noisy tabulations. An analysis function $\randalg$ could be applied to these results -- for example, it could calculate the age distribution in each state, or it could combine the output of $\mech$ with additional data sources to produce a hierarchical data model. We use the notation $\randalg\circ\mech$ to denote the process of running $\mech$ on the data and $\randalg$ on the result. A privacy-accounting framework should \emph{not} claim that $\randalg(\mech(\data))$ has a higher privacy cost than $\mech(\data)$ because $\randalg$ has no direct access to $\data$, and $\randalg$ can be performed by external users once the output of $M(\data)$ is published. Simply put, post-processing invariance means that the privacy loss attributed to $\randalg\circ\mech$ should not be larger than that of $\mech$. Violations of post-processing invariance can be likened, in the financial world, to hidden costs or the practice of surprise billing.

Post-processing invariance is a fundamental and useful property because it rules out seemingly intuitive quantities as the basis for privacy-loss accounting.
%Post-processing invariance, however, is more difficult to achieve and several variations of differential privacy have been proposed without this property, and later discarded in favor of variations that do have post-processing invariance. %
%The reason for this difficulty is that post-processing does not correspond to a nice property of privacy-loss random variables.
For instance, it is well known that the tail probabilities of privacy-loss random variables are not post-processing invariant \cite{whatisdelta,meiseradp}. That is, for a fixed $t$, $P(\plrv[\data_1][\data_2][\mech_1]> t)$ may increase or decrease with postprocessing as the following example shows:
\begin{example}\label{ex:tailpost}
Let $\mech_1$ be the randomized response mechanism from Example \ref{ex:plrv:rr} (that asks if \target has cancer) and let $\mech_2$ be a probabilistic copy of $\mech_1$ (that is, $\mech_2$ is the same as $\mech_1$ except that its source of randomness is independent of $\mech_1$). Let $\mech^*$ be the mechanism that, on input $\data$, outputs the results of both mechanisms: $\mech^*(\data)=(\mech_1(\data),~ \mech_2(\data))$. From Examples  \ref{ex:plrv:rr} and \ref{ex:comp}, if \target has a cancer record in $\data_1$ but not in $\data_2$ then:
\begin{align*}
\plrv[\data_1][\data_2][\mech^*]&=
\begin{cases}
2\epsilon &\text{ with probability } \frac{e^{2\epsilon}}{(1+e^\epsilon)^2}\\
0 & \text{ with probability } 2\frac{e^\epsilon}{(1+e^\epsilon)^2}\\
-2\epsilon&\text{ with probability }\frac{1}{(1+e^\epsilon)^2}
\end{cases}
\end{align*}
Now consider two post-processing functions $\randalg_1$ and $\randalg_2$. The function $\randalg_1$ ignores its input and always outputs the number 1. The function $\randalg_2$ takes the output of $\mech^*$ and ignores the second part (so that $\randalg_2(\mech^*(\data))=\mech_1(\data)$). This results in the following privacy-loss random variables:
\begin{align*}
\plrv[\data_1][\data_2][\randalg_1\circ\mech^*] &=  0 \text{ with probability 1}
\\
\plrv[\data_1][\data_2][\randalg_2\circ\mech^*] &=  
\begin{cases}
\epsilon & \text{ with probability }\frac{e^\epsilon}{1+e^\epsilon}\\
-\epsilon & \text{ with probability }\frac{1}{1+e^\epsilon}\end{cases}
\end{align*}

From these equations,  we can see that $P(\plrv[\data_1][\data_2][\randalg_1\circ\mech^*] > \epsilon/2)=0$, which is less than  $P(\plrv[\data_1][\data_2][\mech^*] > \epsilon/2)=\frac{e^{2\epsilon}}{(1+e^\epsilon)^2}$, which is less than $P(\plrv[\data_1][\data_2][\randalg_2\circ\mech^*] > \epsilon/2) = \frac{e^{\epsilon}}{1+e^\epsilon}$ and thus different choices for post-processing can increase or decrease the tail probabilities. Note that the mechanisms and the post-processing in this example continue to satisfy Definition \ref{def:dp}. The privacy-loss random variables $\plrv[\data_1][\data_2][\mech^*]$, $\plrv[\data_1][\data_2][\randalg_1\circ\mech^*]$, and $\plrv[\data_1][\data_2][\randalg_2\circ\mech^*]$ are all bounded by $2\epsilon$ with probability 1, so the mechanisms $\mech^*, \randalg_1\circ\mech^*$, and $\randalg_2\circ\mech^*$ all satisfy $(2\epsilon)$-differential privacy.  This illustrates that pure differential privacy satisfies post-processing invariance, but the tail-areas of privacy-loss random variables do not.

\end{example}

%let $\mech_1$ be a mechanism and let $\mech_2$ be a post-processed version (i.e., $\mech_2 = f\circ \mech_1$ for some function $f$). One might expect that $\plrv[\data_1][\data_2][\mech_1]$ stochastically dominates $\plrv[\data_1][\data_2][\mech_2]$, however this is not always the case (see Appendix \ref{sec:nondominance}).  
%Thus, one must carefully choose a privacy cost function that satisfies the data processing inequality (i.e., the cost of $f\circ\mech$ is at most the cost of $\mech$).

\textbf{Composition.} 
Composition refers to the combined privacy loss due to the release of the outputs of multiple mechanisms $\mech_1,\dots, \mech_k$ and hence covers, as a special case, the privacy loss due to the release of multiple privacy-protected datasets.  For example, if $\mech_1,\dots, \mech_k$ satisfy pure differential privacy with parameters $\epsilon_1,\dots,\epsilon_k$, respectively, then their combined privacy cost is at most the sum of the individual costs: $\sum_{i=1}^k \epsilon_i$ -- in other words, the mechanism that releases the outputs of $\mech_1,\dots,\mech_k$ satisfies $(\sum_{i=1}^k\epsilon_i)$-differential privacy (although often the actual privacy parameter is even smaller than this) \cite{dmns06}. This is a desirable property because it shows that differential privacy can account for the interactions from output released using different mechanisms. Note that this composition is linear and can be likened to an actual budget of $\sum_{i=1}^k\epsilon_i$ that is allocated across different mechanisms.

On the other hand, many SDL methods including $k$-anonymity \cite{kanonymity}, cell suppression \cite[Ch.~4]{duncan:etal:2011}, and other legacy disclosure avoidance methods do not have suitable composition properties. That is,  each data release $\mech_i(\data)$ may \emph{individually} satisfy some SDL confidentially requirements, but the combination of all of the data releases can fail to satisfy any of those requirements (under any parameter settings), in many cases completely revealing confidential information such as individual records \cite{gantacomposition}. This failure of composition is similar to the idea that three linear equations on three variables can determine them uniquely even though any individual equation retains a large amount of uncertainty about their exact values. On the other hand, formal privacy methods consider future possible interaction between the outputs of different mechanisms. 

There are several important types of composition to consider.
\begin{itemize}
\item \emph{Independent composition.} Suppose $\mech_1,\dots,\mech_k$ are mechanisms whose only inputs are the data $\data$ (they are unaware of the outputs of the other mechanisms). Independent composition refers to their overall privacy cost. For example, as discussed earlier, if each $\mech_i$ satisfies $\epsilon_i$-differential privacy, then the combined release of the outputs of $\mech_1,\dots,\mech_k$ satisfies $\epsilon$-differential privacy for an $\epsilon$ that is guaranteed to satisfy $\epsilon\leq \sum_i\epsilon_i$. This is a linear composition property. Some privacy definitions support a sublinear composition property, which would be referred to as   \emph{advanced composition} \cite{dpbook,dworkboosting,kairouz15}. 
\item \emph{Adaptive Composition.}
Next, suppose the input to each $\mech_i$ is the dataset $\data$ along with the output $\outp_{i-1}$ of the previous mechanism. This allows a mechanism to adapt its behavior based on the history of the previous outputs. 
Now, for each fixed $\outp_{i-1}$, we can view $\mech_i(\outp_{i-1}, \cdot)$ to be a mechanism whose input is $\data$ (the dot  is a placeholder for where to insert the data).
Now, suppose that for each $i$, and for each $\outp_{i-1}$, the mechanism $\mech_{i}(\outp_{i-1},\cdot)$ satisfies $\epsilon_i$-differential privacy. Adaptive composition refers to the overall privacy cost of releasing those outputs $\outp_1,\dots,\outp_k$. For the case of pure differential privacy, again we have the guarantee that the privacy cost $\epsilon$ satisfies $\epsilon\leq \sum_i\epsilon_i$ \cite{dpbook}, so pure differential privacy also has adaptive composition.
\item \emph{Fully Adaptive Composition.} Next, suppose further that each mechanism can adjust its privacy cost based on the output history. That is the privacy cost of $\mech_{i}(\outp_{i-1},\cdot)$ may differ from that of $\mech_{i}(\outp^\prime_{i-1},\cdot)$ for $\outp_{i-1}\neq \outp^\prime_{i-1}$. Fully adaptive composition refers to the overall privacy cost of this scenario \cite{odometers,feldman2021individual}. It turns out that under pure differential privacy, if for all possible sequences $\outp_1,\dots,\outp_k$, the sum of the individual privacy costs is at most some value $\epsilon^*$ with probability 1, then the combined release of all of the outputs satisfies $\epsilon^*$-differential differential privacy, so that pure differential privacy satisfies fully adaptive composition \cite{odometers}. This is a highly desirable property because it allows policy makers to make their future decisions on data releases and privacy budgets based on what has already been released.
\item \emph{Convexity.} Convexity is a degenerate case of adaptive composition that is useful as an easy-to-prove sanity check for proposals of new privacy accounting frameworks. Suppose $\mech_1$ and $\mech_2$ satisfy the same privacy definition with the same privacy-loss parameters. Consider the mechanism $\mech^*$ such that $\mech^*(\data)$ outputs $\mech_1(\data)$ with probability $p$, and otherwise outputs $\mech_2(\data)$ (in other words, it randomly chooses which mechanism to run). Does $\mech^*$ also satisfy the same privacy definition with the same parameters? If yes, the privacy definition is convex. If not, it is not convex \cite{linaxioms}. This is a special case of adaptive composition in which an initial mechanism ignores the data (hence has 0 privacy cost) and outputs 0 or 1, and a subsequent mechanism uses that bit to decide whether it acts like $\mech_1$ or $\mech_2$. Note that this definition of convexity is applicable when $\mech_1$ and $\mech_2$ satisfy a privacy definition under the same parameters, it is not a statement about what happens when the privacy parameters are different.
\end{itemize}

Such composition properties of formal privacy accounting frameworks allow policymakers to treat privacy cost as a resource, controlled by something called the \emph{privacy-loss budget}.
The privacy-loss budget can be allocated across the different mechanisms $\mech_1,\dots, \mech_k$ -- just like the overall $\epsilon$ in pure differential privacy can be split into parts, with each part $\epsilon_i$ allocated to a mechanism $\mech_i$.
Furthermore, the privacy-loss budget and its allocation across different mechanisms can be turned into semantic privacy guarantees, which we explain in Sections \ref{sec:freq}, \ref{sec:bayes}, and \ref{sec:per}. But first, in the Section \ref{sec:privdefs}, we explain the major variants of differential privacy that provide the semantic guarantees.

%% file: privdefs.tex
Pure differential privacy (Definition \ref{def:dp}) was used in Section \ref{sec:plrv} to illustrate how privacy definitions can be formed from privacy-loss random variables and to explain the desirable properties of formal privacy definitions. In this section, we explain the other major variations of differential privacy used in this paper, their relation to the privacy-loss random variables, and discuss the added value they provide. We also, where possible, attempt to provide a minimally technical interpretation of the privacy parameters that can be used for policy discussions.

The starting point is the discussion of the $\epsilon$ parameter in pure differential privacy. Let us consider the situation where an attacker is trying to guess a sensitive property of a target individual \target (e.g., does \target have cancer or not) based on the output $\outp$ of a disclosure avoidance algorithm $\mech$. Even if the attacker has complete information about everyone else, and even if the attacker knows everything except \target's cancer status (so that the only question is whether the input was $\data_1$ in which $\target$ has cancer or its neighbor $\data_2$, in which \target does not have cancer), the odds that the attacker correctly guesses the cancer status of \target's record is at most\footnote{I.e., the odds would change by $\frac{P(\mech(\data_1)=1)}{P(\mech(\data_2)=1)}\leq e^\epsilon$} $e^\epsilon$ times the odds they would be correct if \target's record were replaced with a different record before being processed by $\mech$. In other words, participation in the data changes the odds of a correct guess by a factor of at most $e^\epsilon$.

%following two extreme situation (more realistic scenarios are examined in Section \ref{sec:per}): (1) an attacker is so powerful that they have complete information about every person other than \target, and are interested in determining one property of \target (e.g., whether the datasets is $\data_1$ in which \target has cancer, or $\data_2$ in which \target does not); (2) subject to the restrictions of $\epsilon$-differential privacy, a data curator wishes to reveal the cancer status of \target as accurately as possible (i.e., using the Randomized Response mechanism of Example \ref{ex:plrv:rr} that singles out \target).

%In both cases, after seeing $\outp$, the attacker can determine the odds that \target has cancer (i.e., $\frac{P(\mech(\data_1)=1)}{P(\mech(\data_2)=1)}$). $\epsilon$-Differential privacy guarantees that the odds based on $\outp$ are between $e^{-\epsilon}$ and $e^{\epsilon}$.

This is a strong guarantee, but in many cases the actual odds are significantly lower than the maximum bound.

For instance, consider the mechanism $\mech^*$ of Example \ref{ex:tailpost} which uses randomized response to output \target's cancer status (resulting in the first output $\outp_a$) followed by another randomized response on \target's cancer status (resulting in output $\outp_b$). When $\outp_a=\outp_b=1$, then the odds of an attacker guessing correctly change by a factor of $\frac{P(\mech^*(\data_1)=(1,1))}{P(\mech^*(\data_2)=(1,1))} = e^{2\epsilon}$. However, if $\outp_a\neq\outp_b$ then $\outp_a$ and $\outp_b$ contradict each other, and the odds do not change at all (e.g., $\frac{P(\mech^*(\data_1)=(1,0))}{P(\mech^*(\data_2)=(1,0))}=1$). Finally, if $\outp_a=\outp_b=0$, the odds \emph{decrease} by a factor of $e^{2\epsilon}$ (i.e., get multiplied by $e^{-2\epsilon}$) making the attacker more likely to guess incorrectly.
What we see is that the change in odds is a random variable (because it depends on $\omega$, which is random) and the actual changes in odds are often lower than the worst-case $e^{2\epsilon}$.

This discussion is captured by the privacy-loss random variable $\plrv[\data_1][\data_2][\mech^*]$ of Example \ref{ex:tailpost}, which takes the value $2\epsilon$ (corresponding to an increase in odds of $e^{2\epsilon}$) with probability $\frac{e^{2\epsilon}}{(1+e^{\epsilon})^2}$, takes the value $0$ (no change in odds) with probability $2\frac{e^\epsilon}{(1+e^\epsilon)^2}$, and $-2\epsilon$ (e.g., the odds decrease by $e^{2\epsilon}$) with the remaining probability. In other words, there is a good chance that the privacy loss random variable does not hit its maximum value.

This situation is common when $\mech^*$ is composed of many mechanisms (instead of just \emph{two}  mechanisms), in which case the chance that $\omega$ causes the worst-case change in odds can become vanishingly small.
Thus it is helpful to track the rest of the distribution of the privacy-loss random variable, and not just its extreme value.

%However, although the maximum of the privacy-loss random variable becomes $\sum_{i=1}^k\epsilon_i$ under composition, the probability of large outcomes is vanishingly small. This observation has produced relaxations of pure differential privacy that better summarize the probabilistic behavior of the privacy-loss random variable. These are presented next.

\subsection{Approximate Differential Privacy}

Approximate differential privacy \cite{ourdata}, also known as $(\epsilon,\delta)$-differential privacy, is a  relaxation of pure differential privacy. Its aim is to allow a mechanism to produce a bad outcome $\omega$ (for which the change in odds is large) as long as the probability of bad outcomes is vanishingly small (i.e., harder than winning the lottery). It is formally defined as follows:

\begin{definition}[$(\epsilon,\delta)$-differential privacy \cite{ourdata}]\label{def:dp:approx} Given privacy parameters $\epsilon\geq 0$ and $\delta\in[0,1]$,
a mechanism $\mech$ satisfies $(\epsilon,\delta)$-differential privacy if for all pairs  $\data_1, \data_2$ that are neighbors of each other and for all measurable sets $S$,
\begin{align}
P(\mech(\data_1)\in S) \leq e^\epsilon P(\mech(\data_2)\in S)+\delta. \label{eqn:adp}
\end{align}
Equivalently \cite{BalleW18}, $\mech$ satisfies $(\epsilon,\delta)$-differential privacy if
\begin{align}
P(\plrv[\data_1][\data_2] \geq \epsilon ) - e^{\epsilon}P(\plrv[\data_2][\data_1] \leq -\epsilon) \leq \delta.
\label{eqn:plrvdelta}\end{align}
This relation uses \emph{two} privacy loss random variables: $\plrv[\data_1][\data_2]$ for the first probability and its reverse $\plrv[\data_2][\data_1]$ for the second probability. % Also note that if $(\data_1,\data_2)$ is a pair of neighbors then so is $(\data_2,\data_1)$ and so Equations \ref{eqn:adp} and \ref{eqn:plrvdelta} also hold when the roles of $\data_1$ and $\data_2$ are reversed; this symmetry in $\data_1$ and $\data_2$ holds for all privacy definitions considered here.
\end{definition}

The condition in Equation \ref{eqn:adp} has an important interpretation. When $\epsilon=0$ then $\delta$ is the total variation distance between the output distributions of $\mech(\data_1)$ and $\mech(\data_2)$. Thus, approximate differential privacy is an interpolation between the differential privacy equations (when $\delta=0$) and total variation distance (when $\epsilon=0$).
These conditions on the privacy-loss random variables were discovered after approximate differential privacy was proposed. It is worth noting that Equation \ref{eqn:plrvdelta} does not use the privacy-loss random variable in an intuitive way and portends the difficulties in interpreting the $\delta$ parameter.

The $\delta$ parameter of Definition \ref{def:dp:approx} is often \emph{incorrectly} interpreted as an upper bound on the probability that a privacy-loss random variable will exceed $\epsilon$ (i.e., an upper bound on the probability that the odds of correctly guessing a sensitive piece of information will change by at least $e^\epsilon$). However, this is  known to be false \cite{whatisdelta}, as can be seen from Equation \ref{eqn:plrvdelta}.
However, $\delta$ is an upper bound on something else --  the probability that the privacy-loss random variable is infinite: $\delta \geq P(\plrv[\data_1][\data_2]=\infty)$. The privacy-loss random variable can only be infinite if there is an output $\outp$ such that $P(\mech(\data_1)=\outp)>0$ and $P(\mech(\data_2)=\outp)=0$, in which case $\log\frac{P(\mech(\data_1)=\outp)}{P(\mech(\data_2)=\outp)}=\infty$. However, Equation \ref{eqn:adp} guarantees that this output is only seen with probability at most $\delta$ when $\data_1$ is the input.

Infinite privacy loss is also known as a catastrophic failure. Two well-known $(0,\delta)$-differentially private mechanisms for which this can happen are (1) the mechanism that, with probability $\delta$, releases the entire dataset and (2) the mechanism that releases a random person's record when the input dataset has $\datasize\geq 1/\delta$ people (see Example \ref{ex:plrv:sample}). For this reason, it is often recommended that $\delta$ be a very small quantity, much smaller than $1/\datasize$ \cite{kasiviswanathan2014semantics}.

\begin{figure}
\includegraphics[scale=0.75]{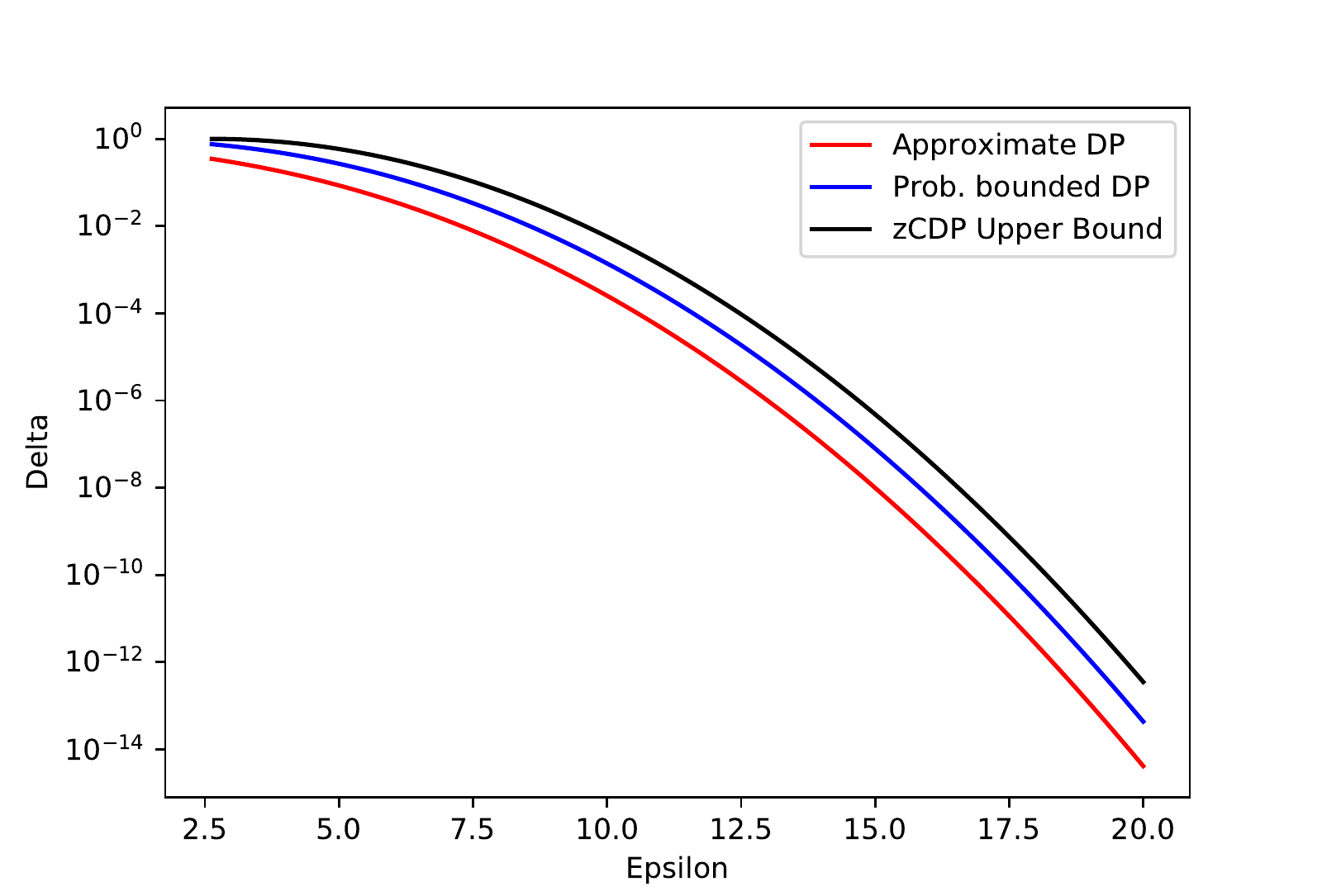}
\caption{The $(\epsilon,\delta)$ curves under Approximate Differential Privacy, \pbdplong, and the zCDP upper bound, had continuous (instead of discrete) Gaussian noise been used in the 2020 redistricting data. This is equivalent to using Gaussian noise variance $\sigma^2=1/2.63$ in Example \ref{ex:plrv:gauss}.}\label{fig:epsdeltacurve}
\end{figure}

In practice, however, most $(\epsilon,\delta)$-differentially private mechanisms cannot have infinite privacy loss, and so the catastrophic-failure interpretation of $\delta$ is not relevant. One example of a mechanism that never catastrophically fails is the Gaussian mechanism from Example \ref{ex:plrv:gauss}. This mechanism satisfies $(\epsilon,\delta)$-differential privacy for infinitely many combinations of $\epsilon$ and $\delta$. Specifically, the Gaussian mechanism satisfies $(\epsilon,\delta)$-differential privacy for every point above the red curve in Figure \ref{fig:epsdeltacurve} (we explain the other curves later in this section).

Every mechanism has its own Pareto curve of $(\epsilon,\delta)$ values and the privacy semantics  depend on the entire curve \cite{privacyprofiles,gaussdp,SommerMM19} -- a single point on the curve is not very informative. We note that as long as this curve is not bounded from below by some positive constant, then there can be no catastrophic output with infinite privacy loss. 

One important reason for considering the entire curve is composition. Using only one point on the $(\epsilon,\delta)$ curve results in weaker composition properties than those achievable when the entire curve is used.
If one uses only individual points on the curve, approximate differential privacy satisfies adaptive composition \cite{dpbook}:
if $\mech_1,\dots, \mech_k$ individually satisfy approximate differential privacy with parameters $(\epsilon_1,\delta_1),\dots,(\epsilon_k,\delta_k)$, the mechanism that releases all of their outputs satisfies $(\sum_{i=1}^k \epsilon_i, \sum_{i=1}^k \delta_i)$-differential privacy. 
This naive composition result can be substantially improved \cite{kairouz15}. However, these results are generic in the sense that they only acknowledge a single point on the $(\epsilon, \delta)$ curve of each mechanism. If the entire curve of each mechanism is used, tighter composition results are achievable. This is the motivation for R\'enyi differential privacy (RDP) \cite{renyidp}, zero-concentrated differential privacy (zCDP) \cite{zcdp}, and $f$-DP \cite{gaussdp} that we discuss later. Privacy accounting is dramatically simplified when using RDP and zCDP, and their parameters can be used to upper bound the parameters of other definitions as needed (e.g., they can generate an upper bound on the $(\epsilon,\delta)$ curve).

Given a value for $\epsilon$, one interpretation of $\delta$, due to Bun and Steinke \cite{zcdp,optcomp}, is the following. If $\data_1$ is the input to $\mech$ and there is an output $\outp$ such that $\frac{P(\mech(\data_1)=\outp)}{P(\mech(\data_2)=\outp)} \leq e^\epsilon$ then outputting $\outp$ was a good event. On the other hand, if $\frac{P(\mech(\data_1)=\outp)}{P(\mech(\data_2)=\outp)} > e^\epsilon$ then this is not necessarily a bad event. Instead, one flips a coin that has $P(\text{heads})=e^\epsilon\Big/\frac{P(\mech(\data_1)=\outp)}{P(\mech(\data_2)=\outp)}$. If that lands heads, it is a good event, and it if lands tails, then it is a bad event. In other words, the probability that observing $\outp$ is considered bad depends on by how much the ratio $\frac{P(\mech(\data_1)=\outp)}{P(\mech(\data_2)=\outp)}$ exceeds $ e^\epsilon$. Then, conditional on a bad event not happening, the probability ratio is $\leq e^\epsilon$, and the overall probability of a bad event is $\leq \delta$ \cite{zcdp,optcomp}. Given the fairly complex nature of a ``bad'' event and the conditional probabilities involved, this interpretation is difficult to explain to policy makers. Desfontaines \cite{desdelta}  gives another interpretation of $\delta$ as ``the mass of all possible bad events, weighted by how likely they are and how bad they are'' that is also fairly technical.

The main difficulty with interpreting $\delta$ is that it appears to be a parameter that was chosen for mathematical convenience. The goal of the next section is to start with a more intuitive quantity and then see where the math leads.

\input{pbdp}

%\subsection{$f$-DP and Gaussian Differential Privacy}\label{subsec:fdp}
%$f$-DP \cite{gaussdp} is a generalization of the concept of an $\epsilon,\delta$ curve. It can be formulated as follows\footnote{Dong et al. \cite{gaussdp} formulate it in a different (equivalent) way, using the language of Neyman-Pearson hypothesis testing, which we discuss in Section \ref{sec:freq}. The reformulation in Definition \ref{def:gaussdpalt} allows for more straightforward comparisons to other flavors of differential privacy.}

\subsection{Zero-Concentrated and R\'enyi Differential Privacy}\label{sec:rzcdp}
Concentrated differential privacy was first introduced by Dwork and Rothblum \cite{firstcdp} and then refined by Bun and Steinke \cite{zcdp} using R\'enyi divergences to create what is now called zero-Concentrated Differential Privacy (zCDP). R\'enyi Differential Privacy (RDP), featuring similar ideas, was developed nearly concurrently by Mironov \cite{renyidp}.

RDP and zCDP have several important properties: 
\begin{itemize}
    \item they have very straightforward adaptive composition rules, 
    \item they allow one to compute conservative privacy parameters for other definitions in situations where exact privacy parameter computation may be intractable (e.g., the Discrete Gaussian Mechanism),
    \item they are instrumental in deriving the Bayesian semantic guarantees in Section \ref{sec:bayes} (those Bayesian guarantees look remarkably similar to \pbdp).
\end{itemize}
\noindent However, their privacy parameters are not very interpretable. This leads us naturally to the definitions that follow.

\begin{definition}[$\alpha$-R\'enyi divergence] \label{def:alpha-renyi-divergence} 
The $\alpha$-R\'enyi divergence of discrete distribution $Q_1$ from discrete distribution $Q_2$, both of whose supports are contained in a set $\Omega$, is
$$\mathfrak{D}_\alpha(Q_1||Q_2)=\frac{1}{\alpha-1}\log\sum_{x\in\Omega} Q_1(x) \frac{Q_1(x)^{\alpha-1}}{Q_2(x)^{\alpha-1}},$$
with summation replaced by integration in the continuous case. Note that the definition is not symmetric for the two input distributions.
\end{definition}

\noindent $(\alpha,\gamma)$-R\'enyi Differential Privacy\footnote{In the original paper \cite{renyidp}, the privacy parameters were $\alpha$ and $\epsilon$, but to avoid confusion with the $\epsilon$ in pure differential privacy, we use the symbol $\gamma$ instead.} is simply a bound on the R\'enyi divergence for a specific $\alpha$:

\begin{definition}[$(\alpha, \gamma)$-R\'enyi Differential Privacy  \cite{renyidp}]
Given an $\alpha>1$ and $\gamma\geq 0$, a mechanism $\mech$ satisfies $(\alpha,\gamma)$-RDP if for all pairs of neighboring datasets $\data_1,\data_2$, the $\alpha$-R\'enyi divergence of the output distribution $\mech(\data_1)$ from the distribution $\mech(\data_2)$ is at most $\gamma$ (or, equivalently,$ E[e^{(\alpha-1)\plrv[\data_1][\data_2][\mech]}]\leq e^{(\alpha-1)\gamma}$).\footnote{Since $E[e^{(\alpha-1)\plrv}]= \sum_{\outp} P(\mech(\data_1)=\outp)\exp\left((\alpha-1)\log\frac{P(\mech(\data_1)=\outp)}{P(\mech(\data_2)=\outp)}\right)=\sum_{\outp} P(\mech(\data_1)=\outp)\left(\frac{P(\mech(\data_1)=\outp)}{P(\mech(\data_2)=\outp)}\right)^{\alpha-1}$ in the discrete case.} 
\end{definition}

\noindent A mechanism can satisfy RDP for many $(\alpha,\gamma)$ pairs and so, in practice, one keeps track of a set of such pairs and then converts them into more interpretable privacy parameters \cite{AbadiDPDL} (we will explain this process for \pbdp).

The idea behind zero-Concentrated Differential Privacy is that for the Gaussian mechanism $\mech$ in Example \ref{ex:plrv:gauss} with variance $\sigma^2$, the $\alpha$-R\'enyi divergence of the output distribution $\mech(\data_1)$ from the distribution $\mech(\data_2)$ equals $\frac{\alpha}{2\sigma^2}$. Therefore, the definition requires that the $\alpha$-R\'enyi divergence be proportional to $\alpha$ for all $\alpha>1$:

\begin{definition}[$\rho$-zero-Concentrated Differential Privacy \cite{zcdp}]
A mechanism $\mech$ satisfies $\rho$-zCDP if for all pairs of neighboring datasets $\data_1,\data_2$ and all $\alpha>1$, the $\alpha$-R\'enyi divergence between the output distributions of $\mech(\data_1)$ and $\mech(\data_2)$ is at most $\alpha\rho$ (or, equivalently, $E[e^{(\alpha-1)\plrv[\data_1][\data_2][\mech]}]\leq e^{\alpha(\alpha-1)\rho}$).
\end{definition}

Clearly satisfying $\rho$-zCDP is equivalent to satisfying $(\alpha, \alpha\rho)$-RDP for all $\alpha>1$.
Both $\rho$-zCDP and $(\alpha, \gamma)$-RDP are invariant under post-processing \cite{zcdp,renyidp}. Furthermore, they satisfy fully adaptive composition \cite{feldman2021individual}. That is, in the case of RDP, for a fixed $\alpha$, the $\gamma$ values of different mechanisms add up linearly. In the case of zCDP, the $\rho$ values add up linearly (just like the $\epsilon$ parameters in pure differential privacy).
The R\'enyi divergences for many different mechanisms have already been computed \cite{zcdp,renyidp}.   Both definitions provide a straightforward privacy-loss accounting framework.

%That is, if $\mech_1,\dots, \mech_k$ each satisfy $\rho_i$-zcdp (resp., $(\alpha,\gamma_i)-rdp$)  then the mechanism $\mech^*$, which releases all of their outputs, satisfies $\sum_i \rho$-zcdp (resp., $(\alpha, \sum_i \gamma_i)$-rdp). Adaptive composition continues to hold even when the input of any $\mech_i$ is the database and the output of $\mech_{i-1}$. That is, adaptive composition holds even if $\mech^*$ dynamically decides how much privacy-loss budget to allocate to $\mech_i$ based on the outputs of the previous mechanisms as long the sum of the $\rho_i$ values (resp., $\gamma_i$ values) is held constant.

RDP and zCDP can be used to obtain upper bounds on the $(\epsilon,\delta)$ curve of approximate differential privacy \cite{NEURIPS2020_b53b3a3d,optrdp}. Existing results can even be leveraged to produce upper bounds on the $(\epsilon,\delta)$-curve of \pbdp.
Specifically, we know that for any point $(\epsilon^*,\delta^*)$ on the $(\epsilon,\delta)$ curve for $\pbdp$ of a mechanism $\mech$, the value of $\delta^*$ is the smallest post-processing invariant quantity such that:
$$P(\plrv[\data_1][\data_2][\mech]> e^{\epsilon^*})\leq \delta^*.$$
Bun and Steinke (Lemma 3.5, \cite{zcdp}) proved that:
\begin{itemize}
\item if $\mech$ satisfies $(\alpha,\gamma)$-RDP with $\alpha>1$, then for $\epsilon>\gamma$, $P(\plrv[\data_1][\data_2][\mech]> e^{\epsilon}) \leq e^{(\alpha-1)(\gamma-\epsilon)}$;
\item if $\mech$ satisfies $\rho$-zCDP, then for $\epsilon>\rho$, $P(\plrv[\data_1][\data_2][\mech]> e^{\epsilon}) \leq e^{-(\epsilon-\rho)^2/(4\rho)}$.
\end{itemize}
Both of these upper bounds are functions of the privacy parameters and are post-processing invariant -- post-processing can only decrease $\rho$ and $\gamma$ (while $\alpha$ stays the same), resulting in smaller probabilities. In other words, post-processing cannot increase the estimates of the tail  probabilities. Therefore, they are both upper bounds on the $\delta$ parameter in $\pbdp$.
Thus, if we know that $\mech$ satisfies $(\alpha,\gamma)$-RDP for several different $(\alpha,\gamma)$ pairs, we can compute the upper bound using each pair and take the smallest one. Alternatively, if we know $\mech$ satisfies $\rho$-zCDP, we can use the zCDP conversion to get an upper bound on $\delta$.

Because of these results, both zCDP and RDP clearly provide a great deal of convenience for developing privacy-loss accounting frameworks, but it should be noted that some precision gets lost in the conversion to the privacy parameters of other definitions. For example, Figure \ref{fig:epsdeltacurve} shows the $(\epsilon,\delta)$ curve for \pbdp of the Gaussian mechanism (blue) and the upper bound on the curve obtained from the zCDP conversion formula (black).

%With this powerful and easy to use composition result, it is preferable to perform all privacy accounting (e.g., for this $\mech^*$) using zcdp or rdp and then to convert the $\rho$ and $\gamma$ parameters to (1) either the $\epsilon,\delta$ curve for approximate differential privacy using the conversions of Canonne et al. \cite{NEURIPS2020_b53b3a3d} or Asoodeh et al. \cite{optrdp}, or (2) to convert it to the interpretable \pbdp $\epsilon,\delta$ curve as following Example \ref{ex:renyi}: if a mechanism $\mech$ satisfies $(\alpha,\gamma)$-rdp for all $(\alpha,\gamma)$ in some set $\Theta$ then $\delta\leq \min_{\alpha,\gamma\in\Theta} e^{(\alpha-1)\gamma}e^{-(\alpha-1)\epsilon}$. In particular, this means that if $\gamma=\rho\alpha$ for all $\alpha$ (i.e., $\mech$ satisfies $\rho$-zcdp) then, using the results of \cite{zcdp} (as in Example \ref{ex:renyi}), one gets $\delta\leq e^{-(\epsilon-\rho)^2/(4\rho)}$.

%% file: pbdp.tex
\subsection{\Pbdplong and \texorpdfstring{$f$}{f}-DP}\label{subsec:pbdp}

%The difficulty of interpreting the $\delta$ parameter in approximate differential privacy is problematic when explaining privacy definitions to executive decision-makers charged with setting privacy policy. 
Executive decision-makers charged with setting privacy policy need interpretable privacy parameters to support their decisions. 
%In order to obtain parameters that are more interpretable than the $\delta$  in approximate differential
This has led us to create an alternative to approximate differential privacy called $(\epsilon,\delta)$-\pbdplong (\pbdp). To the best of our knowledge, this privacy definition has not been proposed before, but it turns out to have natural connections to and serve as good motivation for $f$-DP, zCDP, and RDP.
 Its $\delta$ parameter is much more interpretable, and the privacy definition is post-processing invariant. As with approximate differential privacy, a mechanism satisfies  $\pbdp$  for a continuum of $(\epsilon,\delta)$ values. If one acknowledges only one point on this curve, the privacy definition turns out to be not convex (see Section \ref{sec:desiderata}). However, when the entire curve is considered, not only does the privacy definition become convex, but it also becomes equivalent to a re-parametrization of $f$-DP. Thus, we view our result as a demonstration that $f$-DP is a natural consequence of addressing the interpretability issues of approximate differential privacy.
 
Hence, the ultimate goal of this section is to define \pbdp as a way of explaining and motivating $f$-DP. Later, in Section \ref{sec:rzcdp}, we discuss zCDP and RDP, which solve a computational problem -- the exact privacy parameters of \pbdp and $f$-DP are difficult to compute -- but are easy to upper bound with zCDP and RDP.

The prerequisite to explaining $\pbdp$ to policy makers is to first have them understand the meaning of the odds $\frac{P(\mech(\data_1)=\outp)}{P(\mech(\data_2)=\outp)}$ -- i.e., first they need to understand pure differential privacy.

%It is post-processing invariant, but not convex. Fixing the convexity issue leads directly to $f$-dp (in other words, fixing the interpretability of $\delta$ leads inevitably towards $f$-dp). However, computing the interpretable parameters for \pbdp is easiest when using tools provided by zcdp and rdp (but note that by themselves, the parameters of zcdp and rdp are not very interpretable). In this section we motivate \pbdp and along the way provide a new justification for $f$-dp. Then, in Section \ref{sec:rzcdp}, we discuss rdp and zcdp and how they help with the \pbdp privacy parameter computation. 

We begin by considering a flawed approach for modifying approximate differential privacy to make $\delta$ interpretable -- setting $\delta$ to be the tail probability of the privacy-loss random variables. This is the same as the probability of observing an $\outp$ for which the odds $\frac{P(\mech(\data_1)=\outp)}{P(\mech(\data_2)=\outp)}$ are too large (greater than $e^\epsilon$). Formally, suppose we require that for all pairs of neighbors $\data_1$ and $\data_2$, that $P(\plrv[\data_1][\data_2][\mech]> \epsilon)\leq \delta$. This version of $\delta$ is the upper bound on the probability that the odds of guessing a sensitive attribute correctly change by more than $e^{\epsilon}$. Another way of saying this is that such a $\delta$ is a bound on the probability that the $\epsilon$-differential privacy constraints (Equation \ref{eq:puredp}) fail to hold.\footnote{This is equivalent to something that is called probabilistic differential privacy \cite{meiseradp}.} 
%One could also consider  $P(\plrv[\data_1][\data_2][\mech]> \epsilon)\leq \delta$, if one prefers strict inequalities. 
However, as noted in \cite{meiseradp} and shown in Example \ref{ex:tailpost} the tail probability %$P(\plrv[\data_1][\data_2][\mech]\geq \epsilon)$ and 
$P(\plrv[\data_1][\data_2][\mech]> \epsilon)$ is not post-processing invariant and so this first attempt is not viable.
Post-processing invariance 
%for the strict and non-strict versions 
can be added as follows, by considering all possible post-processing functions.

\begin{definition}[\Pbdplong]\label{def:pbdpplrv}
Given privacy parameters $\epsilon> 0$ and $\delta\in[0,1]$, a mechanism $\mech$ satisfies $(\epsilon,\delta)$-\pbdp if for all (possibly randomized) post-processing functions $\randalg$ (whose domain is the range of $\mech$) and all pairs $\data_1,\data_2$ of datasets that are neighbors of each other,
%$$P(\plrv[\data_1][\data_2][\randalg\circ\mech]\geq\epsilon)\leq \delta$$
%and $\mech$ satisfies $(\epsilon,\delta)$-\pbdpg if 
$$P(\plrv[\data_1][\data_2][\randalg\circ\mech]>\epsilon)\leq \delta.$$
%(with the strict inequality involving $\epsilon$).
\end{definition}

%\begin{definition}[\Pbdplongg]\label{def:pbdpg}
%Given privacy parameters $\epsilon> 0$ and $\delta\in[0,1]$, a mechanism $\mech$ satisfies $(\epsilon,\delta)$-\pbdpg if for all (possibly randomized) post-processing functions $\randalg$ (whose domain is the range of $\mech$) and all pairs $\data_1,\data_2$ of datasets that are neighbors of each other,
%$$P(\plrv[\data_1][\data_2][\randalg\circ\mech]>\epsilon)\leq \delta$$
%\end{definition}

Definition \ref{def:pbdpplrv} is post-processing invariant by construction. In fact, Definition \ref{def:pbdpplrv}  provides the smallest $\delta$ that is both post-processing invariant and an upper bound on the privacy-loss tail probabilities. One can think of the post-processing algorithms $\randalg$ as secondary analyses of the output of $\mech$ and the definition ensures that no secondary analysis can raise the odds (of correctly guessing a piece of sensitive information) more than $e^{\epsilon}$, except with probability bounded by $\delta$. %that the probability that any of them produce a high privacy loss is bounded be $\delta$. 
%Thus this $\delta$ parameter is interpretable as a post-processing invariant tail bound on the privacy loss. 
However, even though this $\delta$ now has a very concrete interpretation,
Definition \ref{def:pbdpplrv} is not easy to work with because it is defined directly in terms of privacy-loss random variables. One can perform a simplification that removes the reference to the privacy-loss random variable:

\begin{definition}[\Pbdplong,  version 2]\label{def:pbdpv2}
Given privacy parameters $\epsilon> 0$ and $\delta\in[0,1]$, a mechanism $\mech$ satisfies $(\epsilon,\delta)$-\pbdp if for all (possibly randomized) post-processing functions $\randalg$ whose domain is the range of $\mech$ and whose output is $0$ or $1$, and for all pairs $\data_1,\data_2$ of datasets that are neighbors of each other,
%$$\Big(P(\randalg(\mech(\data_1))=1)\geq e^\epsilon P(\randalg(\mech(\data_2))=1) \Big)~ \Rightarrow~ \Big(P(\randalg(\mech(\data_1))=1)< \delta\Big)$$
%$\mech$ satisfies $(\epsilon,\delta)$-\pbdpg if 
$$\Big(P(\randalg(\mech(\data_1))=1)> e^\epsilon P(\randalg(\mech(\data_2))=1) \Big)~ \Rightarrow~ \Big(P(\randalg(\mech(\data_1))=1)\leq \delta\Big).$$
\end{definition}

%\begin{definition}[\Pbdplongg,  version 2]\label{def:pbdpgv2}
%Given privacy parameters $\epsilon> 0$ and $\delta\in[0,1]$, a mechanism $\mech$ satisfies $(\epsilon,\delta)$-\pbdpg if for all (possibly randomized) post-processing functions $\randalg$ whose domain is the range of $\mech$ and whose output is $0$ or $1$, and for all pairs $\data_1,\data_2$ of datasets that are neighbors of each other,
%$$\Big(P(\randalg(\mech(\data_1))=1)> e^\epsilon P(\randalg(\mech(\data_2))=1) \Big)~ \Rightarrow~ \Big(P(\randalg(\mech(\data_1))=1)\leq \delta\Big)$$
%\end{definition}

\begin{theoremEnd}[category=privdefs]{theorem}\label{thm:pbdpfirst}
Definitions \ref{def:pbdpplrv} and \ref{def:pbdpv2} are equivalent for $\delta\in[0,1]$. 
\end{theoremEnd}
\begin{proofEnd}
First, we note that when $\delta=0$, both definitions are equivalent to pure $\epsilon$-differential privacy.

Let $\randalg$ be a post-processing algorithm as in Definition \ref{def:pbdpplrv}. Define $\randalg^*$ to be the algorithm such that $\randalg^*(\outp)=1$ if $P(\randalg(\mech(\data_1))=\outp)> e^\epsilon P(\randalg(\mech(\data_2))=\outp)$  and $\randalg^*(\outp)=0$ otherwise (for continuous distributions interpret these as  Radon-Nikodym densities). Then, by definition of the privacy loss random variables,
$P(\plrv[\data_1][\data_2][\randalg\circ\mech]> e^\epsilon)=P(\plrv[\data_1][\data_2][\randalg^*\circ\mech]> e^\epsilon)$.
So we can restrict Definition \ref{def:pbdpplrv} to binary-valued post-processing functions.

Also note that if $\randalg$ is a binary-valued algorithm, then considering both $\randalg$ and $1-\randalg$, we see that Definition \ref{def:pbdpv2} does not change if we require both
\begin{align*}
\Big(P(\randalg(\mech(\data_1))=1)&> e^\epsilon P(\randalg(\mech(\data_2))=1) \Big)~ \Rightarrow~ \Big(P(\randalg(\mech(\data_1))=1)\leq \delta\Big)\\
\Big(P(\randalg(\mech(\data_1))=0)&> e^\epsilon P(\randalg(\mech(\data_2))=0) \Big)~ \Rightarrow~ \Big(P(\randalg(\mech(\data_1))=0)\leq \delta\Big).
\end{align*}

Let $\randalg$ be a binary-valued post-processing algorithm, and let $\data_1$ and $\data_2$ be neighbors. Consider the following cases:
\begin{itemize}
\item If $P(\randalg(\mech(\data_1))=1)\geq e^\epsilon P(\randalg(\mech(\data_2))=1)$ and $P(\randalg(\mech(\data_1))=0)\geq e^\epsilon P(\randalg(\mech(\data_2))=0)$ or if  $P(\randalg(\mech(\data_1))=1)\leq e^\epsilon P(\randalg(\mech(\data_2))=1)$ and $P(\randalg(\mech(\data_1))=0)\leq e^\epsilon P(\randalg(\mech(\data_2))=0)$ then, since $\randalg$ either outputs 0 or 1,  the probabilities do not depend on whether the input was $\data_1$ or $\data_2$ (e.g., $P(\randalg(\mech(\data_2))=1)=P(\randalg(\mech(\data_1))=1)$), $\epsilon$ must therefore be $0$, and the privacy loss is $0$, so both definitions are vacuously satisfied (since they are active for $\epsilon>0$).
\item If $P(\randalg(\mech(\data_1))=1)> e^\epsilon P(\randalg(\mech(\data_2))=1)$ but $P(\randalg(\mech(\data_1))=0) \leq e^\epsilon P(\randalg(\mech(\data_2))=0)$
then,  by definition of the privacy-loss random variable,  $P(\plrv[\data_1][\data_2][\randalg\circ\mech]> \epsilon)=P(\randalg(\mech(\data_1))=1)$ since 1 is the only output under $\data_1$ with sufficient privacy loss. Thus, either
both $P(\plrv[\data_1][\data_2][\randalg\circ\mech]> \epsilon)\leq\delta$ and $P(\randalg(\mech(\data_1))=1)\leq \delta$ or both $P(\plrv[\data_1][\data_2][\randalg\circ\mech]> \epsilon)>\delta$ and $P(\randalg(\mech(\data_1))=1)> \delta$. Therefore, with this setting  of $\randalg, \data_1, \data_2$  the constraints of Definitions \ref{def:pbdpv2} and \ref{def:pbdpplrv} are either both satisfied or both not satisfied.
\item Similarly, if $P(\randalg(\mech(\data_1))=1)\leq e^\epsilon P(\randalg(\mech(\data_2))=1)$ but $P(\randalg(\mech(\data_1))=0) > e^\epsilon P(\randalg(\mech(\data_2))=0)$
then, by definition of the privacy-loss random variable,  $P(\plrv[\data_1][\data_2][\randalg\circ\mech]> \epsilon)=P(\randalg(\mech(\data_1))=0)$ since 0 is the only output under $\data_1$ with sufficient privacy loss. Thus  either
both $P(\plrv[\data_1][\data_2][\randalg\circ\mech]> \epsilon)\leq\delta$ and $P(\randalg(\mech(\data_1))=0)\leq \delta$ or both $P(\plrv[\data_1][\data_2][\randalg\circ\mech]> \epsilon)>\delta$ and $P(\randalg(\mech(\data_1))=0)> \delta$. Therefore, with this setting  of $\randalg, \data_1, \data_2$  the constraints of Definitions \ref{def:pbdpv2} and \ref{def:pbdpplrv} are either both satisfied or both not satisfied.
\end{itemize}
Thus the two definitions are equivalent. 
\end{proofEnd}

The binary-valued post-processing algorithms $\randalg$ in Definition \ref{def:pbdpv2} can be thought of as attack algorithms (does the algorithm predict \target has some property or not), or, equivalently, as a test between two hypotheses ($H_0:$ \target has the property vs. $H_1:$ \target does not).

The next simplification is the easiest to work with:

\begin{definition}[\Pbdplong, version 3]\label{def:pbdpv3}
Given privacy parameters $\epsilon> 0$ and $\delta\in(0,1]$, a mechanism $\mech$ satisfies $(\epsilon,\delta)$-\pbdp if for all (possibly randomized) post-processing functions $\randalg$ whose domain is the range of $\mech$ and whose output is $0$ or $1$, and for all pairs $\data_1,\data_2$ of datasets that are neighbors of each other,
%$$\Big(P(\randalg(\mech(\data_1))=1)\leq e^{-\epsilon}\delta \Big)~ \Rightarrow~ \Big(P(\randalg(\mech(\data_2))=1)< \delta\Big)$$
%and $\mech$ satisfies $(\epsilon,\delta)$-\pbdpg if
$$\Big(P(\randalg(\mech(\data_1))=1)\leq e^{-\epsilon}\delta \Big)~ \Rightarrow~ \Big(P(\randalg(\mech(\data_2))=1)\leq \delta\Big).$$
\end{definition}
%\noindent As we shall soon see, if one specifies an entire $(\epsilon,\delta)$ curve for Definition \ref{def:pbdpv3}, and replaces $\leq\delta$ with $<\delta$, then one obtains $f$-dp. But first, we show that this formulation is equivalent to the previous formulations of \pbdp.

\begin{remark} Note that Definition \ref{def:pbdpv3} excludes the case of $\delta=0$ while the others do not. The first reason is that setting $\delta=0$ would be equivalent to saying that if $\mech(\data_1)$ cannot produce a certain output, then neither can $\mech(\data_2)$ and vice versa. This is clearly not equivalent to $\epsilon$-differential privacy. However, it can be shown (see the Appendix) that if a mechanism $\mech$ satisfies Definition \ref{def:pbdpv3} for a fixed $\epsilon$ and all $\delta>0$, then it satisfies pure $\epsilon$-differential privacy and vice versa. 
\end{remark}

\begin{theoremEnd}[category=privdefs, all end]{theorem}\label{thm:pbdpcont}
Given an $\epsilon>0$, a mechanism $\mech$ satisfies pure $\epsilon$-differential privacy if and only if it satisfies Definition \ref{def:pbdpv3} for all $\delta>0$.
\end{theoremEnd}
\begin{proofEnd}
If $\mech$ satisfies pure differential privacy, then by postprocessing invariance, if $e^{-\epsilon}\delta \geq P(\randalg(\mech(\data_1))=1)$, we have:
\begin{align*}
e^{-\epsilon}\delta &\geq P(\randalg(\mech(\data_1))=1)\geq e^{-\epsilon}P(\randalg(\mech(\data_2))=1)\\
&\text{ and so } \delta \geq P(\randalg(\mech(\data_2))=1)
\end{align*}
for any $\delta\geq 0$.

For the other direction suppose $\mech$ satisfies Definition \ref{def:pbdpv3} for the given $\epsilon$ and all $\delta>0$. If $\mech$ does not satisfy $\epsilon$ differential privacy then there exists an $\randalg$ and neighboring $\data_1,\data_2$ such that $P(\randalg(\mech(\data_2))=1) > e^{\epsilon} P(\randalg(\mech(\data_1))=1)$. Setting $\delta=e^{\epsilon}P(\randalg(\mech(\data_1))=1)$, we get
\begin{align*}
P(\randalg(\mech(\data_1))=1) &\leq e^{-\epsilon}\delta \quad\text{ by choice of $\delta$}\\
P(\randalg(\mech(\data_2))=1) &\leq \delta\quad\text{ since $\mech$ satisfies Def \ref{def:pbdpv3} for this $\epsilon,\delta$ combination}\\
&=e^{\epsilon}P(\randalg(\mech(\data_1))=1)
\end{align*}
contradicting the assumption that $\mech$ does not satisfy pure $\epsilon$-differential privacy.
\end{proofEnd}

\begin{theoremEnd}[category=privdefs]{theorem}\label{thm:pbdpsecond}
Definitions \ref{def:pbdpv2} and Definition \ref{def:pbdpv3} are equivalent for $\delta\in(0,1]$.
\end{theoremEnd}
\begin{proofEnd}
%We first prove it for \pbdpg.
We note that, due to symmetry of neighbors, if $\data_1$ is a neighbor of $\data_2$ then $\data_2$ is a neighbor of $\data_1$, thus, the condition in Definition \ref{def:pbdpv2} can be replaced with:
$$\Big(P(\randalg(\mech(\data_2))=1)> e^\epsilon P(\randalg(\mech(\data_1))=1) \Big)~ \Rightarrow~ \Big(P(\randalg(\mech(\data_2))=1)\leq \delta\Big).$$

If there exists a $\mech$ that does not satisfy Definition \ref{def:pbdpv3}, then there exist a binary-valued post-processing algorithm $\randalg$, neighbors $\data_1$ and $\data_2$ such that $P(\randalg(\mech(\data_1))=1)\leq e^{-\epsilon}\delta$ but $P(\randalg(\mech(\data_2))=1)>\delta$. In this case, $P(\randalg(\mech(\data_2))=1)> e^\epsilon P(\randalg(\mech(\data_1))=1) $ but $P(\randalg(\mech(\data_2))=1)>\delta$ and so $\mech$ does not satisfy Definition \ref{def:pbdpv2}. Hence Definition \ref{def:pbdpv2} implies Definition \ref{def:pbdpv3}.

For the other direction, suppose $\mech$ does not satisfy Definition \ref{def:pbdpv2}. Then there exists a binary-valued post-processing algorithm $\randalg$, neighbors $\data_1$ and $\data_2$ such that $P(\randalg(\mech(\data_2))=1)> e^\epsilon P(\randalg(\mech(\data_1))=1)$ and $P(\randalg(\mech(\data_2))=1)>\delta$.
Suppose, for the sake of contradiction, that $\mech$ satisfies Definition \ref{def:pbdpv3}. Since $P(\randalg(\mech(\data_2))=1)>\delta$ this would mean that $P(\randalg(\mech(\data_1))=1)> e^{-\epsilon}\delta$. Define $\gamma=\frac{e^{-\epsilon}\delta}{P(\randalg(\mech(\data_1))=1)}$ and note that $\gamma<1$. Define the post-processing function $\randalg^*$ such that $\randalg^*(\outp)$ returns $\randalg(\outp)$ with probability $\gamma$ and otherwise returns $0$. Then $P(\randalg^*(\mech(\data_1))=1)=\gamma P(\randalg(\mech(\data_1))=1)= e^{-\epsilon}\delta$. Since we assumed (by way of contradiction) that $\mech$ satisfies Definition \ref{def:pbdpv3}, we must have:
\begin{align*}
\delta &\geq P(\randalg^*(\mech(\data_2))=1)\\
&=\gamma P(\randalg(\mech(\data_2))=1)\\
&= \frac{e^{-\epsilon}\delta P(\randalg(\mech(\data_2))=1)}{P(\randalg(\mech(\data_1))=1)}\\
&\Rightarrow \frac{ P(\randalg(\mech(\data_2))=1)}{P(\randalg(\mech(\data_1))=1)}\leq e^\epsilon,
\end{align*}
which contradicts the starting requirements on $\mech$ and $\randalg$ that $P(\randalg(\mech(\data_2))=1)> e^\epsilon P(\randalg(\mech(\data_1))=1)$.  Thus, if $\mech$ does not satisfy Definition \ref{def:pbdpv2} then it cannot satisfy Definition \ref{def:pbdpv3} either. Hence, Definition \ref{def:pbdpv3} implies Definition \ref{def:pbdpv2}.
\end{proofEnd}
Even though the form of Definition \ref{def:pbdpv3} is now very different from the original motivation, the chain of equivalences maintains the original post-processing-invariant interpretations of $\epsilon$ and $\delta$ that, with probability at most $\delta$, \pbdp $\mech$ produces a bad output $\outp$ (i.e., one that changes the odds of correctly guessing sensitive information by more than $e^{\epsilon}$).
Nevertheless, there is still one drawback. Namely, if one only considers a single $(\epsilon,\delta)$ pair instead of the entire curve, then, as the following example shows, the privacy definition is not convex.

\begin{example}
Let $\mech_1$ be a variation of the randomized response mechanism such that:
\begin{align*}
P(\mech_1(\data_1)=2) &= \frac{e}{1+e}\\
P(\mech_1(\data_1)=-2) &= \frac{1}{1+e}\\
P(\mech_1(\data_2)=2) &= \frac{1}{1+e}\\
P(\mech_1(\data_2)=-2) &= \frac{e}{1+e}\\
\end{align*}
Let $\mech_2$ be the mechanism such that:
\begin{align*}
P(\mech_2(\data_1) = 1) &= 0\\
P(\mech_2(\data_1) = 0) &= 1\\
P(\mech_2(\data_2) = 1) &= 0.01\\
P(\mech_2(\data_2) = 0) &= 0.99\\
\end{align*}
First we claim and show that $\mech_1$ and $\mech_2$ both satisfy $(\epsilon=1.01,\delta=0.02)$-\pbdp.

$\mech_1$ satisfies $(\epsilon=1.01,\delta=0.02)$-\pbdp because (1) it satisfies $1$-differential privacy and so does $\randalg\circ\mech_1$ for any (binary-valued) postprocessing algorithm $\randalg$, (2) this means $P(\randalg(\mech_1(\data_1))=1) \geq e^{-1}P(\randalg(\mech_1(\data_2))=1)$, so (3) if $P(\randalg(\mech_1(\data_1))=1)\leq e^{-\epsilon}\delta$ then $e^{-1}P(\randalg(\mech_1(\data_2))=1)\leq e^{-\epsilon}\delta$ and so $P(\randalg(\mech_1(\data_2))=1)\leq\delta$ for $\epsilon=1.01$. The same happens with the roles of $\data_1$ and $\data_2$ reversed.

In the case of $\mech_2$, it also satisfies $(\epsilon=1.01,\delta=0.02)$-\pbdp. To see why, first we note that a binary-valued randomized algorithm $\randalg$ here would be characterized by two numbers $x\equiv P(\randalg(1)=1)$ and $y \equiv P(\randalg(0)=1)$.
Then $P(\randalg(\mech_2(\data_1))=1)=y$ and $P(\randalg(\mech_2(\data_2))=1)=0.99y + 0.01x$, so if  $y=P(\randalg(\mech_2(\data_1))=1)\leq e^{-\epsilon}\delta$ then  $P(\randalg(\mech_2(\data_2))=1)\leq 0.99 e^{-\epsilon}\delta + 0.01x \leq \delta$ when $\epsilon=1.01$ and $\delta=0.02$ (no matter what $x$ is). Furthermore,  $P(\randalg(\mech_2(\data_2))=1)\geq 0.99 P(\randalg(\mech_2(\data_1))=1)$ so if $P(\randalg(\mech_2(\data_2))=1)\leq e^{-\epsilon}\delta$ then $P(\randalg(\mech_2(\data_1))=1)\leq e^{-\epsilon}\delta/0.99 \leq \delta$ for our choice of $\epsilon$ and $\delta$. Hence $\mech_2$ also satisfies the privacy definition with these parameters.

Now, consider algorithm $\mech^*$ that runs $\mech_1$ with probability $1/2$ and otherwise runs $\mech_2$. Consider the postprocessing algorithm $\randalg$ such that $\randalg(1)=1$ with probability 1, $\randalg(-2)=1$ with probability $0.05$ and in all other settings, $\randalg$ outputs $0$. Then:
\begin{align*}
P(\randalg(\mech^*(\data_1))=1) &=
\frac{1}{2}P(\mech_1(\data_1)=-2)P(\randalg(-2)=1)    +    \frac{1}{2}P(\mech_2(\data_1)=1)P(\randalg(1)=1)\\
&= \frac{1}{2}*\frac{1}{1+e}* 0.05 < 0.0068 < e^{-1.01}*0.02=e^{-\epsilon}\delta\\
P(\randalg(\mech^*(\data_2))=1) &=
\frac{1}{2}P(\mech_1(\data_2)=-2)P(\randalg(-2)=1)    +    \frac{1}{2}P(\mech_2(\data_2)=1)P(\randalg(1)=1)\\
&= \frac{1}{2}*\frac{e}{1+e}* 0.05 
+ \frac{1}{2}*0.01 > 0.023 > \delta
\end{align*}
and so $\mech^*$ does not satisfy $(\epsilon=1.01,\delta=0.02)$-\pbdp.
\end{example}

Convexity of the privacy definition can be fixed by considering a curve, which we can do by introducing a function $f$. Then the conditions of Definition \ref{def:pbdpv3} can be turned into the conditions:
$$\forall a:~P(\randalg(\mech*(\data_1))=1) \leq a \Rightarrow P(\randalg(\mech*(\data_2))=1) \leq 1-f(a)$$
Note that $a \leq 1-f(a)$ since that also occurs in Definition \ref{def:pbdpv3} because $e^{-\epsilon}\delta \leq \delta$. This type of generic constraint was first studied by Kifer and Lin \cite[Theorem 2.1.3 ]{Kifer_Lin_2012} who showed that for postprocessing invariance and convexity, a necessary condition on $1-f$ is that it is concave and non-decreasing (so $f$ is convex and non-increasing). Throwing in continuity at $0$ for $f$ (continuity elsewhere is guaranteed by convexity and monotonicity), 
%
%It turns out that requiring $\mech$ to satisfy \pbdp for a curve of $(\epsilon,\delta)$ values results in a privacy definition that satisfies convexity, and is  
this becomes equivalent to $f$-DP, recently proposed by Dong et al. \cite{gaussdp}, which can be defined as:

\begin{definition}[$f$-DP \cite{gaussdp}]\label{def:gaussdpalt}
Let $f: [0, 1]\rightarrow[0,1]$ be a continuous, convex, non-increasing function such that $f(x)\leq 1-x$. A mechanism $\mech$ satisfies $f$-DP if for pairs of neighboring datasets $\data_1, \data_2$ and all binary-valued post-processing algorithms $\randalg$ (whose domain contains the range of $\mech$), 
\begin{align*}
1-f(P(\randalg(\mech(\data_1))=1)) &\geq P(\randalg(\mech(\data_2))=1). 
\end{align*}
\end{definition}

The privacy parameter for $f$-DP is this function $f$ (selecting it is an area of current research \cite{gaussdp}). It is known as a trade-off function \cite{gaussdp}, a name that comes from the interpretation of $\randalg$ as a hypothesis test of $H_0:$ the input is $\data_1$ vs. $H_1:$ the input is $\data_2$, performed on the output of $\mech$. Here $\randalg$ outputs $1$ to reject the null hypothesis and $0$ to fail to reject. Under this interpretation, the expression  $P(\randalg(\mech(\data_1))=1) $ is the probability of a Type I  error (significance level) and $P(\randalg(\mech(\data_2))=1)$ is the power of the test. The function $f$ provides the trade-off of the maximum  power one can achieve for a given significance level. 

From such a function $f$, we can recover the $\epsilon$ and $\delta$ parameters as follows. For a chosen $\delta$, the corresponding $\epsilon$ is $\log\frac{\delta}{f^{-1}(1-\delta)}$, where, to account for the possibility that $f$ is not one-to-one, $f^{-1}$ is defined\footnote{Essentially, recalling that $f$ is non-increasing, $f^{-1}(z)$ selects the smallest pre-image of $z$} as in \cite{gaussdp} (i.e., $f^{-1}(z) = \inf\{y\in[0,1]~:~f(y)\leq z\}$).  The continuity and non-increasing property of $f$ implies that $f(f^{-1}(z))\geq z$ with equality when $f$ is invertible.

The $(\epsilon,\delta)$ parameter conversion works because $P(\randalg(\mech(\data_1))=1)\leq e^{-\epsilon}\delta\equiv f^{-1}(1-\delta)$ implies $P(\randalg(\mech(\data_2))=1) \leq 1-f(P(\randalg(\mech(\data_1))=1)) \leq  1-f(e^{-\epsilon}\delta)=1-f(f^{-1}(1-\delta))\leq\delta$. This $(\epsilon,\delta)$ curve also has a Bayesian interpretation that we discuss in Section \ref{sec:bayes}.

\begin{remark}
$f$-DP is postprocessing invariant and the adaptive composition results of Dong et al.  \ref{def:gaussdpalt} show that it is also convex. Hence, aside from continuity at $0$, the conditions on $f$ are both necessary and sufficient (this also follows from \cite[Theorem 2.1.4 ]{Kifer_Lin_2012}). The function $f$ can even be replaced with the function $\max(f, f^{-1})$ without altering the privacy definition \cite{gaussdp}. Furthermore, Dong et al. \cite{gaussdp} showed that for every $\data_1,\data_2$ there exists a mechanism $\mech$ and a worst-case post-processing algorithm $\randalg$ for which the inequality in Definition \ref{def:gaussdpalt} is satisfied with equality.
%Finally, the adaptive composition results of \cite{gaussdp} imply that the privacy definition is convex.
\end{remark}

\begin{remark}
Following an observation of Desfontaines and Pejó \cite{sokdps}, one can use the convexity of $f$ to eliminate $\randalg$ from Definition \ref{def:gaussdpalt} and require, for all measurable sets $S$, $1-f(P(\mech(\data_1))\in S) \geq P(\mech(\data_2)\in S)$. The definition then becomes very similar to a previously introduced  convex, post-processing invariant, abstract (and harder to interpret) version differential privacy \cite{Kifer_Lin_2012}.
\end{remark}

It is interesting to note that the chosen direction for resolving an interpretability issue with approximate differential privacy has   $f$-DP as its consequence.
%is  a natural consequence of making the $\delta$ in approximate DP interpretable.

\subsubsection{Computing the privacy parameters}
Computing the $(\epsilon,\delta)$-\pbdp parameters of a mechanism $\mech$ (or, equivalently, computing the trade-off function $f$) is not always easy. Here we discuss how to do this for the Gaussian Mechanism and in Section \ref{sec:rzcdp} we explain how to use RDP and zCDP to obtain a conservative bound for more complex mechanisms.

\begin{example}[Gaussian Mechanism and Gaussian Differential Privacy \cite{gaussdp}]\label{ex:gaussmechdp}
The Gaussian Mechanism \cite{dpbook,BalleW18,XiaoDWZK21,gaussdp} can be applied more generally than  was done in Example \ref{ex:plrv:gauss}.
Let $\query$ be a function whose input is a dataset and whose output is a vector (i.e., a collection of query answers). Suppose multivariate Gaussian noise with a \emph{diagonal} covariance matrix $\Sigma$ is added to the query answers. That is, the mechanism is $\mech(\data)=\query(\data) + N(\vec{0}, \Sigma)$. This gives us a noisy vector as the output. The privacy loss random variable  for component $i$ of this vector, denoted \plrv[\data_1][\data_2][\query_i], can be shown to have the distribution $N(\frac{(\query(\data_1)_i - \query(\data_2)_i)^2}{2\sigma^2_i}, ~\frac{(\query(\data_1)_i - \query(\data_2)_i)^2}{\sigma^2_i})$, where the $\sigma^2_i$ are the diagonal elements of $\Sigma$. By composing the privacy-loss random variables (see Example \ref{ex:comp}), the overall privacy-loss random variable has the univariate Gaussian 
%\todo{No reason to call it Normal here.} 
distribution $N(\frac{\mu^2_{1,2}}{2}, \mu^2_{1,2})$, where $\mu^2_{1,2}=(q(\data_1)-q(\data_2))^T\Sigma^{-1}(q(\data_1)-q(\data_2))$. Let $\mu$ be the supremum of $\mu_{1,2}$, taken over all pairs of neighboring datasets $\data_1$,$\data_2$. Then:
\begin{itemize}
\item The exact $(\epsilon,\delta)$-differential privacy parameters can be computed as: $\delta=\Phi\left(-\frac{\epsilon}{\mu}+\frac{\mu}{2}\right)-e^\epsilon\Phi\left(-\frac{\epsilon}{\mu}-\frac{\mu}{2}\right)$ \cite{BalleW18,gaussdp}, where $\Phi$ is the CDF of the standard normal distribution.
\item The trade-off function $f$ is $f(x)=\Phi(\Phi^{-1}(1-x)-\mu)$. When such a trade-off function is used with $f$-DP, one calls it \emph{Gaussian Differential Privacy}.
\item The $(\epsilon,\delta)$-\pbdp  parameters can then be computed as: $$\epsilon=\log\frac{\delta}{f^{-1}(1-\delta)}=\log\frac{\delta}{1-\Phi(\Phi^{-1}(1-\delta)+\mu)}=\log\frac{\delta}{\Phi(-\Phi^{-1}(1-\delta)-\mu)}.$$
\end{itemize}
This result can be extended to covariance matrices $\Sigma$ that are not diagonal by using standard techniques that diagonalize general Gaussian distributions (see, for example, \cite{XiaoDWZK21}).
\end{example}

For other mechanisms, such as the Discrete Gaussian Mechanism used by the TopDown Algorithm in the 2020 Census Disclosure Avoidance System \cite{tdahdsr}, it is very difficult to compute the exact $(\epsilon,\delta)$ curves. Instead, we use zCDP accounting to conservatively approximate (upper-bound) the privacy parameters for \pbdp (see Section \ref{sec:rzcdp}). 

For the purposes of comparison,
consider the hypothetical situation in which the (continuous) Gaussian Mechanism  is used in the production settings of the 2020 redistricting data. The $\mu$ parameter (of Example \ref{ex:gaussmechdp}) would have been $\sqrt{2\rho}$ (where $\rho=2.63$ is the final production $\rho$-zCDP parameter for the redistricting application of the TopDown Algorithm) \cite{tdahdsr}. Under this setting, Figure \ref{fig:epsdeltacurve} plots the $(\epsilon,\delta)$-curves for approximate differential privacy (which is difficult to interpret) and the corresponding curve for \pbdp/$f$-DP (which has a more natural interpretation). It also plots the upper bound on the \pbdp  parameter $\delta$ using $\rho$-zCDP privacy-loss accounting. We believe, based on the results in Section \ref{sec:freq}, that the  \pbdp curve of the \emph{Discrete} Gaussian Mechanism is almost identical to that of the continuous Gaussian Mechanism (instead of being close to the $\rho$-zCDP upper bound). 

We conclude this part of the discussion with an open question: how does one choose a trade-off function $f$ or an $(\epsilon,\delta)$-curve? We do not have a definitive answer to this question, but the approach taken by the Disclosure Avoidance Team for the 2020 Census was to use the curve that was implicitly provided by the $\rho$ parameter of zCDP and to base recommendations on the semantic consequences of different $\rho$-values (discussed in Sections \ref{sec:rzcdp}, \ref{sec:freq}, \ref{sec:bayes}, and \ref{sec:per}). Afterwards, we derived these tighter guarantees based on \pbdp.

%% file: freq.tex
The hypothesis testing interpretation of differential privacy is based on how well an attacker could succeed in the following experiment: $\data_1$ and $\data_2$ are arbitrary datasets that differ on the contents of the record of one individual (say, \target). One of them is selected as an input to $\mech$ and the output $\outp$ is provided to the attacker. The success of an attacker in deciding between $\data_1$ and $\data_2$ is then equivalent to the success of the attacker in inferring a piece of sensitive information about \target (success strictly due to the use of \target's record, since everything else is held the same). While a hypothesis test does not directly quantify the causal contribution of the use of \target's record to make inferences about \target (i.e., it doesn't tell us what the inference about \target would have been if the record had been scrubbed), it does measure a related quantity -- to what extent could an attacker determine that the record had been scrubbed. 

In the frequentist hypothesis testing framework, one dataset would be designated as the null hypothesis: $H_0=\data_1$ and the other as the alternative hypothesis: $H_1=\data_2$. Due to symmetry in the differential privacy definitions, the roles of $\data_1$ and $\data_2$ can also be switched.

The uniformly most powerful test in this case is the likelihood ratio test \cite{nptest}. One would compare the likelihood ratio test statistic $\log\frac{P(\mech(\data_1))=\outp)}{P(\mech(\data_2)=\outp)}$
 to a threshold $t$. If the test statistic is smaller, one rejects the null hypothesis. If it is larger, one fails to reject the null hypothesis, and if the test statistic is equal to $t$, then one can randomize the decision rule by rejecting the null hypothesis with probability $c\in[0,1]$.

%Since $\plrv[\data_1][\data_2][\mech]$ has the  distribution of $\log\frac{P(\mech(\data_1)=\omega)}{P(\mech(\data_2)=\omega)}$ under the null hypothesis, a most powerful test is defined by a threshold $\tau$ and a probability $c$. If $\log\frac{P(\mech(\data_1)=\omega)}{P(\mech(\data_2)=\omega)}>\tau$ then we fail to reject the null hypothesis, and if $\log\frac{P(\mech(\data_1)=\omega)}{P(\mech(\data_2)=\omega)}=\tau$ then we fail to reject with probability $1-c$. Otherwise, we reject the null hypothesis.

%To avoid conflicts with the greek letters used previously, we use the following notation for statistical tests:
The important properties of this (or any other) statistical test are:
\begin{itemize}
\item \textbf{Power}, denoted by $\power$, which is the probability of correctly rejecting the null hypothesis (i.e., rejecting the null hypothesis when the true input to $\mech$ was $\data_2$). 
\item \textbf{Type II error probability}, denoted by $\typetwo$, is the probability of failing to reject the null hypothesis when the alternative hypothesis is true.
\item \textbf{Significance level}, denoted by $\level$ is the probability of incorrectly rejecting the null hypothesis (i.e., rejecting the null hypothesis when the true input to $\mech$ was $\data_1$). Significance level is also known as the probability of a Type I error.
\end{itemize}

We note, from Section \ref{sec:plrv}, that the distribution of the privacy-loss random variable $\plrv[\data_1][\data_2][\mech]$ has the distribution of the likelihood ratio test statistic under the null hypothesis. Similarly, the distribution of $-\plrv[\data_2][\data_1][\mech]$ has the corresponding distribution under the alternate hypothesis. Thus one can write expressions for the power $\power$  and significance level $\level$ of the likelihood ratio test that uses the threshold $t$ and tie-breaking probability $c$ as follows:

%\noindent Since $\plrv[\data_1][\data_2][\mech]$ has the null distribution and $\plrv[\data_2][\data_1][\mech]$ has the alternative distribution, we have:
\begin{align*}
%%%%\level &= cP(\plrv[\data_1][\data_2][\mech]= t) + P(\plrv[\data_1][\data_2][\mech]< t)\\
%%%%\power &= cP(-\plrv[\data_2][\data_1][\mech]= t)+ P(-\plrv[\data_2][\data_1][\mech]< t)
%1-\level &= P(\plrv[\data_1][\data_2][\mech]>t)+(1-c)P(\plrv[\data_1][\data_2][\mech]=t) = cP(\plrv[\data_1][\data_2][\mech]>t)+(1-c)P(\plrv[\data_1][\data_2][\mech]\geq t)\\
\level &= cP(\plrv[\data_1][\data_2][\mech]\leq t)+(1-c)P(\plrv[\data_1][\data_2][\mech]< t)\\
%\typetwo &= cP(\plrv[\data_2][\data_1][\mech]<-t)+(1-c)P(\plrv[\data_2][\data_1][\mech]\leq -t)\\
\power &= cP(\plrv[\data_2][\data_1][\mech]\geq -t)+(1-c)P(\plrv[\data_2][\data_1][\mech]> -t)
\end{align*}

%\begin{itemize}
%\item If $\alpha$ is the significance level, then $1-\alpha$ is:
%$P(\plrv[\data_1][\data_2][\mech]>t)+(1-c)P(\plrv[\data_1][\data_2][\mech]=t)$, which also equals $cP(\plrv[\data_1][\data_2][\mech]>t)+(1-c)P(\plrv[\data_1][\data_2][\mech]\geq t)$.
%\item The probability of Type I error is: $cP(\plrv[\data_1][\data_2][\mech]\leq t)+(1-c)P(\plrv[\data_1][\data_2][\mech]< t)$. 
%\item The Type II error is $cP(\plrv[\data_2][\data_1][\mech]<-t)+(1-c)P(\plrv[\data_2][\data_1][\mech]\leq -t)$.
%\item The power is $cP(\plrv[\data_2][\data_1][\mech]\geq -t)+(1-c)P(\plrv[\data_2][\data_1][\mech]> -t)$.
%\end{itemize}
%
%\noindent We next relate these quantities to the variations of differential privacy and explain how they are used to define the Gaussian Differential Privacy and $f$-DP \cite{gaussdp}, which we have deferred to this section.

This discussion shows that there is a deep connection between differential privacy and hypothesis testing. Each definition provides a trade-off between the power and significance level of \emph{any} hypothesis test of $H_0: \data_1$ vs. $H_1: \data_2$ that is based on the output of a mechanism satisfying the privacy definition.
Furthermore, the trade-off  can be directly computed  from the privacy parameters. That is, the power for any significance level $\level$ for (say) pure $\epsilon$-differential is upper bounded by a function of $\level$ and $\epsilon$. This upper bound applies to any $\mech$ that satisfies $\epsilon$-differential privacy. However, some of these mechanisms might have a tighter trade-off (i.e., lower power at a given significance level). Computing that tighter tradeoff for a specific mechanism $\mech$ would require reasoning about the privacy loss random variables $\plrv$ for all neighbors $\data_1,\data_2$.

\subsection{Frequentist semantics of pure differential privacy}

One can think of a hypothesis test as a randomized algorithm $\randalg$ whose input is $\outp$ and whose output is either 1 (reject the null hypothesis) or 0 (fail to reject). Thus the significance level is $\level=P(\randalg(\mech(\data_1))=1)$ and the power is $\power=P(\randalg(\mech(\data_2))=1)$. If $\mech$ satisfies $\epsilon$-differential privacy then so does $\randalg\circ\mech$ and so the pure differential privacy constraints apply to both $P(\randalg(\mech(\data_1))=1)$ and  $P(\randalg(\mech(\data_2))=1)$. Noting the symmetry of $\data_1$ and $\data_2$ in the definition of differential privacy, and noting that if $\randalg$ is a hypothesis test, then so is $1-\randalg$, then we obtain the following result due to Wasserman and Zhou \cite{WassermanZhou}: 

\begin{theorem}[\cite{WassermanZhou}]\label{thm:dp:freq}
Let $\data_1$ and $\data_2$ be neighboring datasets, let $\mech$ be an $\epsilon$-differential privacy mechanism, and let $\outp$ denote the output of $\mech$. Then, any hypothesis test of $H_0=\data_1$ vs. $H_1=\data_2$ that is a function of $\outp$ has the following relation between power $\power$, Type II error probability $\typetwo$ and significance level $\level$:
\begin{align*}
\level &\leq e^\epsilon (\power)\\
\power &\leq e^\epsilon \level\\
(1-\level) & \leq e^\epsilon\typetwo\\
\typetwo &\leq e^\epsilon (1-\level)
\end{align*}
\end{theorem}
%\noindent The last equation is an uncertainty principle that states that the probabilities of Type I and Type II errors cannot be simultaneously small.

\noindent Combining these inequalities, we get the following explicit limitations on power at a given significance level for any hypothesis testing procedure:
\begin{align}
\max(e^{-\epsilon}\level, 1-e^{\epsilon}(1-\level))\leq \power \leq \min(e^\epsilon\level, 1-e^{-\epsilon}(1-\level)). \label{eqn:purepowerlevel}
\end{align}
We note that a non-informative test that cannot distinguish between $H_0$ and $H_1$ at all would be rejecting the null hypothesis at the same rate regardless of the truth of $H_0$ or $H_1$; the power and significance levels would always be equal to each other in such a test. From Equation \ref{eqn:purepowerlevel},  we see that as $\epsilon$ approaches 0, the  power approaches the significance level. When $\epsilon=0$ only the  non-informative tests are possible.
Table \ref{tab:epspowertable} uses the power-bounds formula above to plot the maximum achievable power in inferring personal information for a given significance level and $\epsilon$ parameter.
\begin{table}[h!]
\begin{tabular}{|c|rrrrr|}\cline{2-6}
\multicolumn{1}{c}{} & \multicolumn{5}{|c|}{$\epsilon$}\\\cline{1-1}
Significance Level & 0.1 & 0.5 & 1 & 2 & 4\\\hline
0.01 & 0.011 & 0.016 & 0.027 & 0.074 & 0.550\\
0.05 & 0.055 & 0.820 & 0.136 & 0.370 & 0.983\\
0.10  & 0.111 & 0.165 & 0.272 & 0.739 & 0.984\\\hline
\end{tabular}
\caption{Maximum power for inferring personal information under pure differential privacy at a given significance level, for different values of $\epsilon$.}\label{tab:epspowertable}
\end{table}

We note that at $\epsilon=4$, these frequentist guarantees do not at first appear meaningful because of the large achievable power. However, this is the power that is achieved by mechanisms that use all of their privacy-loss budget to try to reveal \target's personal information. Since most differentially private algorithms would spread a large privacy-loss budget over many queries, it is often possible to use additional properties of the mechanism to reduce these upper bounds. For example, one can consider the entire $(\epsilon,\delta)$ curve of a mechanism under $f$-DP/\pbdp, which we demonstrate later in this section. Another approach is to apply the ideas developed in this section while examining how the algorithm allocates privacy-loss budget among queries, which we consider in Section \ref{sec:per}.

\subsection{Frequentist Semantics of Approximate Differential Privacy}

To get frequentist guarantees for approximate differential privacy,  Kairouz et al. \cite{kairouz15} used ideas similar to those used for pure differential privacy to get the following result:

\begin{theorem}[\cite{kairouz15}]\label{thm:adp:freq}
Let $\data_1$ and $\data_2$ be neighboring datasets, let $\mech$ be an $(\epsilon,\delta)$-differential privacy mechanism, and let $\outp$ denote the output of $\mech$. Then, any hypothesis test of $H_0=\data_1$ vs. $H_1=\data_2$ that is a function of $\outp$ has the following relation between power $\power$, Type II error probability $\typetwo$, and significance level $\level$:
\begin{align*}
\level &\leq e^\epsilon (\power) + \delta\\
\power &\leq e^\epsilon \level + \delta\\
(1-\level) & \leq e^\epsilon\typetwo + \delta\\
\typetwo &\leq e^\epsilon (1-\level) + \delta.
\end{align*}
\end{theorem}

\noindent Combining these inequalities, we get the following explicit limitations on power at a given significance level for any hypothesis testing procedure:
\begin{align*}
\max(e^{-\epsilon}(\level-\delta), 1-e^{\epsilon}(1-\level)-\delta)\leq \power \leq \min(e^\epsilon\level+\delta, 1-e^{-\epsilon}(1-\level-\delta)).
\end{align*}
\noindent However, we note that a mechanism $\mech$ satisfies approximate differential privacy for a curve $C$ of $(\epsilon,\delta)$ values, and so the upper bound on the power is 
actually $\min\limits_{(\epsilon,\delta)\in C}\min(e^\epsilon\level+\delta, 1-e^{-\epsilon}(1-\level-\delta))$. This is another reason to use the entire $(\epsilon,\delta)$ curve rather than rely on a single point and the ensuing inaccurate summary of the privacy protections.

\subsection{Frequentist semantics of \pbdp, \texorpdfstring{$f$}{f}-DP, RDP, and zCDP}
Frequentist semantics for the remaining privacy definitions can be derived in the same manner as those of pure and approximate DP. We group them here so that we can compare them using the production settings of the 2020 Census redistricting data.

We start with the significance level/power trade-offs for zCDP and RDP, which were studied by Balle et al. \cite{balle20a}. Then, we follow the reasoning used for pure and approximate DP, combined with the definition of the R\'enyi divergence to complete the semantics.

\begin{theorem}[\cite{balle20a}]\label{thm:rdp:freq}
Let $\data_1$ and $\data_2$ be neighboring datasets, let $\mech$ satisfy $(\alpha,\gamma)$-RDP. Let $\outp$ denote the output of $\mech$. Then, any hypothesis test of $H_0=\data_1$ vs. $H_1=\data_2$ that is a function of $\outp$ has the following relation between power, $\power$, Type II error probability, $\typetwo$, and significance level, $\level$:
\begin{align*}
\level^\alpha (\power)^{1-\alpha} + (1-\level)^\alpha\typetwo^{1-\alpha} &\leq e^{\gamma(\alpha-1)}\\
(\power)^\alpha \level^{1-\alpha} + \typetwo^\alpha (1-\level)^{1-\alpha} & \leq e^{\gamma(\alpha-1)}.
\end{align*}
\end{theorem}
\noindent Since satisfying $\rho$-zCDP is equivalent to satisfying $(\alpha, \rho\alpha)$-RDP for all $\alpha>1$, the relations for $\rho$-zCDP are:
\begin{align*}
\level^\alpha (\power)^{1-\alpha} + (1-\level)^\alpha\typetwo^{1-\alpha} &\leq e^{\rho\alpha(\alpha-1)}~ \text{ for all } \alpha>1\\
(\power)^\alpha \level^{1-\alpha} + \typetwo^\alpha (1-\level)^{1-\alpha} & \leq e^{\rho\alpha(\alpha-1)}~\text{ for all } \alpha>1.
\end{align*}

Because the constraints are nonlinear, it is not possible to present an exact algebraic bound on the power as a function of level and $\rho$ (or $\gamma$ and $\alpha$). However, an upper bound on power can be estimated numerically by considering a large number of $\alpha$ values and finding points on the boundaries of the constraints.

\begin{figure}[h!]
\includegraphics[scale=0.8]{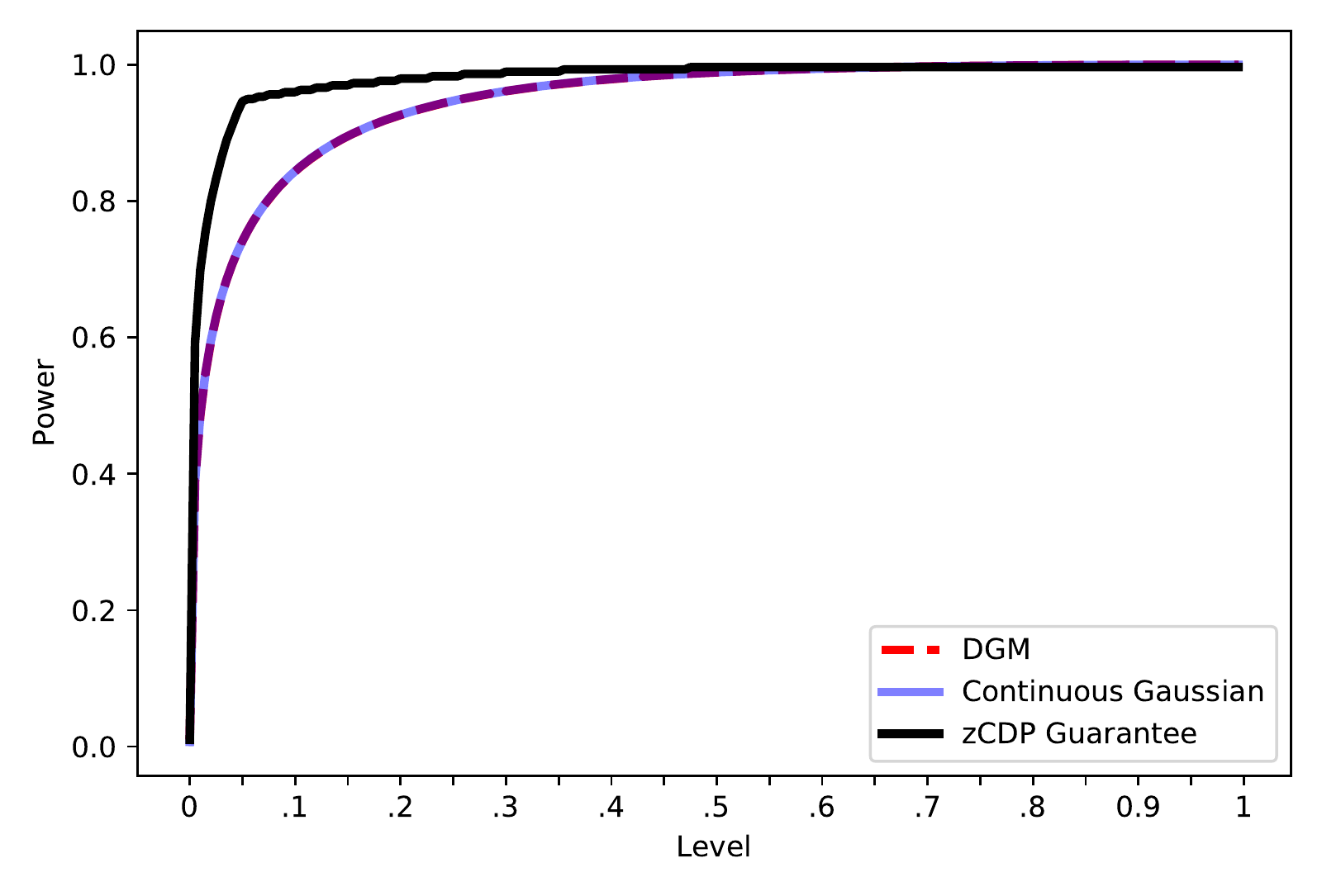}
\caption{Level (x-axis) vs. power (y-axis) curves for (1) the Gaussian mechanism, which is the same as the test of $H_0=N(0, \frac{1}{2\rho})$ vs. $H_1=N(1, \frac{1}{2\rho})$ for $\rho=2.63$, (2) the likelihood ratio test of the discrete Gaussian noisy queries having the same privacy-loss budget allocation as in the production settings for 2020 Census redistricting data. }
\label{fig:gauss263}.
\end{figure}

The noise-generating component of the disclosure avoidance system for the 2020 redistricting data satisfies $\rho$-zCDP with $\rho=2.63$ \cite{tdahdsr}. In Figure \ref{fig:gauss263} we plot the numerically estimated upper bound on power for each significance level, and in Table \ref{tab:gauss263} we list the upper bounds for a selection of significance levels.

\begin{table}[h]
\begin{tabular}{|c||c|c|c|}\hline
\textbf{Significance Level} & \textbf{Power (Gaussian)} & \textbf{Power (DGM)} &
\textbf{zCDP Upper Bound}\\\hline
0.01
&0.49 
&0.49 
&0.70\\\hline
0.05
&0.74
&0.74
&0.95\\\hline
0.10
&0.84
&0.84
&0.96\\\hline
\end{tabular}
\caption{Likelihood ratio test significance level/power trade-off for 2020 Census redistricting data production privacy-loss budget allocations (1) if Gaussian noise is used, (2) if discrete Gaussian noise is used, (3) guaranteed upper bound if an arbitrary $\rho$-zCDP mechanism with $\rho=2.63$ is used.}\label{tab:gauss263}
\end{table}

We note  that these upper bounds on power appear quite high, but this is because the upper bounds are guarantees for \emph{all} possible mechanisms that satisfy $\rho$-zCDP for $\rho=2.63$ (not just the TopDown Algorithm). In particular, this guarantee would also bind
mechanisms that specifically target \target and use their entire privacy-loss budget in an attempt to reveal \target's personal information. Furthermore, even though $\rho$-zCDP can produce an entire $(\epsilon,\delta)$ curve for $f$-DP/\pbdp, the number of curves one can obtain by changing $\rho$ is limited and hence may not be the tightest representation of a mechanism.

A more precise analysis and tighter bounds can be achieved by considering $f$-DP (for the purposes of this section, the reparametrization offered by \pbdp does not affect the frequentist semantics). The guarantees can be read directly off Definition \ref{def:gaussdpalt} by noting that the randomized algorithm $\randalg$ it uses can be viewed as a hypothesis test (in fact, $f$-DP was originally introduced directly as the trade-off between the Type I and Type II error probabilities \cite{gaussdp}). Thus we get:
\begin{align*}
\power & \leq 1-f(\level)\\
\level &\leq 1-f(\power)
\end{align*}

In the case of the Gaussian Mechanism (Example \ref{ex:gaussmechdp}), the function $f$ is known exactly and so we can compare its trade-off with the upper bound provided by zCDP as follows. Let us consider some query $\query$ such that $\query(\data_1)$ differs from $\query(\data_2)$ by at most 1 for any pair of neighboring $\data_1$, $\data_2$. It is known that the mechanism $\mech$ defined as $M(\data)=q(\data)+ N(0, \sigma^2=\frac{1}{2\rho})$ satisfies $\rho$-zCDP \cite{zcdp} and has the trade-off function  given in Example \ref{ex:gaussmechdp}. We plot the upper bound on power for each significance level for this mechanism in Figure \ref{fig:gauss263} (blue line) and list some of the upper bounds for a selection of significance levels in Table \ref{tab:gauss263} in the column labeled ``Power (Gaussian)''. We note that the $f$-DP trade-off function is the exact trade-off one would get from the likelihood ratio test \cite{gaussdp}. We see that the actual upper bound on power for the continuous Gaussian is less than what zCDP semantics provide for $\rho=2.63$.

But what about the actual TopDown Algorithm that produced redistricting data? Its noise addition mechanism is analogous to the Gaussian Mechanism of Example \ref{ex:gaussmechdp} but it adds  discrete Gaussian (instead of continuous Gaussian) noise to  millions of queries. The details of how noise was allocated to different queries appear in \cite{tdahdsr} and are reproduced in Section \ref{sec:per}. The trade-off function $f$ for the composition of discrete Gaussian mechanisms is not known, but can be estimated using Monte Carlo simulation and the following observations:
\begin{itemize}
\item the maximum possible power for a given significance level is given by the maximum power of the likelihood ratio test (maximized over all pairs of neighboring datasets);
\item one can, without loss of generality, examine only pairs of neighbors $\data_1$, $\data_2$ for which the maximal number of queries have different answers;
\item for the TopDown algorithm, it turns out that all of these maximal neighboring pairs are equivalent for this purpose -- the likelihood ratio test gives exactly the same power for a given significance level.
\end{itemize}

We generated one million samples from the null and one million samples from the alternate hypothesis to empirically estimate the significance level/power trade-off. The results are shown as the dashed red line in Figure \ref{fig:gauss263} and in the column labeled ``Power (DGM)'' in Table \ref{tab:gauss263}. We see that the results are nearly identical to the continuous Gaussian.
Thus, this significance level/power curve represents one of the interpretations of the privacy guarantees of the TopDown Algorithm. We examine Bayesian guarantees in Section \ref{sec:bayes}. We also consider fine-grained privacy semantics that examine the noise added to different queries in Section \ref{sec:per}.

\input{freqextensions}

%% file: freqextensions.tex
\subsection{A Generic Extension of Frequentist Semantics}

While the previous results focussed on significance/level power curves for distinguishing between pairs of neighbors $\data_1$ and $\data_2$, it is possible to  extend the mathematical statements of all of the results to something that more closely matches the informal description of ``can an attacker detect that a record has been altered''? 

Let $\rem[\data]$ refer to the set of records belonging to everyone other than \target and let $F$ be an \emph{arbitrary} distribution for $\rem[\data]$ (e.g., those records do not have to be independent). Given $F$ and two possible candidate records $\record_0,\record_1$ for \target, we can form the hypothesis test of $H_0$: $\rem[\data]$ was sampled from $F$ and $\add{\record_0}{\rem[\data]}$ is the input to $\mech$ vs. $H_1$: $\rem[\data]$ was sampled from $F$ and $\add{\record_1}{\rem[\data]}$ is the input to $\mech$.

We show in this section that if a privacy definition is post-processing invariant,  convex (see Section \ref{sec:desiderata}), and also satisfies a minor regularity condition called \emph{symmetry} then the significance level vs. power trade-off curve is the same as for the earlier cases where one considered pairs of neighboring datasets. Since all the privacy definitions we considered have these properties, this extension applies to them all.

The symmetry property can be described follows. Let $\nu$ be a 1-to-1 and onto function on records that does not alter the id (but the behavior of $\nu$ can depend on the id). For example, $\mu$ might toggle the ethnicity attribute for Alice, toggle both ethnicity and gender for \target, add 5 to Bob's age, etc. One can apply $\nu$ to an entire dataset by defining $\nu(\data)$ to be the result of applying $\nu$ to each record in $\data$. We say that a privacy definition has symmetry if for all mechanisms $\mech$ that satisfy the privacy definition and all such functions $\nu$, then $\mech$ and $\mech\circ\nu$ both satisfy the privacy definition for exactly the same privacy parameter settings (note that $\mech\circ\nu$ is the function that first applies the recoding $\nu$ and then runs $\mech$ on the result). This symmetry property essentially states that no dataset gets extra protections compared to another dataset. Furthermore, these $\nu$ functions preserves the neighborhood structure of datasets: if $\data_1$ and $\data_2$ are neighbors then the pair $\nu(\data_1)$ and $\nu(\data_2)$ are neighbors and same with the pair $\nu^{-1}(\data_1)$ and $\nu^{-1}(\data_2)$.
Symmetry allows us to restrict attention to just one (arbitrary) pair of neighbors $\data_1,\data_2$ when examining the privacy properties of a definition.

The first result in this section is that symmetric, convex,  postprocessing invariant privacy definitions have a concave significance level vs. power trade-off. This is useful for proving the main result of this section and for diagnosing if a privacy definition might violate convexity.

\begin{theoremEnd}[category=freq]{theorem}\label{thm:plconvex} Given a transparent, symmetric, convex and postprocessing invariant privacy definition with privacy parameters $\theta$, let $\mathfrak{M}$ denote the set of all mechanisms that satisfy the privacy definition at those privacy parameter settings. Define the significance level vs. power trade-off function $g$ as follows. For any significance level $\level$, $g(\level)$ is the supremum of achievable power in testing $H_0: \data_1$ vs. $H_1: \data_2$ for all neighboring pairs $\data_1,\data_2$, for all mechanisms $\mech\in \mathfrak{M}$, and for all hypothesis tests whose input is the output of the chosen mechanism $\mech$ and whose significance level is at most $\level$. Then $g$ is a concave function.
\end{theoremEnd}
\begin{proofEnd}
For a mechanism $\mech$, neighboring datasets $\data_1$ and $\data_2$, and significance level $\level$, let $\phi_{(\mech,\data_1,\data_2,\level)}$ denote a hypothesis test on the output of $\mech$ for distinguishing between $H_0:\data_1$ and $H_1: \data_2$ with significance level $\level$. Then we can write:
\begin{align*}
g(\level) &=\sup_{\mech\in \mathfrak{M},}\sup_{\data_1\sim\data_2,}\sup_{\phi_{(\mech,\data_1,\data_2,\level)}} power(\phi_{(\mech,\data_1,\data_2,\level)})
\end{align*}

We can define the significance level vs. power trade-off $g_{\mech}$ for a specific mechanism $\mech$ as follows:
\begin{align*}
g_{\mech}(\level) &=\sup_{\data_1\sim\data_2,}\sup_{\phi_{(\mech,\data_1,\data_2,\level)}} power(\phi_{(\mech,\data_1,\data_2,\level)})
\end{align*}

 Note that $g(\level) = \sup_{\mech\in\mathfrak{M}} g_{\mech}(\level)$.

We then note that postprocessing can only reduce the maximum power. That is, $g_{\mech}(\level) \geq g_{\randalg\circ\mech}(\level)$. This is because any hypothesis test $\phi$ on the output of $\randalg\circ\mech$ can be converted to a hypothesis test on the output of $\mech$ by first running $\randalg$ and then $\phi$.

Next, note that by transparency, $\mech$ can also output its name without affecting privacy properties. 

Now, for any $p\in [0,1]$ and $\mech_1,\mech_2\in\mathfrak{M}$, Consider the mechanism ${\mech_1}\oplus_p{\mech_2}$ such that $({\mech_1}\oplus_p{\mech_2})(\data)$ outputs $\mech_1(\data)$ with probability $p$, and otherwise outputs $\mech_2(\data)$ (in other words, it randomly chooses which mechanism to run). By convexity, ${\mech_1}\oplus_p{\mech_2}\in\mathfrak{M}$. Furthermore, the output of ${\mech_1}\oplus_p{\mech_2}$ can include the name of the sub-mechanism that was run ($\mech_1$ or $\mech_2$, by transparency, since $\mech_1$ and $\mech_2$ are in $\mathfrak{M}$) and removing the names can only decrease the maximum power (by postprocessing invariance). Thus, since the identities of the chosen mechanisms are provided, any hypothesis test $\phi^*$ on the output of ${\mech_1}\oplus_p{\mech_2}$ produces  a pair of hypothesis tests -- one hypothesis test $\phi_1$ on the output of $\mech_1$  and another test $\phi_2$ on the output of $\mech_2$ (and vice versa). Thus we can use the notation $(\phi_1,\phi_2)$ to represent a hypothesis test on the output of ${\mech_1}\oplus_p{\mech_2}$.

If the significance level of $\phi_1$ is $\level_1$ with power $1-\beta_1$, and the significance level of $\phi_2$ is $\level_2$ with power $1-\beta_2$, then the significance level of $(\phi_1,\phi_2)$ is $p \level_1 + (1-p)\level_2$ and the power is a convex combination of the powers: $p(1- \beta_1) + (1-p)(1-\beta_2)$. Thus we have:
\begin{align*}
\lefteqn{g(p \level_1 + (1-p)\level_2) }\\
&\geq \sup_{\data_1,\data_2}\sup_{\mech_1}\sup_{\mech_2}\sup_{\phi_{(\mech_1,\data_1,\data_2,\level_1)}}\sup_{\phi_{(\mech_2,\data_1,\data_2,\level_2)}} power(\phi_{(\mech_1,\data_1,\data_2,\level_1)}, \phi_{(\mech_2,\data_1,\data_2,\level_2)})\\
&=\sup_{\data_1,\data_2}\sup_{\mech_1}\sup_{\mech_2}\sup_{\phi_{(\mech_1,\data_1,\data_2,\level_1)}}\sup_{\phi_{(\mech_2,\data_1,\data_2,\level_2)}} p* power(\phi_{(\mech_1,\data_1,\data_2,\level_1)}) + (1-p)power( \phi_{(\mech_2,\data_1,\data_2,\level_2)})\\
&=\sup_{\mech_1}\sup_{\mech_2}\sup_{\phi_{(\mech_1,\data_1,\data_2,\level_1)}}\sup_{\phi_{(\mech_2,\data_1,\data_2,\level_2)}} p* power(\phi_{(\mech_1,\data_1,\data_2,\level_1)}) + (1-p)power( \phi_{(\mech_2,\data_1,\data_2,\level_2)})\\
\intertext{The symmetry property allows us to fix any neighboring pair $\data_1,\data_2$, since optimizing over mechanisms and neighbors becomes redundant. This is because  power$(\phi_{(\mech^\prime,\data_3,\data_4,\level_1)})$ is the same as power$(\phi_{(\mech^\prime\circ\nu^{-1},\data_1,\data_2,\level_1)}$ for the recoding $\nu$ that maps dataset $\data_3$ to $\data_1$ and $\data_4$ to $\data_2$}
&= p g(\ell_1) + (1-p)g(\ell_2)
\end{align*}
%
%
%
%
%
%$\leq p g(\level_1) + (1-p) g(\level_2)$ since $\mech^*\in \mathfrak{M}$. Taking the supremum over all possible $\mech_1$ and $\mech_2$ (including the case where $\mech_1=\mech_2$) and all hypothesis tests on the output of the resulting $\mech^*$
\end{proofEnd}

Next is the main result of this section, which shows that generalizing the hypothesis testing setup does not change the significance level vs. power curve.

\begin{theoremEnd}[category=freq]{theorem}\label{thm:plequal} Given a transparent, symmetric, convex and postprocessing invariant privacy definition with privacy parameters $\theta$, let $\mathfrak{M}$ denote the set of all mechanisms that satisfy the privacy definition at those privacy parameter settings.
\begin{itemize}
\item Define the \emph{pointwise} significance level vs. power tradeoff function $g$ as follows. For any significance level $\level$, $g(\level)$ is the supremum of achievable power in testing $H_0: \data_1$ vs. $H_1: \data_2$ for all neighboring pairs $\data_1,\data_2$, for all mechanisms $\mech\in \mathfrak{M}$, and for all hypothesis tests whose input is the output of the chosen mechanism $\mech$ and whose significance level as at most $\level$. 
\item Define the \emph{compound} significance level vs. power tradeoff function $g^*$ as follows. For any significance level $\level$, $g^*(\level)$ is the supremum of achievable power in testing $H^*_0$: $\rem[\data]$ was sampled from $F$ and $\add{\record_0}{\rem[\data]}$ is the input to $\mech$ vs. $H^*_1$: $\rem[\data]$ was sampled from $F$ and $\add{\record_1}{\rem[\data]}$ is the input to $\mech$, maximizing over all distributions $F$, pairs of records $\record_1,\record_2$, mechanisms $\mech\in \mathfrak{M}$, and for all hypothesis tests whose input is the output of the chosen mechanism $\mech$ and whose significance level is at most $\level$. 
\end{itemize}
Then, $g=g^*$.
\end{theoremEnd}
\begin{proofEnd}
First, we note that $g^*(\level)\geq g(\level)$ for all $\level$ because testing between neighbors $\data_1$ and $\data_2$ is a special case where $F$ puts probability 1 on the set of records on which $\data_1$ and $\data_2$ do not differ. 

For the other direction, note that a hypothesis test $\phi_\mech$ for $H^*_0$ vs $H^*_1$ (defined by $F, \record_1,\record_2$) based on the output of $\mech$ can be converted to a test $\phi_{\mech,\data_1,\data_2}$ for $H_0$ vs $H_1$ for neighboring pairs $\data_1,\data_2$ (in which the record that they differ by has value $\record_1$ or $\record_2$) just by taking its recommended action (reject or fail to reject the null hypothesis). $F$ induces a distribution over pairs of neighbors (since it is a distribution over the records shared by $\data_1$ and $\data_2$ and the remaining record is either specified by the null or alternate hypothesis). Letting $\level$ be the significance level of $\phi_\mech$ when testing $H^*_0$ vs. $H^*_1$ and $\level_{\data_1,\data_2}$ be the significance level of $\phi_{\mech,\data_1,\data_2}$, we have that 
\begin{align*}
\level &= E_{(\data_1,\data_2)\sim F}[\level_{\data_1,\data_2}]\\
power(\phi_\mech) &= E_{(\data_1,\data_2)\sim F}[power(\phi_{\mech,\data_1,\data_2})]\\
&\leq E_{(\data_1,\data_2)\sim F}[g(\level_{\data_1,\data_2})]\\
&\leq g(E_{(\data_1,\data_2)\sim F}[\level_{\data_1,\data_2}])\\
&=g(\level)
\end{align*}
Maximizing over all $F$, $\record_1$,$\record_2$, $\mech$ and $\phi_{\mech}$ that have level $\level$
we therefore get $g^*(\level)\leq g(\level)$. 

Combining this with our first result and applying them for all $\ell$, we have $g=g^*$.

\end{proofEnd}

%% file: bayes.tex
In this section, we present the Bayesian privacy semantics for pure differential privacy, zCDP, and RDP. The semantics for zCDP and RDP are new and resemble those derivable from \pbdp. We also provide a direct Bayesian interpretation of the $(\epsilon,\delta)$ curve of \pbdp. While some results about the Bayesian semantics for approximate differential privacy are known \cite{kasiviswanathan2014semantics}, we do not review the exact details here because they only apply to a small subset of privacy-loss parameters ($\epsilon$ must be smaller than 1/4) and do not use the full $(\epsilon,\delta)$ curve.

We consider an attacker using a prior $\prior$ over possible datasets $\data_1,\data_2,\dots$. This prior is independent of the randomness in $\mech$ (since the attacker does not have access to the bits of $\mech$). Otherwise, we make no assumptions about this prior -- in particular, records do not have to be independently and identically distributed (i.i.d.). Thus, in the attacker's view, the dataset is a random variable $\datarv$ having distribution $\prior$ (we often use notation such as $\prior(\data)$ and $\prior(\cdot~|~\data)$ as shorthand for $\prior(\datarv=\data)$ and the conditional distribution $\prior(\cdot~|~\datarv=\data)$).
A mechanism $\mech$ operates on the data and produces an output $\outp$ (e.g., privacy-protected microdata or tabulations). This output $\outp$ along with all  details of $\mech$, except for the specific values of the random bits it uses, are released publicly and are known to the attacker. 
The attacker is trying to infer information about the record  belonging to a specific target individual \target (for records, we use the notation $\recordrv$ to denote the random variable  and $\record$ to represent a specific value). What can be guaranteed about the posterior $\prior(\recordrv~|~\mech(\datarv)=\outp)$?
Before answering this question, it is important to note that we use $\prior$ for expressions in which the attacker's belief plays a role and $P$ for probabilities in which only the randomness in $\mech$ plays a role. In other words, $P(\mech(\data)=\outp)$ is the probability of $\outp$ when the input $\data$ is given and $\prior(\mech(\datarv)=\outp)$ is the probability of $\outp$ when the input $\datarv$ is also random.

\subsection{The Unsuitability of Prior to Posterior Comparisons}

Traditional disclosure avoidance methodology attempts to provide semantic guarantees based on comparisons between the prior $\prior(\recordrv)$ and posterior $\prior(\recordrv~|~\mech(\datarv)=\outp)$ distributions.
However, in light of privacy research over the past two decades that studies the consequences of prior-posterior comparisons, the suitability of these comparisons has been thrown into question.
First, the difference between the prior and posterior can be arbitrarily large (rendering the comparison vacuous) unless assumptions are made about the prior \cite{dwork:2006,Dwork:Naor:2010,nfl}. Obtaining consensus about the right prior(s) to use will always remain an unsettled issue, and the approach will never be future-proof -- knowledge learned in the future may result in a different prior from which a different semantic conclusion will be reached.
Second, even more importantly, prior-posterior comparisons do not distinguish between changes that are due to gaining scientific knowledge (such as learning that smoking causes cancer) and changes that are specifically due to participation in the data \cite{dworkpottenger}. In particular, prior-posterior comparisons may show large changes even for people whose information is not recorded in the data.

As discussed in Section \ref{sec:informal}, gaining generalizable scientific/statistical knowledge is the whole point of data release. Disclosure avoidance methodologies that penalize a data release for providing statistical knowledge run the risk of severely damaging the usefulness of privacy-protected data.
However, an alternative approach \cite{kasiviswanathan2014semantics}, known as \emph{posterior-posterior} comparisons provides a clean way of separating inference about individuals due to generalizable knowledge from inference that is possible specifically because an individual's record is in the input to $\mech$. Such posterior-posterior semantics are the focus of this section.

\subsection{The Posterior-Posterior framework}

The main idea behind the posterior-to-posterior framework is to compare (1) what an attacker would learn about \target if she participated in the data collection process (e.g., submitted survey responses)  to (2) what would be learned about \target in a counterfactual world in which \target's data were not collected or not used.
This counterfactual world represents the privacy-preserving gold standard for \target. 
Furthermore, unlike the frequentist setting, the attacker knows if we are operating in the actual world or the counterfactual world (since the goal is not to detect whether a change has occurred, but to directly compare how the contents of a record affect inference about it). On a technical note, the attacker must use the same prior in both worlds, as the hypothetical choice of the world would not depend on the records in the actual data.

There are several ways of formalizing the counterfactual world.
\begin{itemize}
\item \textbf{Option 1}: Drop \target's record from the dataset before running $\mech$. Although this is a very appealing option, it is not always applicable. For example, when using a differential privacy variant with \emph{bounded neighbors}, the size $\datasize$ of the input to $\mech$ is not allowed to change; therefore, the counterfactual world must respect that constraint and also use a dataset of size $\datasize$.
\item \textbf{Option 2}: Replace  \target's record with a pre-determined default record \cite{kasiviswanathan2014semantics}. This option creates a counterfactual world in which the \emph{contents} of \target's survey responses are not used at all. There are many possible choices for default records, and the default record could even be sampled from a distribution that does not depend on \target's actual responses. There is one specific choice that is very appealing both mathematically and conceptually, and that is the basis of the next option.
\item \textbf{Option 3}: Replace  \target's record with a random sample from the posterior distribution the attacker would use, if the attacker knew the records of everyone else. This posterior distribution is arguably the statistical information that the attacker would learn from the rest of the data records. Hence, this option can be thought of as replacing \target's actual record with statistical information.\footnote{This is like saying that in the actual world, the record used for \target provides information about how \target differs from her community, but the substituted record in the counterfactual world does not.} This is the option we adopt for the posterior-to-posterior semantics in this section.
\end{itemize}

We now formalize the posterior-to-posterior comparisons using Option 3 as the counterfactual world.
For any dataset $\data$, let $\add{\record}{\data}$ be the result of adding record $\record$ to $\data$. Let  $\rem[\data]$ be the result of removing the record of the target person \target from $\data$.  Let $\psample[\data]$ represent the result of replacing \target's record with a random record sampled from $\prior(\recordrv~|~\rem[\datarv]=\rem[\data])$, which is the same as creating $\repl[\data][\record]$ where $\record\sim \prior(\recordrv~|~\rem[\datarv]=\rem[\data])$.\footnote{In other words, $\psample[\data]$ is a random variable that equals $\repl[\data][\record^\prime]$ with probability $\prior(\record^\prime~|~\rem[\datarv]=\rem[\data])$.} Finally, let $\dataspace$ represent the set of possible datasets (assumed discrete to simplify the discussion) that include survey information from \target, and let $\rem[\dataspace]$ represent the  datasets in $\dataspace$ after \target's record has been removed.
The goal is to compare the posterior distribution in the actual world after the output $\outp$ of $\mech$ has been observed (i.e.,  $\prior(\recordrv~|~\mech(\datarv)=\outp)$) to the corresponding posterior in the counterfactual world: 
$\prior(\recordrv~|~\mech(\psample[\datarv])=\outp)$.
The formulas for these posterior distributions are given below.
\begin{align*}
\intertext{\textbf{Actual World:}}
\prior(\recordrv=\record~|~\mech(\datarv)=\outp) &= 
\frac{\sum_{\rem[\data] \in \rem[\dataspace]} \prior(\rem[\data])\prior(\record ~|~\rem[\data])P(\mech(\add{\record}{\rem[\data]})=\outp)}{\sum_{\rem[\data] \in \rem[\dataspace]}\sum_{\record^\prime} \prior(\rem[\data])\prior(\record^\prime ~|~\rem[\data])P(\mech(\add{\record^\prime}{ \rem[\data]})=\outp)}.\\
\intertext{\textbf{Counterfactual World:}}
\prior(\recordrv=\record ~|~\mech(\psample( \datarv))=\outp) &= 
\frac{\sum_{\rem[\data] \in \rem[\dataspace]} \prior(\rem[\data])\prior(\record ~|~\rem[\data])P(\mech(\psample(\add{\record}{\rem[\data]}))=\outp)}{\sum_{\rem[\data] \in \rem[\dataspace]}\sum_{\record^\prime} \prior(\rem[\data])\prior(\record^\prime ~|~\rem[\data])P(\mech(\psample(\add{\record^\prime}{ \rem[\data]}))=\outp)}\\
&=\frac{\sum_{\rem[\data] \in \rem[\dataspace]} \prior(\rem[\data])\prior(\record ~|~\rem[\data])\Big(\sum_{\record^{\prime\prime}}\prior(\record^{\prime\prime}~|~\rem[\data])P(\mech(\add{\record^{\prime\prime}}{\rem[\data]})=\outp)\Big)}{\sum_{\rem[\data] \in \rem[\dataspace]}\sum_{\record^\prime} \prior(\rem[\data])\prior(\record^\prime ~|~\rem[\data])\Big(\sum_{\record^{\prime\prime}}\prior(\record^{\prime\prime}~|~\rem[\data])P(\mech(\add{\record^{\prime\prime}}{\rem[\data]})=\outp)\Big)}\\
&=\frac{\sum_{\rem[\data] \in \rem[\dataspace]} \prior(\rem[\data])\prior(\record ~|~\rem[\data])\Big(\sum_{\record^{\prime\prime}}\prior(\record^{\prime\prime}~|~\rem[\data])P(\mech(\add{\record^{\prime\prime}}{\rem[\data]})=\outp)\Big)}
{
\sum_{\rem[\data] \in \rem[\dataspace]}
%\sum_{\record^{\prime\prime}} 
\prior(\rem[\data])\sum_{\record^{\prime\prime}}\prior(\record^{\prime\prime}~|~\rem[\data])P(\mech(\add{\record^{\prime\prime}}{\rem[\data]})=\outp)}.\\
\end{align*}

We are interested in the ratio of the posterior in the actual world to the one from the counterfactual world, with a ratio close to one indicating that the two posteriors are similar. Noting that the denominators in both expressions are the same, the ratio of the posteriors is:

\begin{align*}
\frac{
    \prior(\recordrv=\record~|~\mech(\datarv)=\outp)
    }
    {
    \prior(\recordrv=\record ~|~\mech(\psample( \datarv))=\outp)
    } &=
\frac{
\sum_{\rem[\data] \in \rem[\dataspace]} \prior(\rem[\data])\prior(\record ~|~\rem[\data])P(\mech(\add{\record}{\rem[\data]})=\outp)
}
{
\sum_{\rem[\data] \in \rem[\dataspace]} \prior(\rem[\data])\prior(\record ~|~\rem[\data])\Big(\sum_{\record^{\prime\prime}}\prior(\record^{\prime\prime}~|~\rem[\data])P(\mech(\add{\record^{\prime\prime}}{\rem[\data]})=\outp)\Big)
}.
\end{align*}
The ratio of posteriors is a function of $\omega$ and can be viewed as the Bayesian analogue of the privacy-loss random variable.
We next consider what can be said about this ratio of posteriors when $\mech$ satisfies different versions of differential privacy.

\subsection{Posterior-to-Posterior Semantics of Pure Differential Privacy}

It is easy to bound the posterior-to-posterior ratio under pure differential privacy. For all priors and all $\outp$, the ratio of actual to counterfactual posteriors is always between  $e^{-\epsilon}$ and $e^\epsilon$. 

\begin{theoremEnd}[category=bayes]{theorem}\label{thm:pure_dp_bayes}
If $\mech$ satisfies $\epsilon$-differential privacy, then for any attacker prior $\prior$ and all $\omega$ and every $\record$ that has nonzero probability, 
$$e^{-\epsilon} \leq 
\frac{
    \prior(\recordrv=\record~|~\mech(\datarv)=\outp)
    }
    {
    \prior(\recordrv=\record ~|~\mech(\psample( \datarv))=\outp)
    }
    \leq e^\epsilon.$$
\end{theoremEnd}
\begin{proofEnd}
\begin{align*}
\frac{
    \prior(\recordrv=\record~|~\mech(\datarv)=\outp)
    }
    {
    \prior(\recordrv=\record ~|~\mech(\psample( \datarv))=\outp)
    } &=
\frac{
\sum_{\rem[\data] \in \rem[\dataspace]} \prior(\rem[\data])\prior(\record ~|~\rem[\data])P(\mech(\add{\record}{\rem[\data]})=\outp)
}
{
\sum_{\rem[\data] \in \rem[\dataspace]} \prior(\rem[\data])\prior(\record ~|~\rem[\data])\Big(\sum_{\record^{\prime\prime}}\prior(\record^{\prime\prime}~|~\rem[\data])P(\mech(\add{\record^{\prime\prime}}{\rem[\data]})=\outp)\Big)
}\\
&\leq 
\frac{
\sum_{\rem[\data] \in \rem[\dataspace]} \prior(\rem[\data])\prior(\record ~|~\rem[\data])P(\mech(\add{\record}{\rem[\data]})=\outp)
}
{
\sum_{\rem[\data] \in \rem[\dataspace]} \prior(\rem[\data])\prior(\record ~|~\rem[\data])\Big(\sum_{\record^{\prime\prime}}\prior(\record^{\prime\prime}~|~\rem[\data])e^{-\epsilon}P(\mech(\add{\record}{\rem[\data]})=\outp)\Big)
}\\
&= 
\frac{
\sum_{\rem[\data] \in \rem[\dataspace]} \prior(\rem[\data])\prior(\record ~|~\rem[\data])P(\mech(\add{\record}{\rem[\data]})=\outp)
}
{
\sum_{\rem[\data] \in \rem[\dataspace]} \prior(\rem[\data])\prior(\record ~|~\rem[\data])e^{-\epsilon}P(\mech(\add{\record}{\rem[\data]})=\outp)
}\\
&= e^\epsilon.
\end{align*}
The $\geq e^{-\epsilon}$ part of the theorem statement is proved analogously using the fact that $P(\mech(\add{\record^{\prime\prime}}{\rem[\data]}) \leq e^\epsilon P(\mech(\add{\record}{\rem[\data]})$.
\end{proofEnd}
\noindent It is worth noting that Theorem \ref{thm:pure_dp_bayes} is quite strong: no matter what prior the attacker uses (even if it is completely wrong), and no matter what output $\mech$ produces, any posterior probability about \target (or any other person) is within a factor of $e^\epsilon$ compared to purely statistical inference about that person.\footnote{Prior work (e.g, \cite{kasiviswanathan2014semantics}) and folklore results in which the counterfactual world is obtained by replacing \target's record with a pre-specified  default record can achieve similar guarantees but with a factor of $e^{2\epsilon}$ instead of $e^\epsilon$.}

Clearly, if an attacker's prior is completely unreasonable (e.g., all people have 3 arms) then any inference performed with such a prior is suspect. A large difference between an actual posterior and counterfactual posterior would only be a \emph{perceived} privacy breach, not a real one. Pure differential privacy always provides protections even against perceived privacy breaches. 
However, this is not always the case for other variants of differential privacy. To see why, we adapt a well-known example discussed by Kasiviswanathan and Smith  and credited to Dwork and McSherry \cite{kasiviswanathan2014semantics}. 
\begin{example}\label{ex:wrongattacker}
Suppose the true dataset $\data$ contains records from \target and 99 other individuals, with everyone in the data being Hispanic, but the attacker (incorrectly) believes that the other 99 individuals are not Hispanic, that \target's ethnicity is independent of the other people, and that $P(\text{\target is Hispanic})=0.01$.
Thus, in the attacker's view,  there are only 2 possible datasets and
\begin{align*}
\prior(\datarv = \text{Only \target is Hispanic}) &= 0.01\\
\prior(\datarv = \text{No one is Hispanic}) &=0.99
\end{align*}
%Note that in the attacker's view, $\prior(\text{\target is Hispanic} ~|~\text{all other records})=0.01$.
Let $\mech$ be a mechanism that adds $N(0,1)$ noise to the number of Hispanic individuals in the dataset (such a mechanism would satisfy approximate differential privacy, zCDP, RDP, and $f$-DP under appropriate parameter settings). Thus $\mech(\data)$ outputs $100+N(0,1)$. Suppose the realized value is $\outp=101.3$. 

In the actual world (i.e., \target's record is part of the input to $\mech$),  the attacker believes that $N(0,1)$ was either added to 1 or to 0, resulting in the unlikely outcome $101.3$ and the posterior inference:
\begin{align*}
\prior(\text{\target is Hispanic}~|~\outp=101.3) &= \frac{0.01 \frac{1}{\sqrt{2\pi}}e^{-(101.3-1)^2/2}}{0.01 \frac{1}{\sqrt{2\pi}}e^{-(101.3-1)^2/2} + 0.99 \frac{1}{\sqrt{2\pi}}e^{-(101.3-0)^2/2}}\\
&=\frac{1}{1 + 99 e^{-101.3 +1/2}}\approx 1.
\end{align*}
In the counterfactual world, \target's record would be replaced with a Hispanic record with probability $0.01$ and a non-Hispanic record with probability $0.99$. Simple computation shows that:
$$\prior(\text{\target is Hispanic} |~\mech(\psample(\datarv))=\outp)=0.01.$$
Intuitively, in the counterfactual world, $\outp$ only depends on the rest of the records, but the attacker believes the rest of the people are all non-Hispanic with probability 1 and so $\outp$ carries no new information.
Thus, the ratio of the two posteriors is almost 100, and we could make it as large as we want simply by choosing a prior where \target has an even smaller probability of being Hispanic (regardless of the privacy parameters associated with $\mech$).
\end{example}

The key in Example \ref{ex:wrongattacker} is to note that under the attacker's wildly incorrect prior, the output $\outp$ that caused this perceived privacy breach is virtually impossible and so, in such a situation, the attacker should question the correctness of his prior. Thus, Kasiviswanathan and Smith \cite{kasiviswanathan2014semantics} recommend examining how posteriors in the actual and counterfactual worlds differ when the attacker's prior is correct (i.e., what is the probability of a large difference when $\outp$ is generated according to the randomness of $\mech$ and the prior). This idea is the basis for our semantic results for zCDP and RDP.

\subsection{Posterior-to-Posterior Semantics of RDP and zCDP}

We consider two types of Bayesian posterior-to-posterior semantics for RDP and zCDP. In the first case, the attacker knows $\rem[\data]$  (i.e., has complete information about everyone else in the dataset). In the second type of semantics, the attacker is uncertain about $\rem[\data]$. This situation is more general, but the bounds we can prove are slightly weaker. In both settings, we get an $(\epsilon,\delta)$ curve with the guarantee that for any $(\epsilon,\delta)$ pair on that curve, the probability that the ratio of posteriors is at least $e^\epsilon$ is bounded by $\delta$. Furthermore, this $(\epsilon,\delta)$ curve is computed using R\'enyi divergences and, hence, has desirable properties under post-processing and composition.

\subsubsection{Bayesian semantics when the rest of the data are known.}\label{subsec:bayes1}

In this setting, the true dataset is $\data$ and the attacker knows $\rem[\data]$, so that  $\prior(\rem[\datarv]=\rem[\data])=1$. Under the assumption that the prior is correct, the observed output $\outp$ is a random variable whose distribution is a function of  both $\prior$ and the randomness in $\mech$. The data  generating model for $\outp$ in the actual world is:
\begin{enumerate}
\item sample $\recordrv\sim\prior(\recordrv ~|~\rem[\data])$,
\item set $\data = \add{\recordrv}{\rem[\data]}$,
\item produce $\outp=\mech(\data)$.
\end{enumerate}
In the counterfactual world, $\mech(\data)$ is replaced with $\mech(\psample(\data))$.

In the actual world, the attacker's posterior belief about $\record$, conditioned on $\outp$, is:
\begin{align*}
\prior(\recordrv=\record~|~\rem[\datarv]=\rem[\data], \mech(\datarv)=\outp) &= \frac{\prior(\recordrv=\record~|~\rem[\data])P(\mech(\add{\record}{\rem[\data]})=\outp)}{\sum_{\record^\prime}\prior(\recordrv=\record^\prime~|~\rem[\data])P(\mech(\add{\record^\prime}{\rem[\data]})=\outp)}.
\end{align*}
As noted earlier, it is important to remember that we use $\prior$ for expressions in which the attacker's belief plays a role and $P$ for probabilities in which only the randomness in $\mech$ plays a role.
In the counterfactual world, the posterior is:
\begin{align*}
\prior(\recordrv=\record~|~\rem[\datarv]=\rem[\data], \mech(\psample(\datarv))=\outp) &= \frac{\prior(\recordrv=\record~|~\rem[\data])\sum_{\record^\prime}\prior(\recordrv=\record^\prime~|~\rem[\data])P(\mech(\add{\record^\prime}{\rem[\data]})=\outp)}{\sum_{\record^\prime}\prior(\recordrv=\record^\prime~|~\rem[\data])P(\mech(\add{\record^\prime}{\rem[\data]})=\outp)}\\
&=\prior(\recordrv=\record~|~\rem[\data]),
\end{align*}
which is the same as the prior, since the attacker has complete certainty about the rest of the data.
Thus the actual posterior to counterfactual posterior ratio turns out to be the same as the actual posterior to prior ratio and equals:
\begin{align*}
\frac{P(\mech(\add{\record}{\rem[\data]})=\outp)}{\sum_{\record^\prime}\prior(\recordrv=\record^\prime~|~\rem[\data])P(\mech(\add{\record^\prime}{\rem[\data]})=\outp)}.
\end{align*}
We are interested in the probability of generating an $\outp$ for which this ratio is large. Note that the randomness in the numerator is only with respect to $\mech$.

\begin{figure}[h!]
\includegraphics[scale=0.75]{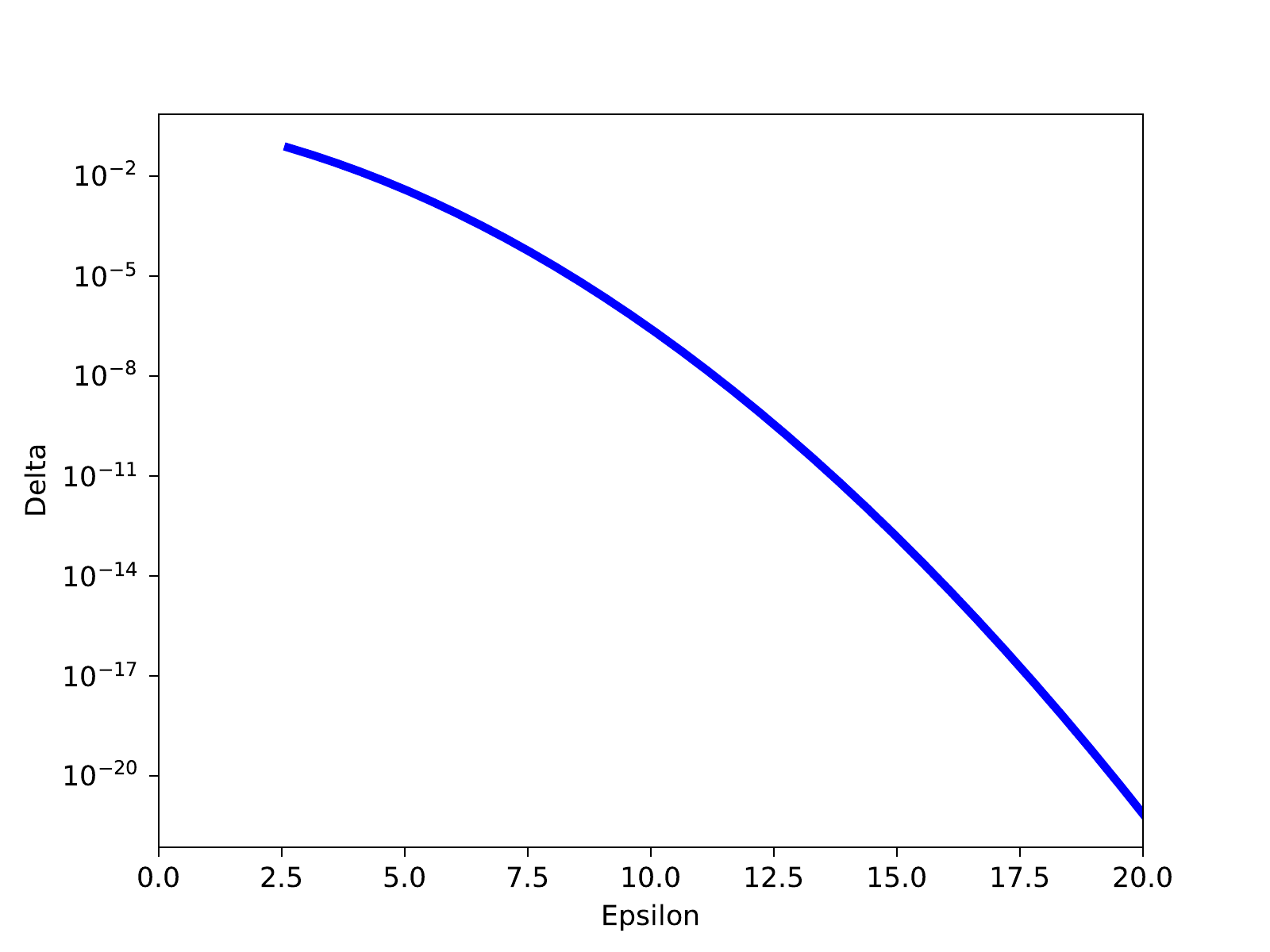}
\caption{Bayesian $(\epsilon,\delta)$ curve under the semantics of Theorem \ref{thm:bayes_1_rdp} for $\rho=2.63$.}\label{fig:bayes1epsdelta}
\end{figure}

\begin{theoremEnd}[category=bayes]{theorem}\label{thm:bayes_1_rdp}
Let $\prior$ be an attacker's prior such that $\prior(\rem[\datarv]=\rem[\data])=1$ for some $\rem[\data]$. Let $\mech$ be a mechanism and let $P_{\outp}$ be the distribution of the output $\outp$ of $\mech$ under the assumption that the attacker's prior $\prior$ is correct (i.e., a record for the target individual is sampled $\recordrv\sim\prior(\recordrv~|~\rem[\data])$ and then $\outp=\mech(\add{\recordrv}{\rem[\data]})$ is produced). 
%and let the true dataset be $\data = \add{\recordrv}{\rem[\data]}$ where $\recordrv\sim\prior(\recordrv~|~\rem[\data])$. Let $\mech$ be a mechanism and $\outp=\mech(\data)$. Let $P_{\outp}$ denote the probability distribution of $\outp$. 
Then for any $\record$,
\begin{itemize}
\item If $\mech$ satisfies $(\alpha,\gamma)$-RDP for $\alpha>1$, then $P_{\outp}\left(\frac{P(\mech(\add{\record}{\rem[\data]})=\outp)}{\sum_{\record^\prime}\prior(\recordrv=\record^\prime~|~\rem[\data])P(\mech(\add{\record^\prime}{\rem[\data]})=\outp)}
\geq e^\epsilon \right) \leq e^{-(\epsilon-\gamma)\alpha - \gamma}$  (i.e., the probability of seeing an $\omega$ for which the posterior-to-posterior ratio exceeds $e^\epsilon$ is at most $\delta=e^{-(\epsilon-\gamma)\alpha - \gamma}$).
\item If $\mech$ satisfies $\rho$-zCDP and $\epsilon>\rho$, then $P_{\outp}\left(\frac{P(\mech(\add{\record}{\rem[\data]})=\outp)}{\sum_{\record^\prime}\prior(\recordrv=\record^\prime~|~\rem[\data])P(\mech(\add{\record^\prime}{\rem[\data]})=\outp)}
\geq e^\epsilon \right) \leq e^{-(\epsilon+\rho)^2/(4\rho)}$.  
\end{itemize}
\end{theoremEnd}
\begin{proofEnd}

\begin{align*}
\lefteqn{P_{\outp}\left(\frac{P(\mech(\add{\record}{\rem[\data]})=\outp)}{\sum_{\record^\prime}\prior(\recordrv=\record^\prime~|~\rem[\data])P(\mech(\add{\record^\prime}{\rem[\data]})=\outp)}
\geq e^\epsilon \right) }\\
&=P_{\outp}\left(\left(\frac{P(\mech(\add{\record}{\rem[\data]})=\outp)}{\sum_{\record^\prime}\prior(\recordrv=\record^\prime~|~\rem[\data])P(\mech(\add{\record^\prime}{\rem[\data]})=\outp)}
\right)^{\alpha}
\geq e^{\epsilon\alpha}\right)\\
&\leq e^{-\epsilon\alpha}E_{\outp}\left[\left(\frac{P(\mech(\add{\record}{\rem[\data]})=\outp)}{\sum_{\record^\prime}\prior(\recordrv=\record^\prime~|~\rem[\data])P(\mech(\add{\record^\prime}{\rem[\data]})=\outp)}\right)^{\alpha}\right] \quad\text{by Markov's inequality}\\
&= \sum_{\outp} e^{-\epsilon\alpha}\left(\frac{P(\mech(\add{\record}{\rem[\data]})=\outp)}{\sum_{\record^\prime}\prior(\recordrv=\record^\prime~|~\rem[\data])P(\mech(\add{\record^\prime}{\rem[\data]})=\outp)}\right)^{\alpha} \left(\sum_{\record^{\prime\prime}}\prior(\recordrv=\record^{\prime\prime}~|~\rem[\data])P(\mech(\add{\record^{\prime\prime}}{\rem[\data]})=\outp)\right)\\
&= \sum_{\outp} e^{-\epsilon\alpha}\frac{\left(P(\mech(\add{\record}{\rem[\data]})=\outp)\right)^\alpha}{\left(\sum_{\record^\prime}\prior(\recordrv=\record^\prime~|~\rem[\data])P(\mech(\add{\record^\prime}{\rem[\data]})=\outp)\right)^{\alpha-1}} \\
&\leq \sum_{\outp} e^{-\epsilon\alpha}
\sum_{\record^\prime}\prior(\recordrv=\record^\prime~|~\rem[\data])
\frac{\left(P(\mech(\add{\record}{\rem[\data]})=\outp)\right)^\alpha}{\left(P(\mech(\add{\record^\prime}{\rem[\data]})=\outp)\right)^{\alpha-1}} \quad\text{ by convexity of $f(z)=1/z^{\alpha-1}$ when $\alpha> 1$}\\
%
%&=e^{-\epsilon\alpha} \sum_{\record^\prime}\prior(\recordrv=\record^\prime~|~\rem[\data])\sum_{\outp} 
%\frac{\left(P(\mech(\add{\record}{\rem[\data]})=\outp)\right)^\alpha}{\left(P(\mech(\add{\record^\prime}{\rem[\data]})=\outp)\right)^{\alpha-1}} \\
%
&=e^{-\epsilon\alpha} \sum_{\record^\prime}\prior(\recordrv=\record^\prime~|~\rem[\data])\sum_{\outp} 
\frac{\left(P(\mech(\add{\record}{\rem[\data]})=\outp)\right)^\alpha}{\left(P(\mech(\add{\record^\prime}{\rem[\data]})=\outp)\right)^{\alpha-1}} \\
\intertext{Note that $\frac{1}{\alpha-1}$ times $\log$ of the inner summation over $\outp$ is a R\'enyi divergence.}
&\leq e^{-\epsilon\alpha} \sum_{\record^\prime}\prior(\recordrv=\record^\prime~|~\rem[\data])e^{(\alpha-1)\gamma} \quad\text{by definition of RDP}\\
&= e^{-\epsilon\alpha} e^{(\alpha-1)\gamma} = e^{-(\epsilon-\gamma)\alpha - \gamma}.
\end{align*}

Now we note that if $\mech$ satisfies $\rho$-zCDP then it satisfies $(\alpha, \rho \alpha)$-RDP for every $\alpha>1$, and we can choose the $\alpha$ that minimizes the above expression when $\gamma$ is replaced by $\rho\alpha$, which is $\alpha^*=\frac{\epsilon+\rho}{2\rho}$. Note that $\alpha^*>1$ when $\epsilon>\rho$. Plugging in $\rho\alpha^*$ for $\gamma$, we get:

\begin{align*}
P_{\outp}\left(\frac{P(\mech(\add{\record}{\rem[\data]})=\outp)}{\sum_{\record^\prime}\prior(\recordrv=\record^\prime~|~\rem[\data])P(\mech(\add{\record^\prime}{\rem[\data]})=\outp)}
\geq e^\epsilon \right) 
&\leq e^{-(\epsilon - \frac{\epsilon+\rho}{2})\frac{\epsilon+\rho}{2\rho} - \frac{\epsilon+\rho}{2}}\\
&= e^{-\frac{\epsilon-\rho}{2}\frac{\epsilon+\rho}{2\rho}-\frac{\epsilon+\rho}{2}}\\
&=e^{-\frac{\epsilon+\rho}{2}(1+\frac{\epsilon-\rho}{2\rho})}\\
&=e^{-\frac{\epsilon+\rho}{2}\frac{\epsilon+\rho}{2\rho}}=e^{-(\epsilon+\rho)^2/(4\rho)}.
\end{align*}
\end{proofEnd}
We note that the probabilities that the ratios exceed $e^\epsilon$ only depend on $\mech$ through the RDP privacy parameters $(\alpha, \gamma)$ or the zCDP privacy parameter $\rho$. Post-processing $\mech$ cannot increase $\gamma$ or $\rho$, and the upper bounds given in Theorem \ref{thm:bayes_1_rdp} are non-increasing in $\rho$ and $\gamma$, thus making these post-processing invariant upper bounds. In other words, we can replace $\mech$ with $\randalg\circ\mech$, for any $\randalg$ whose domain contains the range of $\mech$, and the bounds will still hold. 
The Bayesian $(\epsilon,\delta)$ curve for the semantics provided by Theorem \ref{thm:bayes_1_rdp}  for the production $\rho$ setting for the 2020 Census redistricting data is shown in Figure \ref{fig:bayes1epsdelta}. More fine-grained results are shown in Section \ref{sec:per}.

We can sharpen this result even more by considering how the posterior-to-posterior ratio changes for the true record. That is, we consider the ratio for a record $\record$ when $\outp$ is drawn from the distribution $\prior(\mech(\datarv)=\outp~|~\recordrv=\record)$. In this case, it is easy to see that $\prior(\mech(\datarv)=\outp~|~\recordrv=\record)=P(\mech(\add{\record}{\rem[\data]})=\outp)$ where the randomness comes only from $\mech$. So, the curve of posterior-to-posterior semantics coincides with the $(\epsilon,\delta)$-\pbdp/$f$-DP curve,\footnote{Recall, the conversion from $f$ to the $(\epsilon,\delta)$ curve is provided after Definition \ref{def:gaussdpalt}.} as shown by the following theorem. 

\begin{theoremEnd}[category=bayes]{theorem}\label{thm:bayes_pbdp} Let $\mech$ be a mechanism that satisfies $f$-DP. Let $\prior$ be an attacker's prior such that $\prior(\rem[\datarv]=\rem[\data])=1$ for some $\rem[\data]$. Let  $P_{\outp~|~\record}$ be the distribution of the output $\outp$ of $\mech$ under the assumption that the attacker's prior $\prior$ is correct and the true record is value $\record$. Define  $f^{-1}(z) = \inf\{y\in[0,1]~:~f(y)\leq z\}$.
Then for any $\record$ and $\delta\in[0,1]$, if $\epsilon \geq\log\frac{\delta}{f^{-1}(1-\delta)}$, then
  \begin{align*}
 P_{\outp|\record}\left(\frac{P(\mech(\add{\record}{\rem[\data]})=\outp)}{\sum_{\record^\prime}\prior(\recordrv=\record^\prime~|~\rem[\data])P(\mech(\add{\record^\prime}{\rem[\data]})=\outp)}
> e^\epsilon \right) \leq \delta.
\end{align*}
Furthermore, this same bound holds if $\mech$ is replaced by $\randalg\circ\mech$.
\end{theoremEnd}
\begin{proofEnd}
First, note that the only uncertainty in $\prior$ is in the record $\record$ of the target person. Hence $\prior(\mech(\datarv)=\outp~|~\recordrv=\record)=P(\mech(\add{\record}{\rem[\data]})=\outp)$. 

Let $\record^{\prime\prime}$ be a record different from $\record$ and for any $\record^\prime$, let $\mech_{\record^\prime}$ be the algorithm that replaces $\record^{\prime\prime}$ with $\record^\prime$. Then it is easy to see that $\mech_{\record^\prime}$ also satisfies $f$-DP. Now, let $\mech^*$ be an algorithm that, on any input $\data^*$ (it does not have to be related to the attacker's $\rem[\data]$ in any way), chooses among the $\mech_{\record^\prime}$ with probability $P(\record^{\prime}~|~\rem[\data])$ and runs the chosen algorithm on $\data^*$. By convexity of $f$-DP, $\mech^*$ also satisfies $f$-DP.

Now, let $\data_1=\add{\record}{\rem[\data]}$, and let $\data_2=\add{\record^{\prime\prime}}{\rem[\data]}$. Then the distribution of $\log\frac{P(\mech(\add{\record}{\rem[\data]})=\outp)}{\sum_{\record^\prime}\prior(\recordrv=\record^\prime~|~\rem[\data])P(\mech(\add{\record^\prime}{\rem[\data]})=\outp)}$ for $\outp\sim P_{\outp~|~\record}$ is the same as for the privacy loss random variable $\plrv[\data_1][\data_2][\mech^*]$. By the equivalence between $f$-DP and the \pbdp reparametrization, $P(\plrv[\mech^*][\data_1][\data_2] > \epsilon)\leq \delta$ when $\epsilon \geq\log\frac{\delta}{f^{-1}(1-\delta)}$. The postprocessing invariance of \pbdp finishes the proof.
\end{proofEnd}

\begin{figure}[h!]
\includegraphics[scale=0.75]{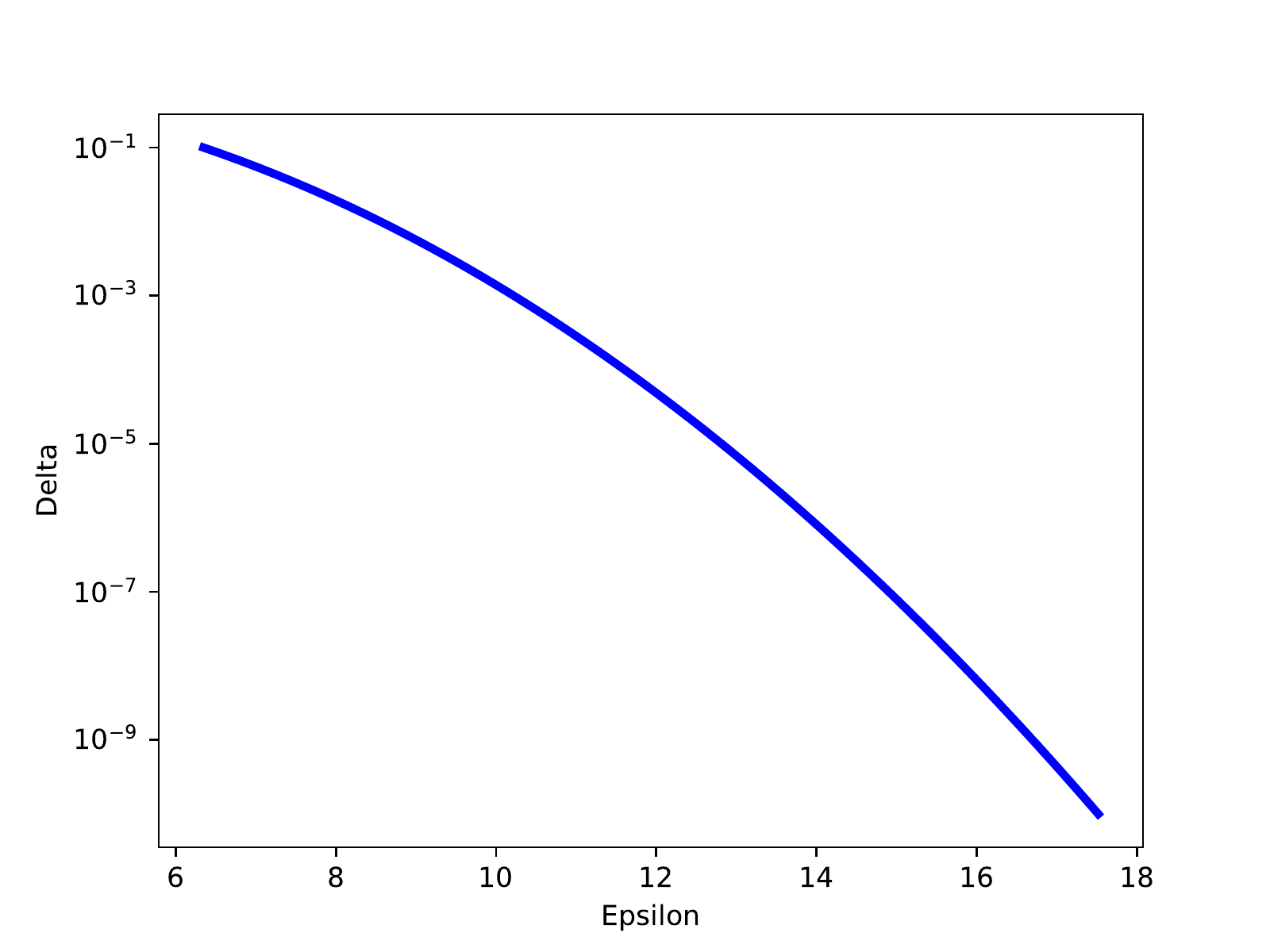}
\caption{Bayesian $(\epsilon,\delta)$ curve for the continuous Gaussian Mechanism satisfying $\rho$-zCDP under the semantics of Theorem \ref{thm:bayes_pbdp}, for $\rho=2.63$.}\label{fig:pdbp_bayes_production}
\end{figure}

The \pbdp curve for the discrete Gaussian mechanism is difficult to compute, however, under the assumption that it would be similar to the curve for the continuous Gaussian mechanism (as in Section \ref{sec:freq}), we can instead analyze the curve for the Gaussian Mechanism that satisfies $\rho$-zCDP.\footnote{The Gaussian mechanism that satisfies $\rho$-zCDP also satisfies $f$-DP with $f(x)=\Phi(\Phi^{-1}(1-x)-\sqrt{2\rho})$ \cite{zcdp,gaussdp}.}
The $(\epsilon,\delta)$-curve for the continuous Gaussian mechanism provided by Theorem \ref{thm:bayes_pbdp} for the production $\rho$ settings is shown in Figure \ref{fig:pdbp_bayes_production}. While the Bayesian guarantees for such a $\rho$ value do not appear to be strong, we show in Section \ref{sec:per} that more fine-grained guarantees can be stronger.

\subsubsection{Bayesian semantics for arbitrary priors}\label{subsec:bayes2}
When the prior is arbitrary, one can still provide semantic guarantees, but the bounds are weaker. In Section \ref{subsec:bayes1}, we derive bounds for the posterior-to-posterior ratio when reasoning about an arbitrary record $\record$. In this section, we provide bounds on the posterior-to-posterior ratio for reasoning about the \emph{correct} record value.

For any given record $\record$, we consider the distribution $\prior(\outp~|~\record)$ and the probability that an $\outp$ from this distribution results in a large posterior-to-posterior ratio $\frac{\prior(\record~|~\mech(\datarv)=\outp)}{\prior(\record ~|~\mech(\psample(\datarv))=\outp)}$ for that same $\record$.
Thus the guarantee bounds the inference about the true record between the actual and counterfactual worlds. 

\begin{figure}[h!]
\includegraphics[scale=0.75]{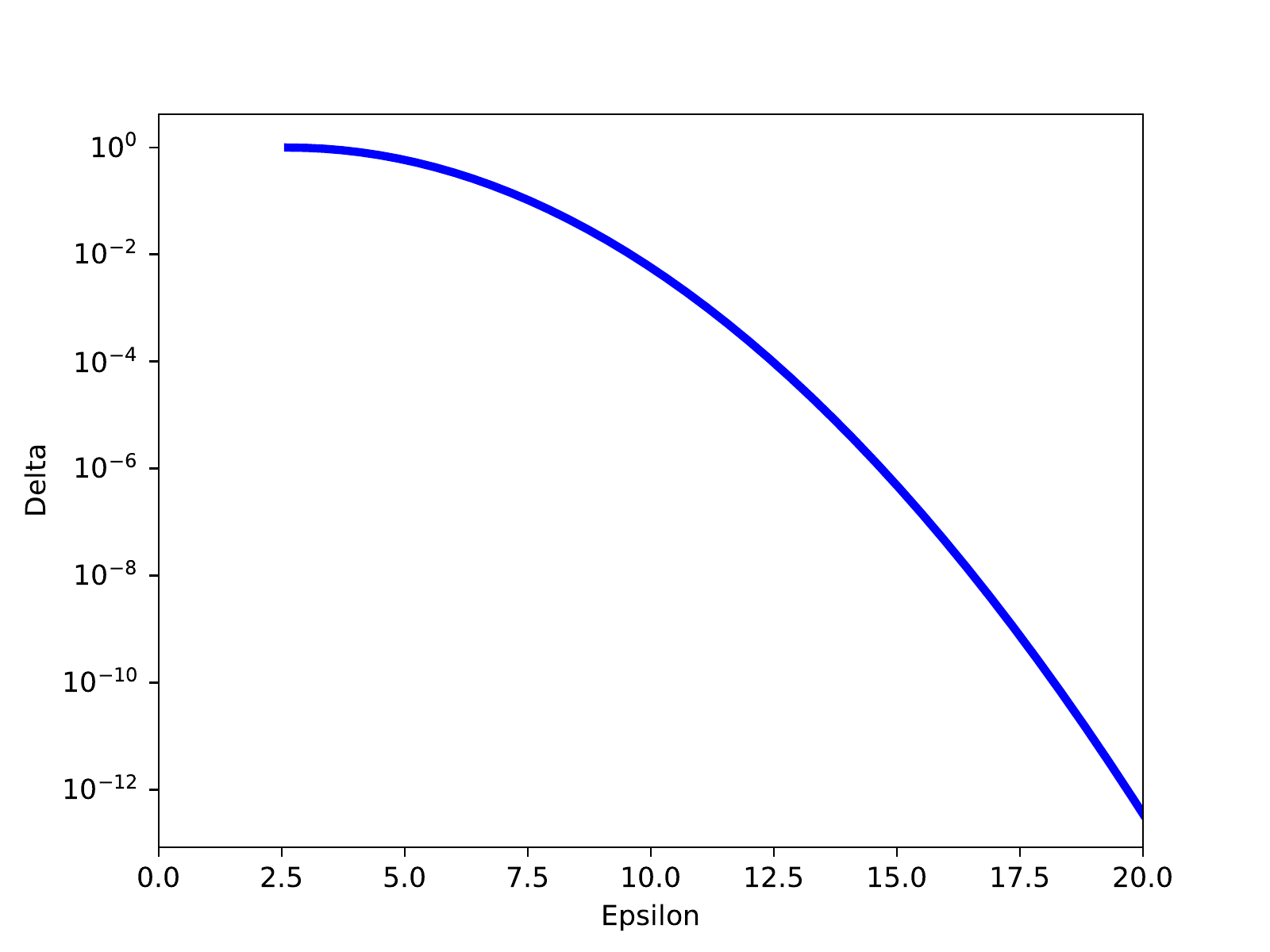}
\caption{Bayesian $(\epsilon,\delta)$ curve under the semantics of Theorem \ref{thm:bayes_2_rdp} for $\rho=2.63$.}\label{fig:bayes2epsdelta}
\end{figure}

\begin{theoremEnd}[category=bayes]{theorem}\label{thm:bayes_2_rdp}
Let $\prior$ be an attacker's prior. Let $\record$ be the target person's true record. Let $\mech$ be a mechanism. Let $P_{\outp~|~\record}$ be the distribution of $\outp$ conditioned on $\record$. That is, a dataset $\data$ is generated from the distribution $\prior(\datarv~|~\record)$ and $\mech(\data)$ produces $\outp$. Then $\mech$ can provide the following guarantees on the posterior-to-posterior ratio $\frac{\prior(\record~|~\mech(\datarv)=\outp)}{\prior(\record ~|~\mech(\psample(\datarv))=\outp)}$ for  the true record $\record$.
\begin{itemize}
\item If $\mech$ satisfies $(\alpha,\gamma)$-RDP for $\alpha>1$ then $P_{\outp~|~\record}\left(\frac{\prior(\record~|~\mech(\datarv)=\outp)}{\prior(\record ~|~\mech(\psample(\datarv))=\outp)} \geq e^\epsilon\right)\leq e^{(\alpha-1)(\gamma-\epsilon)}$.
\item If $\mech$ satisfies $\rho$-zCDP then for $\epsilon>\rho$, $P_{\outp~|~\record}\left(\frac{\prior(\record~|~\mech(\datarv)=\outp)}{\prior(\record ~|~\mech(\psample(\datarv))=\outp)} \geq e^\epsilon\right)\leq e^{-(\epsilon-\rho)^2/(4\rho)}$.
\end{itemize}
\end{theoremEnd}
\begin{proofEnd}
Recall that the target is known to be in the data, that $\dataspace$ is the notation for set of possible datasets, and that $\rem[\dataspace]$ is the set of possible datasets after the target's data have been removed.

We first note that
\begin{align*}
\prior(\mech(\psample(\datarv))=\outp)&=
\sum_{\rem[\data] \in \rem[\dataspace]}\sum_{\record^\prime} \prior(\rem[\data])\prior(\record^\prime ~|~\rem[\data])P(\mech(\psample(\add{\record^\prime}{ \rem[\data]}))=\outp)\\
&=\sum_{\rem[\data] \in \rem[\dataspace]}\sum_{\record^\prime} \prior(\rem[\data])\prior(\record^\prime ~|~\rem[\data])\Big(\sum_{\record^{\prime\prime}}\prior(\record^{\prime\prime}~|~\rem[\data])P(\mech(\add{\record^{\prime\prime}}{\rem[\data]})=\outp)\Big)\\
%
%
%&=\sum_{\rem[\data] \in \rem[\dataspace]}\sum_{\record^{\prime\prime}} \prior(\rem[\data])\sum_{\record^{\prime\prime}}\prior(\record^{\prime\prime}~|~\rem[\data])P(\mech(\add{\record^{\prime\prime}}{\rem[\data]})=\outp)\\
&=\sum_{\rem[\data] \in \rem[\dataspace]} \prior(\rem[\data])\sum_{\record^{\prime\prime}}\prior(\record^{\prime\prime}~|~\rem[\data])P(\mech(\add{\record^{\prime\prime}}{\rem[\data]})=\outp)\\
&=\prior(\mech(\datarv)=\outp)
\end{align*}
So  the marginal probability of $\outp$ is the same in the actual and counterfactual world (but the joint distribution between \target's record and $\outp$ differs since in the actual world it is part of the input to $\mech$). 

Next, 
\begin{align*}
\lefteqn{    P_{\outp~|~\record}\left(\frac{\prior(\record~|~\mech(\datarv)=\outp)}{\prior(\record ~|~\mech(\psample(\datarv))=\outp)} \geq e^\epsilon\right)}\\
&=P_{\outp~|~\record}\left(\left(\frac{\prior(\record~|~\mech(\datarv)=\outp)}{\prior(\record ~|~\mech(\psample(\datarv))=\outp)}\right)^{\alpha-1} \geq e^{\epsilon(\alpha-1)}\right) \quad\text{for any $\alpha > 1$}\\
&\leq e^{-\epsilon(\alpha-1)}E_{\outp~|~\record}\left[\left(\frac{\prior(\record~|~\mech(\datarv)=\outp)}{\prior(\record ~|~\mech(\psample(\datarv))=\outp)}\right)^{\alpha-1} \right] \\
&= e^{-\epsilon(\alpha-1)} \sum_\outp\left[\left(\frac{\prior(\record~|~\mech(\datarv)=\outp)}{\prior(\record ~|~\mech(\psample(\datarv))=\outp)}\right)^{\alpha-1} \right]\prior(\mech(\add{\record}{\rem[\datarv]})=\outp~|~\record) \\
&= e^{-\epsilon(\alpha-1)} \sum_\outp\left[\left(\frac{\prior(\record~,~\mech(\datarv)=\outp)/\prior(\mech(\datarv)=\outp)}{\prior(\record ~,~\mech(\psample(\datarv))=\outp)/\prior(\mech(\psample(\datarv))=\outp)}\right)^{\alpha-1} \right]\prior(\mech(\add{\record}{\rem[\datarv]})=\outp~|~\record) \\
&= e^{-\epsilon(\alpha-1)} \sum_\outp\left[\left(\frac{\prior(\record~,~\mech(\datarv)=\outp)}{\prior(\record ~,~\mech(\psample(\datarv))=\outp)}\right)^{\alpha-1} \right]\prior(\mech(\add{\record}{\rem[\datarv]})=\outp~|~\record) \\
&\quad\text{(Since }\prior(\mech(\datarv)=\outp) = \prior(\mech(\psample(\datarv))=\outp)\text{)}\\
&= e^{-\epsilon(\alpha-1)} \sum_\outp\left[\left(\frac{\prior(\mech(\datarv)=\outp~|~\record)}{\prior(\mech(\psample(\datarv))=\outp~|~\record)}\right)^{\alpha-1} \right]\prior(\mech(\add{\record}{\rem[\datarv]})=\outp~|~\record) \\
&= e^{-\epsilon(\alpha-1)} \sum_\outp\left[\left(\frac{\prior(\mech(\add{\record}{\rem[\datarv]})=\outp~|~\record)}{\prior(\mech(\psample(\add{\record}{\rem[\datarv]}))=\outp~|~\record)}\right)^{\alpha-1} \right]\prior(\mech(\add{\record}{\rem[\datarv]})=\outp~|~\record) \\
&= e^{-\epsilon(\alpha-1)} \sum_\outp\left[\left(\frac{\prior(\mech(\add{\record}{\rem[\datarv]})=\outp~|~\record)}{\prior(\mech(\psample(\add{\record}{\rem[\datarv]}))=\outp~|~\record)}\right)^{\alpha} \right]\prior(\mech(\psample(\add{\record}{\rem[\datarv]}))=\outp~|~\record) \\
&= e^{-\epsilon(\alpha-1)} \sum_\outp\left(\frac
{
\sum_{\rem[\data]\in\rem[\dataspace] }\prior(\add{\record}{\rem[\data]}~|~\record)P(\mech(\add{\record}{\rem[\data]})=\outp)
}{
\sum_{\rem[\data]\in\rem[\dataspace]}\prior(\add{\record}{\rem[\data]}~|~\record)\sum_{\record^\prime} \prior(\record^\prime~|~\rem[\data])P(\mech(\add{\record^\prime}{\rem[\data]})=\outp)
}
\right)^{\alpha}\prior(\mech(\psample(\add{\record}{\rem[\datarv]}))=\outp~|~\record)\\
&= e^{-\epsilon(\alpha-1)} \sum_\outp\left(\frac
{
\sum_{\rem[\data]\in\rem[\dataspace] }\prior(\add{\record}{\rem[\data]}~|~\record)(\sum_{\record^\prime} \prior(\record^\prime~|~\rem[\data]))P(\mech(\add{\record}{\rem[\data]})=\outp)
}{
\sum_{\rem[\data]\in\rem[\dataspace]}\prior(\add{\record}{\rem[\data]}~|~\record)\sum_{\record^\prime} \prior(\record^\prime~|~\rem[\data])P(\mech(\add{\record^\prime}{\rem[\data]})=\outp)
}
\right)^{\alpha}\prior(\mech(\psample(\add{\record}{\rem[\datarv]}))=\outp~|~\record)\\
&\quad\text{Now, use Jensen's inequality ($\alpha>1$), note numerator/denominator only differ on record used by $\mech$.}\\
&\leq e^{-\epsilon(\alpha-1)} \sum_\outp\sum_{\rem[\data]}\sum_{\record^\prime}
\frac
{
\prior(\add{\record}{\rem[\data]}~|~\record) \prior(\record^\prime~|~\rem[\data])P(\mech(\add{\record^\prime}{\rem[\data]})=\outp)
}{
\sum_{\rem[\data]}\prior(\add{\record}{\rem[\data]}~|~\record)\sum_{\record^\prime} \prior(\record^\prime~|~\rem[\data])P(\mech(\add{\record^\prime}{\rem[\data]})=\outp)}
\left(\frac{P(\mech(\add{\record}{\rem[\data]})=\outp)}{P(\mech(\add{\record^\prime}{\rem[\data]})=\outp)}\right)^\alpha\\
&\qquad * \prior(\mech(\psample(\add{\record}{\rem[\datarv]}))=\outp~|~\record)\\
&= e^{-\epsilon(\alpha-1)} \sum_\outp\sum_{\rem[\data]}\sum_{\record^\prime}
\frac
{
\prior(\add{\record}{\rem[\data]}~|~\record) \prior(\record^\prime~|~\rem[\data])P(\mech(\add{\record^\prime}{\rem[\data]})=\outp)
}{
\prior(\mech(\psample(\add{\record}{\rem[\datarv]}))=\outp~|~\record)}
\left(\frac{P(\mech(\add{\record}{\rem[\data]})=\outp)}{P(\mech(\add{\record^\prime}{\rem[\data]})=\outp)}\right)^\alpha\\
&\qquad * \prior(\mech(\psample(\add{\record}{\rem[\datarv]}))=\outp~|~\record)\\
&= e^{-\epsilon(\alpha-1)} \sum_{\rem[\data]}\sum_{\record^\prime}
\prior(\add{\record}{\rem[\data]}~|~\record) \prior(\record^\prime~|~\rem[\data])\left(\sum_\outp P(\mech(\add{\record^\prime}{\rem[\data]})=\outp)
\left(\frac{P(\mech(\add{\record}{\rem[\data]})=\outp)}{P(\mech(\add{\record^\prime}{\rem[\data]})=\outp)}\right)^\alpha\right)\\
&\leq e^{-\epsilon(\alpha-1)} \sum_{\rem[\data]}\sum_{\record^\prime}
\prior(\add{\record}{\rem[\data]}~|~\record) \prior(\record^\prime~|~\rem[\data])
e^{(\alpha-1)\gamma} \quad\text{by $(\alpha,\gamma)$-RDP}\\
&= e^{-\epsilon(\alpha-1)}e^{(\alpha-1)\gamma} = e^{(\alpha-1)(\gamma-\epsilon)}
\end{align*}

In the case of $\rho$-zCDP, this is satisfied with $\gamma=\rho\alpha$ for all $\alpha>1$ and thus we can choose the best $\alpha$.
The optimal choice is $\alpha=\frac{\epsilon+\rho}{2\rho}$, which is greater than 1 whenever $\epsilon> \rho$. Plugging in this value of $\alpha$ and using $\gamma=\rho\alpha$, we see that the probability that the posterior-to-posterior ratio exceeds $e^\epsilon$ is at most:

\begin{align*}
e^{(\frac{\epsilon+\rho}{2\rho}-1)(\frac{\epsilon+\rho}{2}-\epsilon)}&=e^{\frac{\epsilon-\rho}{2\rho}\frac{-\epsilon+\rho}{2}}\\
&=e^{-(\epsilon-\rho)^2/(4\rho)}.
\end{align*}

\end{proofEnd}
\noindent Again note that these bounds are post-processing invariant in that $\mech$ can be replaced with $\randalg\circ\mech$ without increasing the upper bounds.
The  $(\epsilon,\delta)$-curve of Theorem \ref{thm:bayes_2_rdp} for the production settings for the 2020 Census redistricting data is shown in Figure \ref{fig:bayes2epsdelta}.

By comparing Theorem \ref{thm:bayes_2_rdp} with Theorem \ref{thm:bayes_1_rdp}, we see that the $\rho$-zCDP estimate of $\delta=e^{-(\epsilon-\rho^2)/4\rho}$ of Theorem \ref{thm:bayes_2_rdp} is larger than that of Theorem \ref{thm:bayes_1_rdp}.
However, the $\delta$ from Theorem \ref{thm:bayes_2_rdp} is exactly the same as the $\rho$-zCDP estimate of the $\delta$ parameter of $\pbdp$ (see Section \ref{sec:rzcdp}).

%% file: per.tex
In this section, we explain how to use the structure of a mechanism  to provide fine-grained privacy guarantees beyond those based on the overall privacy-loss parameters. This is an illustration of how a statistical agency could make additional choices about the privacy properties of the mechanisms it uses.

The structure of a mechanism allows an agency to consider scenarios with specialized privacy-violating inferences. For example, suppose \target has special concerns about inferences of detailed location such as a census block within a larger area such as a block group or tract but is unconcerned about the other attributes (and thus is also unconcerned about how the other attributes affect fine-grained location inference). In other words, more populous areas may not be as sensitive for \target as an area as small as a census block, which on average house about 50 persons. Or, \target may have special concerns only about inferences regarding race, ethnicity, marital status, or the composition of their household.

In these situations, the inferential focus is not on arbitrary pairs of neighbors, but rather a subset of those pairs of neighbors, and the fine-grained privacy guarantees are obtained by only examining the privacy-loss random variables associated with this subset of pairs of neighbors. In the first example, $\data_1$ and $\data_2$ would only differ on the block in \target's record, but would have the same tract. In another example, a pair of neighbors $\data_1,\data_2$ would only differ on the race and ethnicity in \target's record.

When focusing on such a more limited set of neighboring databases, additional privacy semantics can be provided. These semantics take similar forms to those described in Sections \ref{sec:privdefs}, \ref{sec:freq}, and \ref{sec:bayes} but the appropriate parameters and the resulting trade-off curves are different.

Rather than discuss these issues abstractly, in this section we use the disclosure avoidance system as implemented for the 2020 Census  redistricting data as a case-study to illustrate how such fine-grained semantics are derived. To facilitate this discussion, we first provide a brief overview of the noise module of the TopDown Algorithm, which implements differential privacy, and then we examine the fine-grained privacy semantics it provides. More comprehensive details about the TopDown Algorithm and the data it uses can be found in \cite{tdahdsr}.

\subsection{An Overview of the Noise Module}

The TopDown Algorithm, used to create privacy-protected microdata for the 2020 Census, consists of two stages \cite{tdahdsr}. The first stage computes query answers and adds noise to them. These noisy query answers are called \emph{noisy measurements} and are designed to ensure differential privacy.\footnote{There is a complication due to \emph{invariants} -- properties of the input data that are not known a prior but which must be enforced in the privacy-protected data. One example is that the state-level population counts, which are released as tabulated from the confidential microdata. A study of invariants and how they can alter the privacy guarantees is the subject of a future paper.} The second stage post-processes these noisy measurements to output privacy-protected microdata. The post-processing phase and its consequences are studied in other papers \cite{tdahdsr,neuripsuncertainty}. Post-processing is largely irrelevant to this paper because all semantics we discuss are post-processing invariant. Thus, we only need to focus on the privacy semantics provided by the noisy measurements.

A query $\query$ is specified by a set of attributes, a unit of geography, and zCDP parameter $\rho^*$. One example is VOTINGAGE $\times$ CENRACE for Fairfax County with $\rho^*=0.0175$. This query computes the number of people who are 18 and over for each of the 63 possible race combinations specified in Statistical Policy Directive 15 (add citation to OMB 1997) and the Census Act (add citation to 13 USC Section 191) in Fairfax County, and also the number who are 17 and under for each race combination in Fairfax County. Thus, the query answer is a vector of 126 numbers. Independent Discrete Gaussian noise with scale $\sigma=\sqrt{1/\rho^*}$ is added to each of these numbers. This noise has probability mass function:
\begin{align*}
f(k) = \frac{e^{-k^2/(2\sigma^2)}}{\sum_{j=-\infty}^\infty e^{-j^2/(2\sigma^2)}}.
\end{align*}

\noindent The types of queries are:
\begin{itemize}
\item TOTAL: the total number of people in the geographic unit.
\item CENRACE: the number of people in each of the 63 OMB race combinations in the geographic unit.
\item HISPANIC: the number of people who are Hispanic in the geographic unit and the number who are not Hispanic.
\item VOTINGAGE: the number of people who are 18 and over in the geographic unit, and the number of people who are 17 and under.
\item HHINSTLEVELS: the number of people who live in households, institutionalized group quarters, and non-institutionalized group quarters in the geographic unit.
\item HHGQ: the number of people who live in a household and the number of people who live in each of the 7 major group quarters type, in the geographic region.
\item HISPANIC $\times$ CENRACE: number of people who are Hispanic for each of the 63 OMB race combinations and similarly for non-Hispanic.
\item VOTINGAGE $\times$ CENRACE: the number of people who are 18 and over for each of the 63 possible race combinations and similarly for people who are 17 and under.
\item VOTINGAGE $\times$ HISPANIC: number of people for each combination of the VOTINGAGE and HISPANIC variables.
\item VOTINGAGE $\times$ HISPANIC $\times$ CENRACE: number of people for each combination of the VOTINGAGE, HISPANIC, and CENRACE variables.
\item HHGQ $\times$ VOTINGAGE $\times$ HISPANIC $\times$ CENRACE: number of people for each combination of person-level demographics for each HHGQ value.
\item OCCUPANCY STATUS: Number of occupied housing units in the geographic region and the number of unoccupied housing units.
\end{itemize}

The $\rho^*$ value associated with a query in a geographic unit can be computed from Tables \ref{table:rhovalues}, \ref{table:geographyrho}, \ref{table:personrho}, reproduced from \cite{tdahdsr}. To obtain the $\rho^*$ value for Occupancy Status at a particular geographic location (e.g., Fairfax County), one multiplies the base $\rho$ value for housing characteristics from Table \ref{table:rhovalues} (i.e., 0.07) by the allocation for the geographic level of that location in Table \ref{table:geographyrho} (e.g., $7/82$ for counties, resulting in a $\rho^*=0.07 *7/82\approx 0.0060$). For queries on person characteristics (e.g., VOTINGAGE $\times$ CENRACE in New York), one multiplies the base $\rho$ value for person characteristics from Table \ref{table:rhovalues} (i.e., 2.56) by the allocation for the geographic level of that location in Table \ref{table:geographyrho} (e.g., 1440/4099 for states) by the allocation for that query in Table \ref{table:personrho} (e.g., 12/4097, resulting in a $\rho^*=2.56*1440/4099 * 12/4097\approx 0.0026$).

\begin{table}[h!]
\caption{Base $\rho$ value for Person Characteristics and Housing Unit Characteristic, Production Settings \cite{tdahdsr}}\label{table:rhovalues}
\begin{tabular}{|l|l|}\hline
Person Characteristics & Housing Unit Characteristics\\\hline
2.56 & 0.07\\\hline
\end{tabular}
\end{table}

\begin{table}[h!]
\caption{Per Geographic Level $\rho$ Allocation Proportions for Persons and Housing Units, Production Settings \cite{tdahdsr}}
\begin{tabular}{|l|l|l|}
\hline
Geographic Level & Person $\rho$ Proportions & Housing Units $\rho$ Proportions \\
\hline
US                     & 104/4099 & 1/205                            \\
State                  & 1440/4099  & 1/205                          \\
County                 & 447/4099     & 7/82                       \\
Tract                  & 687/4099    & 364/1025                      \\
Custom Block Group* & 1256/4099   & 1759/4100                      \\
Block                  & 165/4099 & 99/820         \\
\hline
\multicolumn{3}{|p{12.5cm}|}{\footnotesize{*Custom block groups differ from tabulation block groups and are only used by the TDA.}}\\
\hline
\end{tabular} \label{table:geographyrho}
\end{table}

\begin{table}[h!] 
\caption{Per Query $\rho$ Allocation Proportions by Geographic Level for Persons, Production Settings \cite{tdahdsr}}
\begin{tabular}{|p{4.2cm} |p{1.25cm}p{1.5cm}p{1.5cm}p{1.5cm}p{1.5cm}p{1.5cm}|}
\hline
Query & \multicolumn{6}{c|}{Per Query $\rho$ Allocation Proportions by Geographic Level}\\
\hline
& US & State      & County   & Tract    & CBG* & Block    \\
\hline
TOTAL (1 cell)  &      0    & 3773/4097 & 3126/4097 & 1567/4102   & 1705/4099                 & 5/4097   \\
CENRACE (63 cells)        & 52/4097   & 6/4097     & 10/4097   & 4/2051   & 3/4099                   & 9/4097   \\
HISPANIC (2 cells)        & 26/4097   & 6/4097     & 10/4097   & 5/4102   & 3/4099                   & 5/4097   \\
VOTINGAGE (2 cells)        & 26/4097   & 6/4097     & 10/4097   & 5/4102   & 3/4099                   & 5/4097   \\
HHINSTLEVELS (3 cells)    & 26/4097   & 6/4097     & 10/4097   & 5/4102   & 3/4099                   & 5/4097   \\
HHGQ (8 cells)            & 26/4097   & 6/4097     & 10/4097   & 5/4102   & 3/4099                  & 5/4097   \\
HISPANIC$\times$CENRACE (126 cells)    & 130/4097   & 12/4097     & 28/4097   & 1933/4102   & 1055/4099                  & 21/4097  \\
VOTINGAGE$\times$CENRACE (126 cells)  & 130/4097   & 12/4097     & 28/4097   & 10/2051   & 9/4099                   & 21/4097   \\
VOTINGAGE$\times$HISPANIC (4 cells)   & 26/4097   & 6/4097     & 10/4097   & 5/4102   & 3/4099                   & 5/4097  \\
VOTINGAGE$\times$HISPANIC $\times$CENRACE (252 cells)  & 26/241  & 2/241     & 101/4097  & 67/4102  & 24/4099                   & 71/4097  \\
HHGQ$\times$VOTINGAGE $\times$HISPANIC$\times$CENRACE (2,016 cells) & 189/241 & 230/4097   & 754/4097 & 241/2051 & 1288/4099                & 3945/4097 \\
\hline
\multicolumn{7}{|p{14.5cm}|}{\footnotesize{*Custom block groups (CBG) differ from tabulation block groups and are only used by the TDA.}}\\
\hline
\end{tabular} \label{table:personrho}
\end{table}

We note that in this geographic hierarchy, each state containing American Indian or Alaska Native areas is split into the AIAN and non-AIAN portions. Each of these latter portions is at the same level of the geographic hierarchy as states without AIAN areas. Thus, the ``state'' level of the hierarchy has more than 51 entries \cite{tdahdsr}). Furthermore, the block groups for which noisy measurements are computed are custom block groups specific to the disclosure avoidance system and are different from the tabulation block groups.

The overall privacy accounting for zCDP is as follows \cite{tdahdsr,zcdp}. For a single block, we add up the $\rho^*$ values for all queries in that block, and all queries in the block group, tract, county, and state containing that block, as well as the $\rho^*$ values at the national level. The maximum of these values among all blocks is the overall $\rho$ parameter for zCDP (these computed values happen to be the same for each block). In this case, $\rho=2.56+0.07=2.63$.

\subsection{The Fine-Grained Semantics Case Study}

The guarantees for the overall privacy parameters of the algorithm were discussed in Sections \ref{sec:privdefs}, \ref{sec:freq}, and \ref{sec:bayes}.
Given the allocation of privacy-loss budget among different queries,  one can analyze the additional fine-grained privacy protections that are available in a variety of situations. In this section, we consider the following scenarios.

\subsubsection{Scenario A:} \target is concerned about protecting the exact location within a block group (so \target's custom block group would not be very sensitive, but the specific block within it would be). In this setting, the attacker possesses statistical knowledge and combines it with \target's block group. The privacy guarantees are related to  the causal effect of using \target's confidential block as part of the input to the TopDown Algorithm. In the frequentist case, the guarantee determines how well the attacker can distinguish between the setting when the confidential block was part of the input compared to using some other block within the same custom block group. In the Bayesian case, the guarantee determines how much better the attacker's inference about the confidential block is in the actual world compared to the inference the attacker would have in the counterfactual world.

\begin{figure}[h]
\includegraphics[scale=0.8]{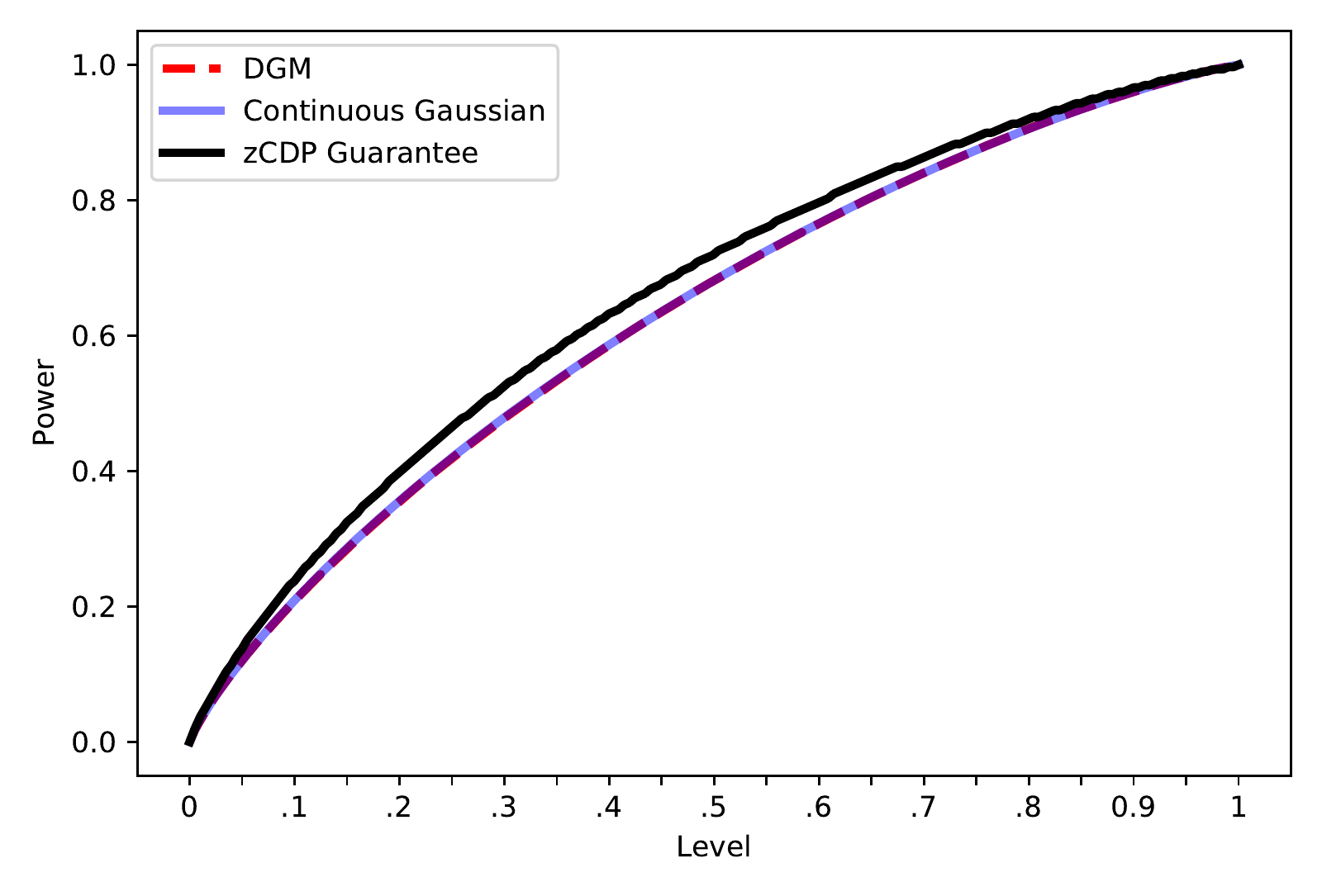}
\caption{\textbf{Block within Custom Block Group:} Level (x-axis) vs. power (y-axis) curves for (1) the Gaussian mechanism over block-level queries at production settings for redistricting data ($\rho=0.1115$), (2) the likelihood ratio test of the discrete Gaussian block-level noisy queries at production settings for redistricting data. }
\label{fig:dgm_block}.
\end{figure}

\begin{table}[h]
\begin{tabular}{|c||c|c|c|}\hline
\textbf{Significance Level} & \textbf{Power (Gaussian)} & \textbf{Power (DGM)} &
\textbf{zCDP Upper Bound}\\\hline
0.01
&0.03 
&0.03 
&0.04\\\hline
0.05
&0.12
&0.12
&0.14\\\hline
0.10
&0.21
&0.21
&0.24\\\hline
\end{tabular}
\caption{\textbf{Block within Custom Block Group:} Likelihood ratio test significance level/power trade-off for block-level queries (1) if Gaussian noise is used, (2) if discrete Gaussian noise is used, (3) guaranteed upper bound if an arbitrary $\rho$-zCDP mechanism with $\rho=0.1115$ is used.}\label{tab:dgm_block}
\end{table}

\begin{figure}[h!]
\includegraphics[scale=0.75]{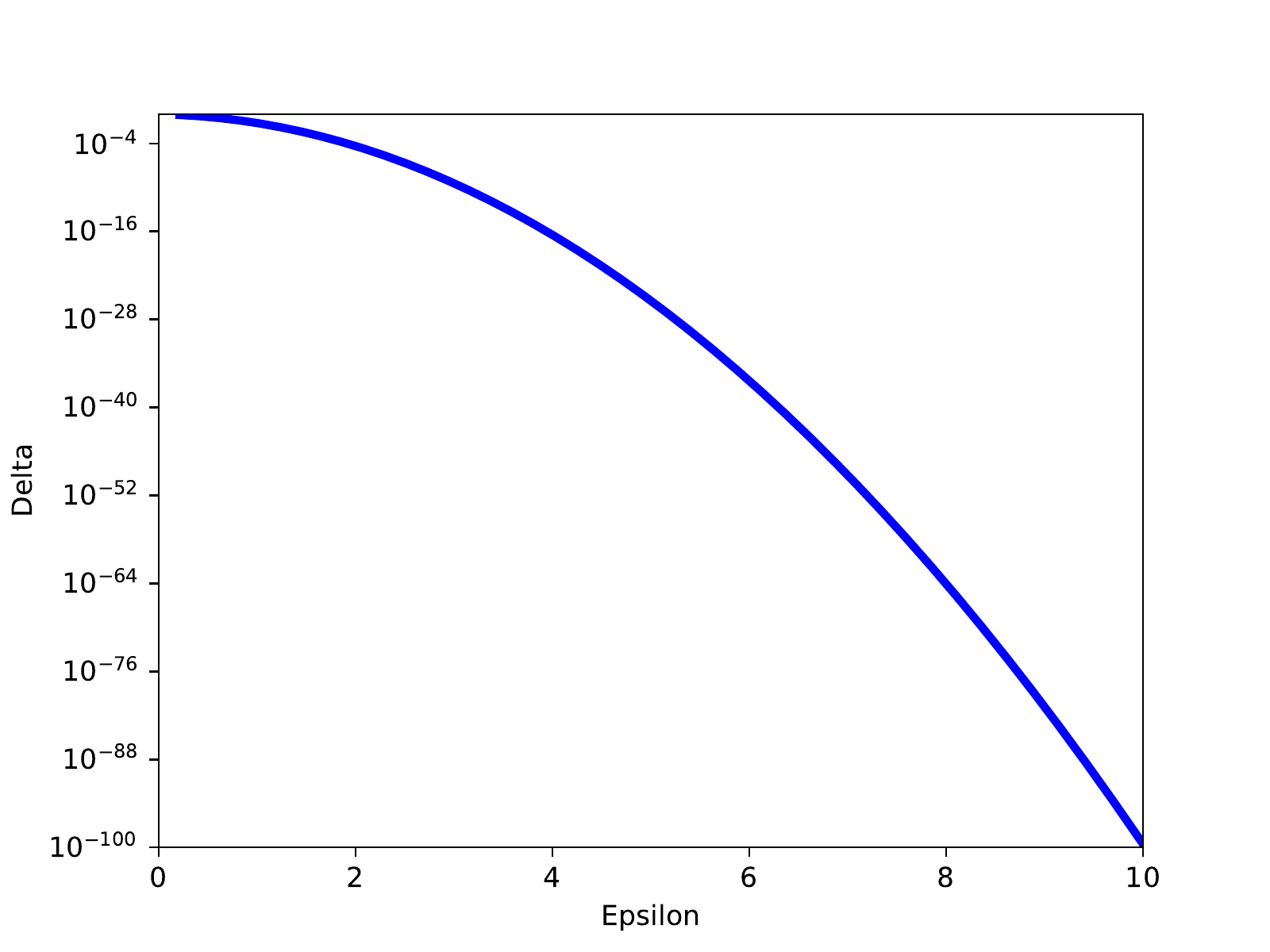}
\caption{Bayesian $(\epsilon,\delta)$ curve under the semantics of Theorem \ref{thm:bayes_1_rdp} (attacker knows all but one person) for block within custom block group inference under the privacy-loss budget allocated to block-level queries under the redistricting data production settings ($\rho=0.1115$).}\label{fig:bayes1epsdelta_block}
\end{figure}

\begin{figure}[h!]
\includegraphics[scale=0.75]{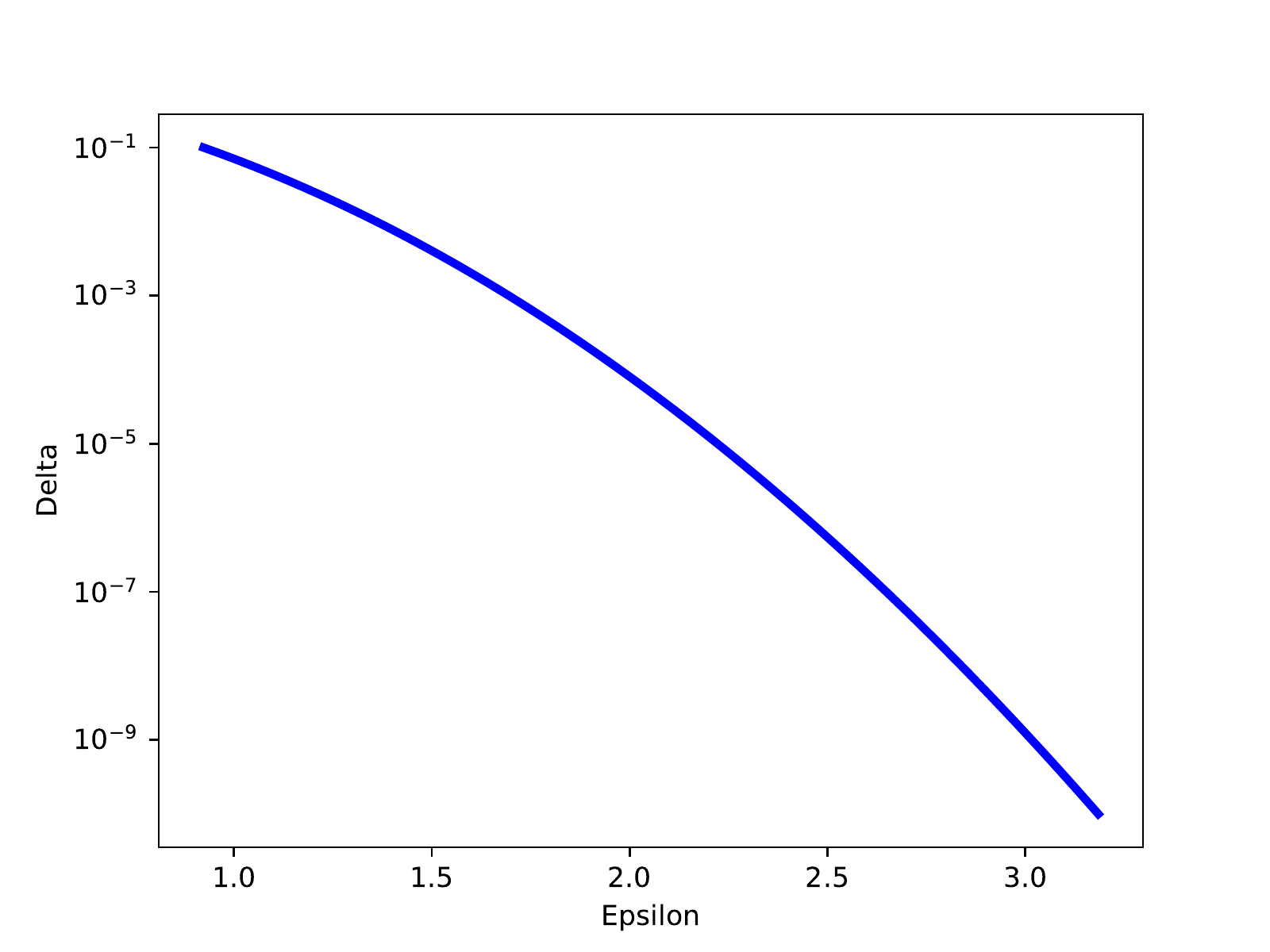}
\caption{Bayesian $(\epsilon,\delta)$ curve under the semantics of Theorem \ref{thm:bayes_pbdp}  for block within custom block group inference under the privacy budget allocated to block-level queries under the redistricting data production settings ($\rho=0.1115$).}\label{fig:pbdpbayesepsdelta_block}
\end{figure}

\begin{figure}[h!]
\includegraphics[scale=0.75]{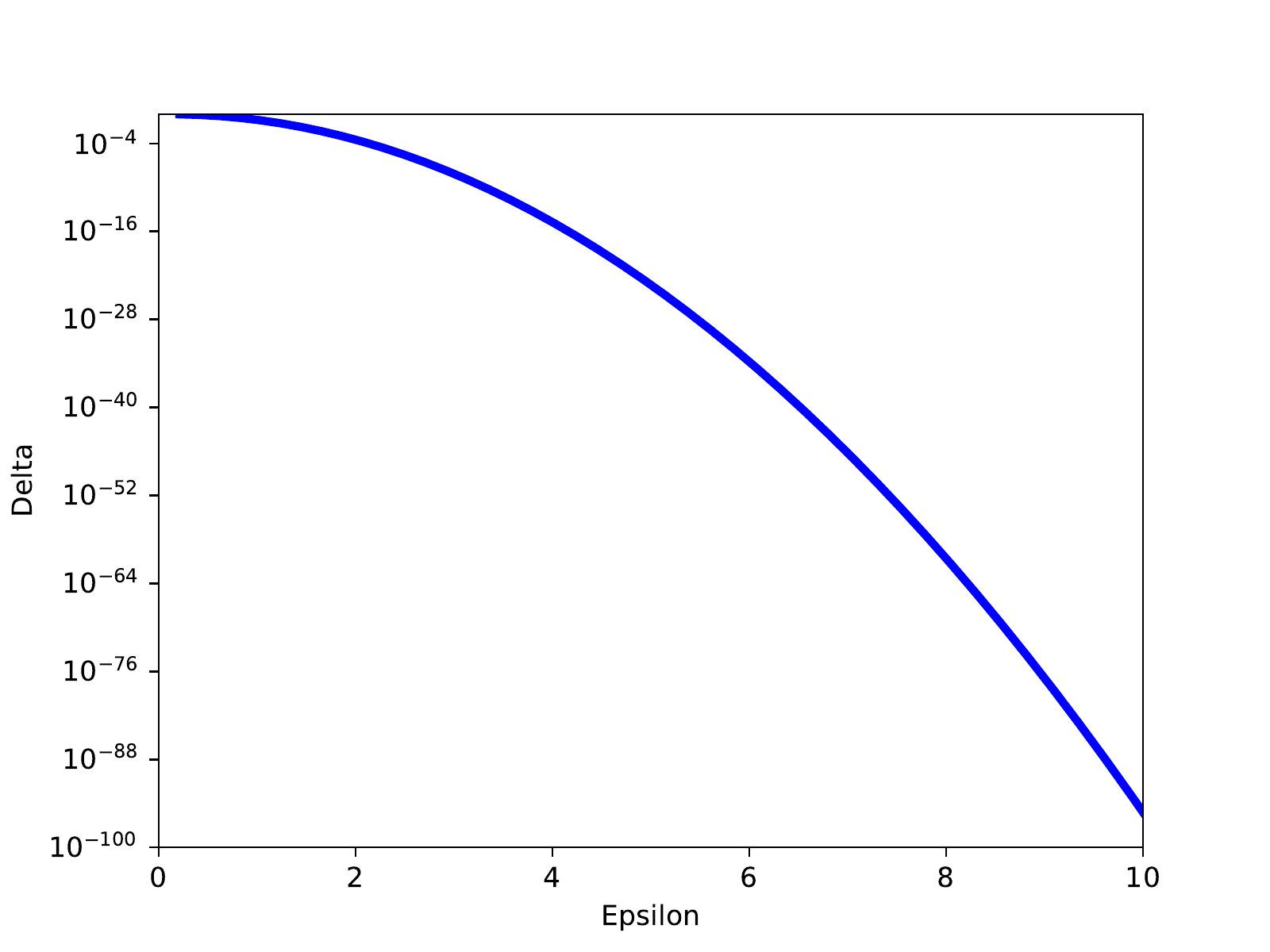}
\caption{Bayesian $(\epsilon,\delta)$ curve under the semantics of Theorem \ref{thm:bayes_2_rdp} (arbitrary priors) for block within custom block group inference under the privacy-loss budget allocated to block-level queries under the redistricting data production settings ($\rho=0.1115$).}\label{fig:bayes2epsdelta_block}
\end{figure}

To explore this scenario, one considers only queries that would be affected by a change in block within the same custom block group. For instance, tract-level, county-level, and state-level queries would not be affected (their answers would not change), and thus we can apply the accounting rules using only to the affected queries (queries about person and housing unit characteristics at the block level). This results in $\rho\approx0.1115$ for this scenario.

Figure \ref{fig:dgm_block} shows the significance level vs. power trade-off for this scenario. The black curve is the upper bound guarantee for any zCDP mechanism for using $\rho=0.1115$. The blue curve is what the result would have been had the continuous Gaussian mechanism been used, and the dashed red line is a Monte Carlo simulation of the curve for the discrete Gaussian mechanism. Again, we see near-perfect alignment between the continuous and discrete Gaussian mechanisms. This suggests that calculations for the continuous Gaussian are a quick and accurate approximate to the discrete Gaussian mechanism used by the TopDown Algorithm.

All three curves in Figure \ref{fig:dgm_block} are near the ideal curve of ``level=power'' that would be achieved by a completely non-informative test indicating strong levels of privacy protection. The power at a selected set of significance levels is given in Table \ref{tab:dgm_block}.

The corresponding Bayesian curves are shown in Figure \ref{fig:bayes1epsdelta_block} for the semantics of Theorem \ref{thm:bayes_1_rdp} (attacker knows all but one person), Figure
\ref{fig:pbdpbayesepsdelta_block} for the semantics of Theorem \ref{thm:bayes_pbdp} for the continuous Gaussian mechanism -- these semantics correspond to the $(\epsilon,\delta)$-\pbdp curve which we believe approximate the curve for the discrete Gaussian Mechanism, and
Figure
\ref{fig:bayes2epsdelta_block} for the semantics of Theorem  \ref{thm:bayes_2_rdp} (arbitrary adversarial priors). Both Theorems \ref{thm:bayes_1_rdp} and \ref{thm:bayes_2_rdp} were derived using Markov's inequality and hence are likely to overestimate $\delta$ for small values of $\epsilon$.

Given the empirical similarity of the significance level vs. power curves between the continuous and discrete Gaussian mechanisms, and given the hypothesized similarity for the $(\epsilon,\delta)$-\pbdp curves for the two mechanisms, in the remaining scenarios we will only present the frequentist semantics and Bayesian semantics of Theorem \ref{thm:bayes_pbdp} for the continuous Gaussian (because of the computational cost associated with estimating the curves for the discrete Gaussian).

\subsubsection{Scenario B:} \target is concerned with protecting the specific block within the tract that \target resides in. Applying the privacy accounting rules to queries at the block and custom block group level, we obtain $\rho\approx 0.926$.

\begin{figure}[h!]
\includegraphics[scale=0.75]{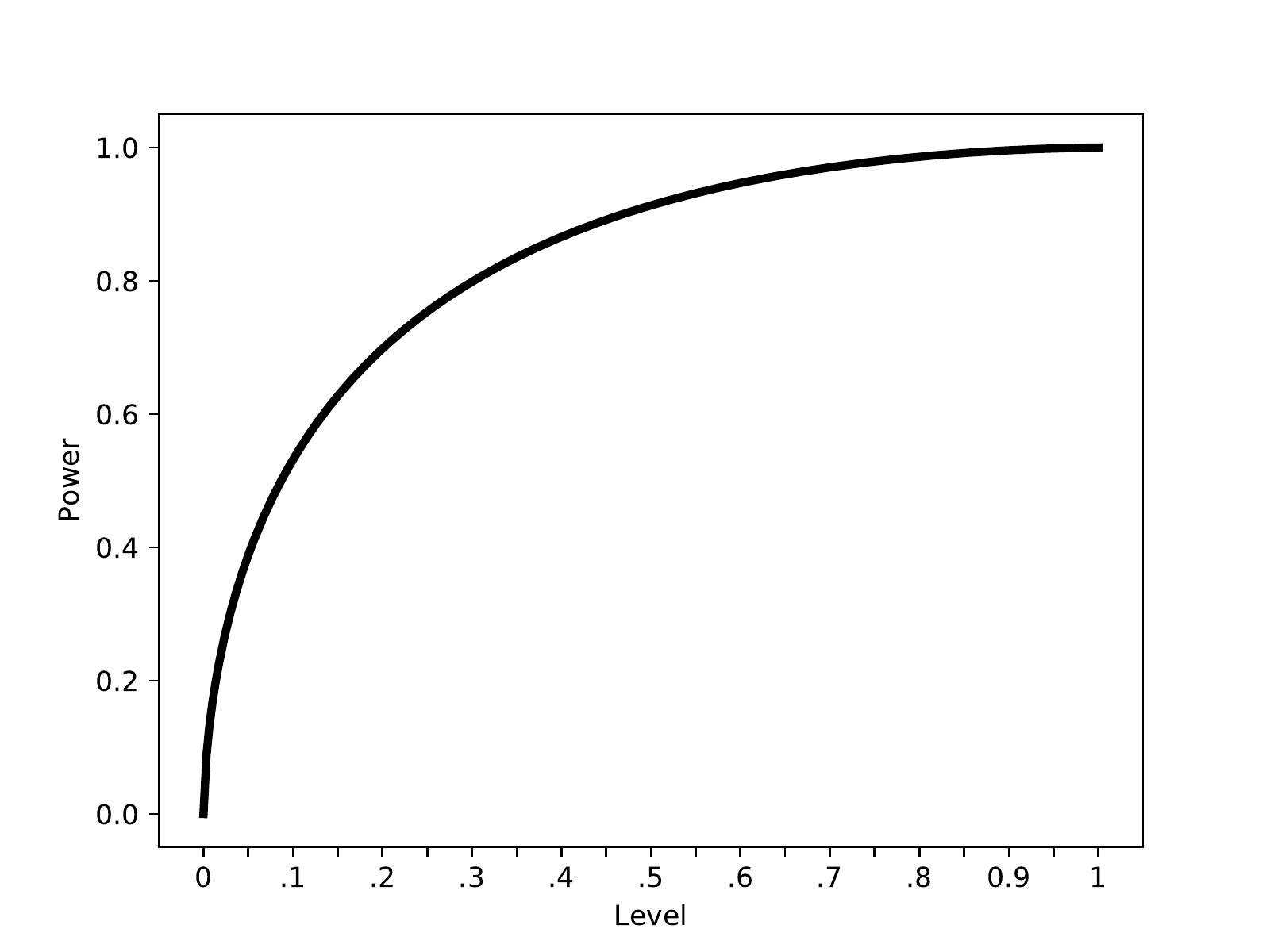}
\caption{Significance level vs. power tradeoff for the Gaussian mechanism at $\rho\approx0.926$.}\label{fig:scenario_b_pl}
\end{figure}

\begin{figure}[h!]
\includegraphics[scale=0.75]{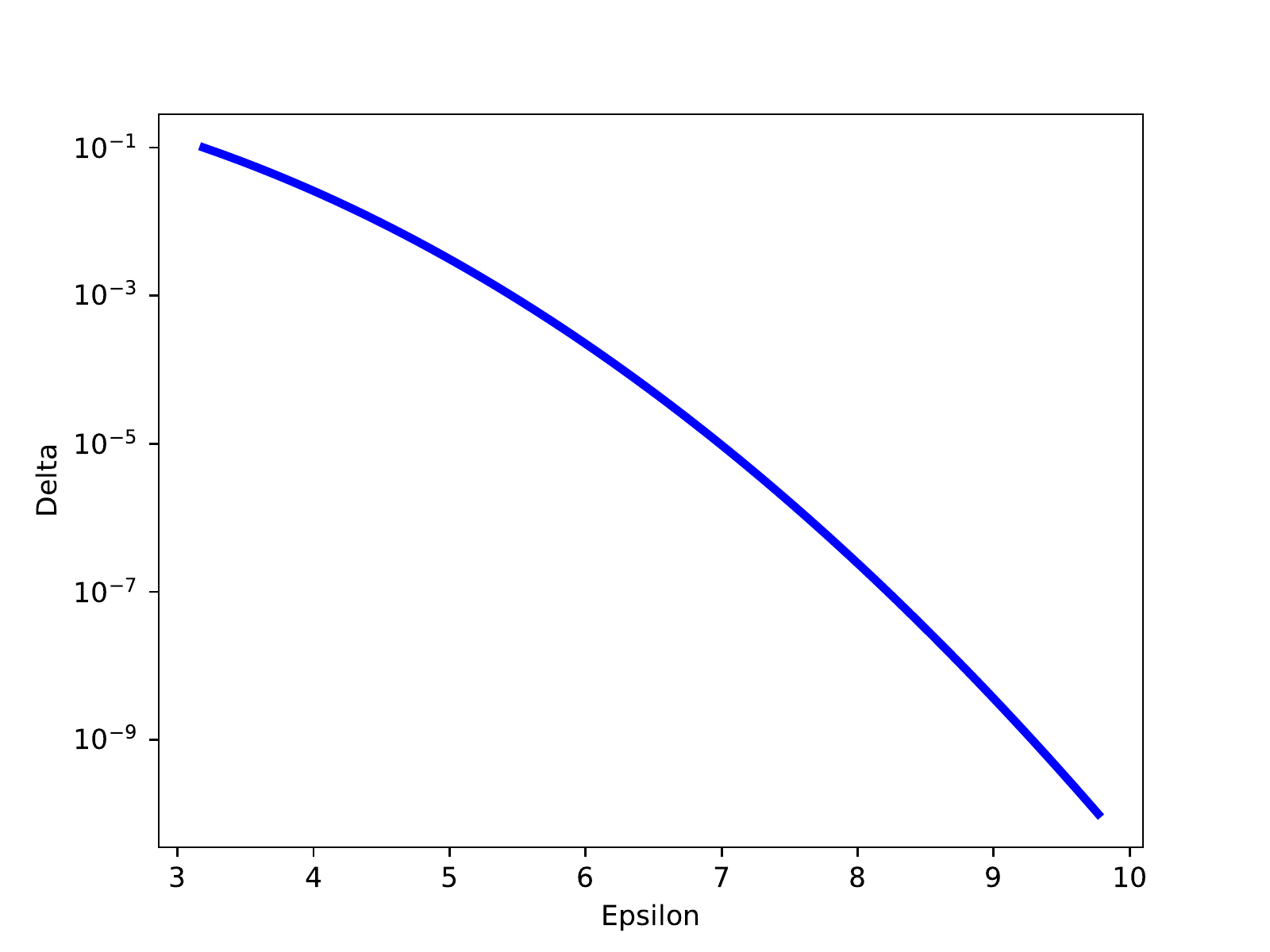}
\caption{Bayesian $(\epsilon,\delta)$ curve of Theorem \ref{thm:bayes_pbdp} for inference about block within tract.}\label{fig:scenario_b_pbdp}
\end{figure}

The frequentist semantics are shown in Figure \ref{fig:scenario_b_pl}. For significance levels $0.01$, $0.05$ and $0.10$, the corresponding power values are $0.17$, $0.39$, and
$0.53$, respectively. The Bayesian $(\epsilon,\delta)$ curve of Theorem \ref{thm:bayes_pbdp}, which is also the $(\epsilon,\delta)$-\pbdp curve is shown in Figure \ref{fig:scenario_b_pbdp}.

% Q2
% total rho: 0.9259579207302197
% At level=0.01 0.16714803700751568
% At level=0.05 0.3882042396706365
% At level=0.10 0.5316028299838497

\subsubsection{Scenario C:} \target is only concerned with protecting inference about race. The total $\rho$ for queries that involve race is $\approx 0.952$ and so the curves would be similar to the previous scenario and hence are not reproduced. The powers for selected significance levels are also almost the same. For significance levels $0.01$, $0.05$ and $0.10$, the corresponding power values are $0.17$, $0.40$, and
$0.54$, respectively.

% Q3
% total rho: 0.9515205932496472
% At level=0.01 0.17186008505411104
% At level=0.05 0.3953714507662144
% At level=0.10 0.5390163929882614

\subsubsection{Scenario D:} 
\target is only concerned about protecting inference about ethnicity. The total $\rho$ for queries that involve ethnicity is $\approx 0.945$ and so the curves would again be similar to the previous scenarios. For significance levels $0.01$, $0.05$ and $0.10$, the corresponding power values are $0.17$, $0.39$, and
$0.54$, respectively.

% Q4 
% total rho: 0.9454429647819133
% At level=0.01 0.17073798134841828
% At level=0.05 0.3936729309418272
% At level=0.10 0.5372640297978697

\subsubsection{Scenario E:} 
\begin{figure}[h!]
\includegraphics[scale=0.75]{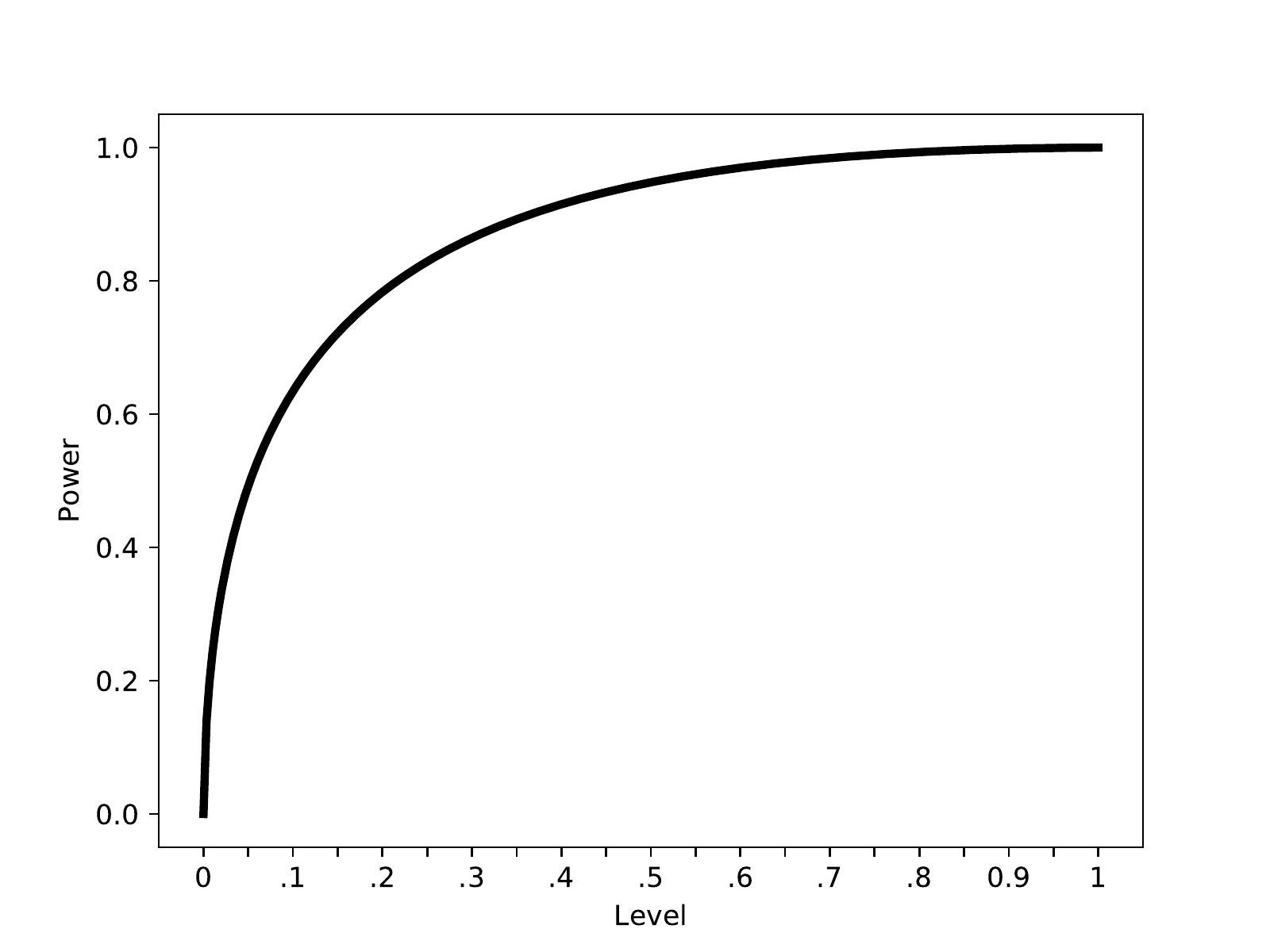}
\caption{Significance level vs. power trade-off for the Gaussian mechanism for Scenario E.}\label{fig:scenario_e_pl}
\end{figure}

\begin{figure}[h!]
\includegraphics[scale=0.75]{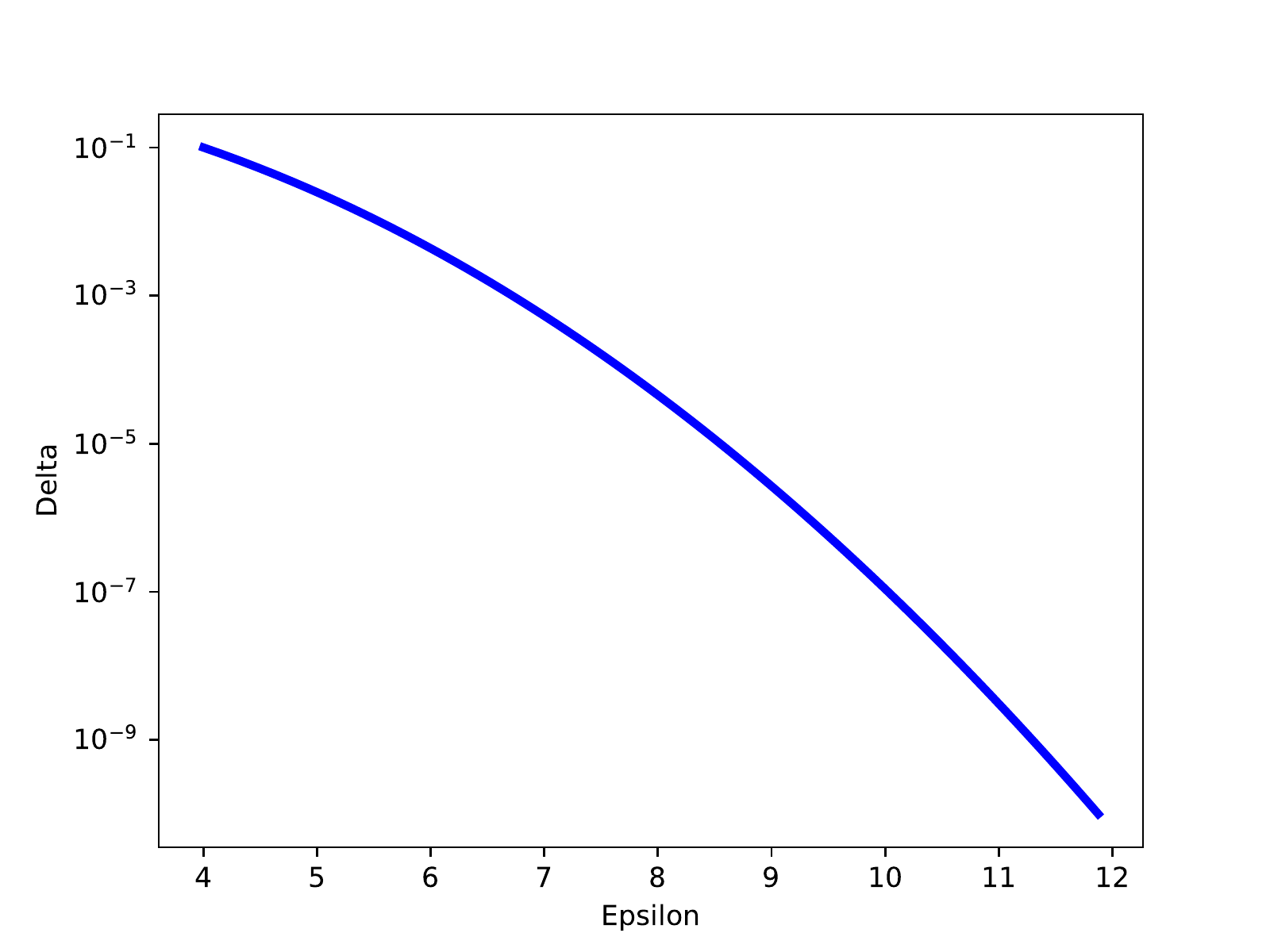}
\caption{Bayesian $(\epsilon,\delta)$ curve of Theorem \ref{thm:bayes_pbdp} for Scenario E.}\label{fig:scenario_e_pbdp}
\end{figure}
\target is concerned about protecting inferences about race, but also believes that knowledge about the specific block within \target's tract would be too revealing about race. In this case, both race and block within tract are the attributes of concern, so all queries at the custom block group and block level are affected as well as all queries involving race at the tract, county, state, and national levels (individuals who do not think race queries at or above the tract level affect inference about them would disregard those queries in their $\rho$ calculations).
The total $\rho$ in this setting is $\approx 1.32$.
At this higher level of $\rho$, the semantics get weaker. For example, for significance levels $0.01$, $0.05$ and $0.10$, the corresponding power values are $0.24$, $0.49$, and
$0.63$, respectively. The full significance level vs. power tradeoff curve is shown in Figure \ref{fig:scenario_e_pl}
 while the Bayesian $(\epsilon,\delta)$-curve is shown in Figure \ref{fig:scenario_e_pbdp}.
% Q5
% total rho: 1.3199144258198976
% At level=0.01 0.2414665524329493
% At level=0.05 0.4919823528375517
% At level=0.10 0.6342772942642629

\subsubsection{Scenario F:}
\begin{figure}[h!]
\includegraphics[scale=0.75]{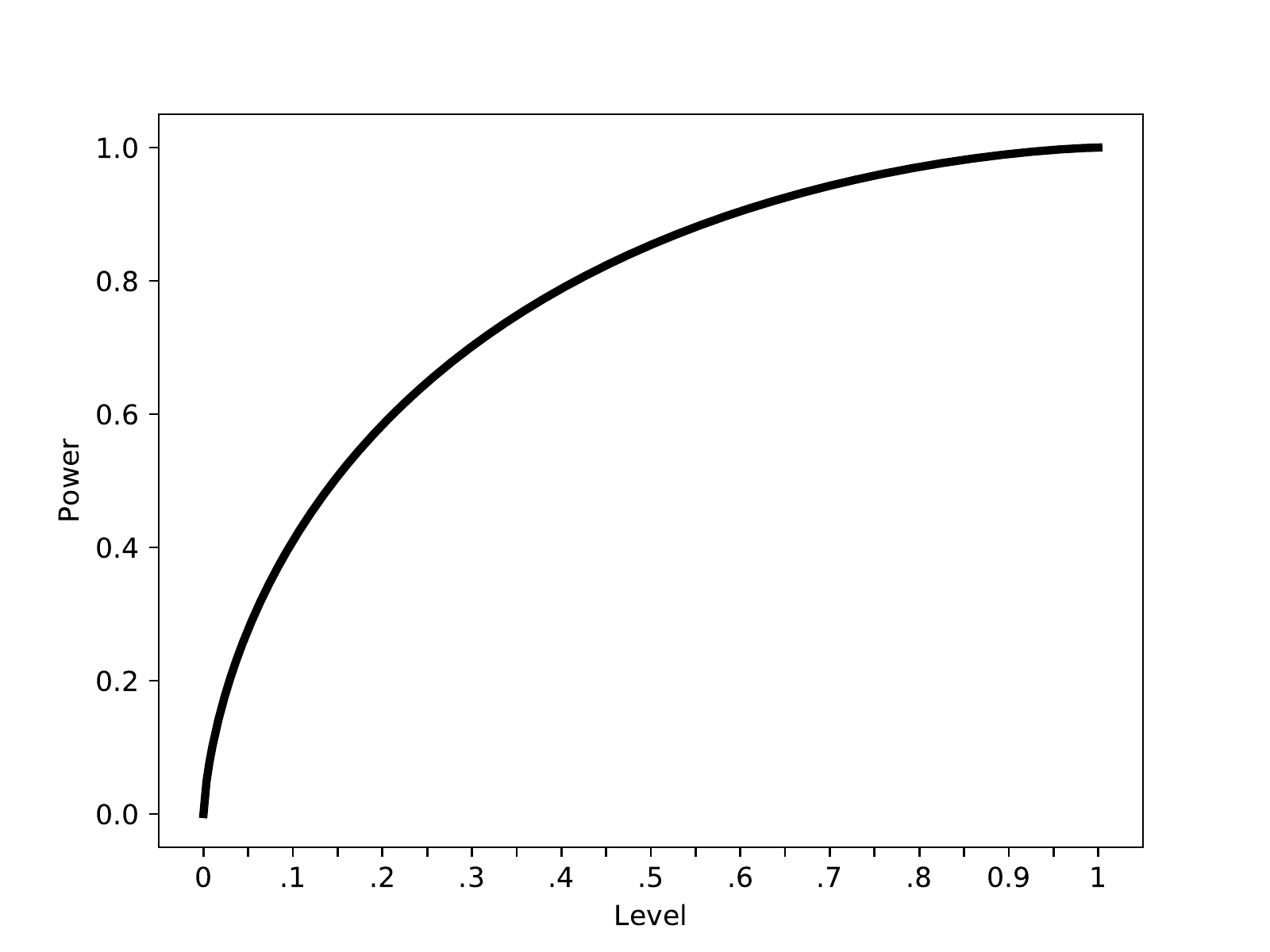}
\caption{Significance level vs. power trade-off for the Gaussian mechanism for Scenario F.}\label{fig:scenario_f_pl}
\end{figure}

\begin{figure}[h!]
\includegraphics[scale=0.75]{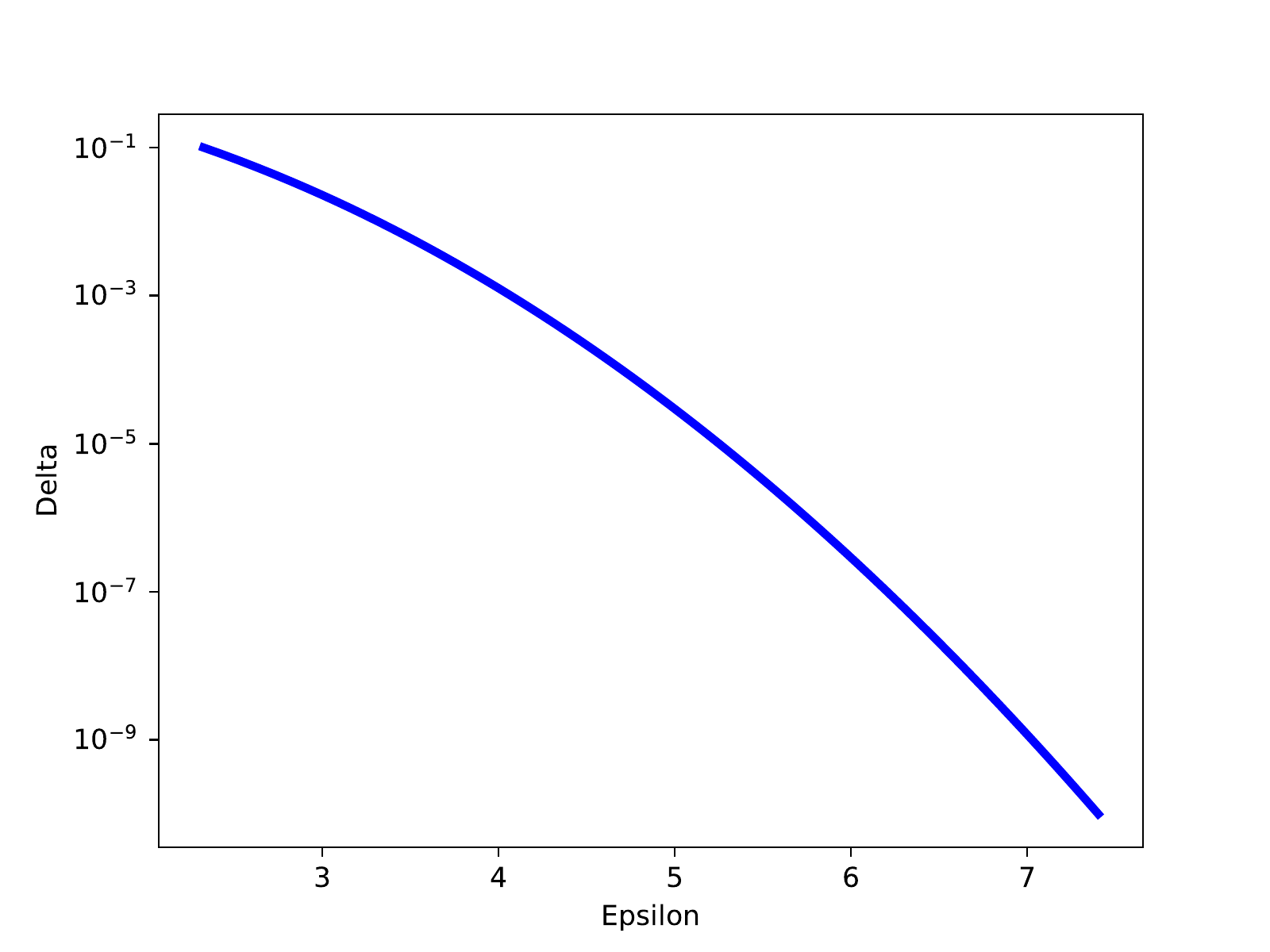}
\caption{Bayesian $(\epsilon,\delta)$ curve of Theorem \ref{thm:bayes_pbdp} for Scenario F.}\label{fig:scenario_f_pbdp}
\end{figure}
\target is concerned about protecting inferences about voting age, but also believes that knowledge about specific block within \target's block group would be too revealing about voting age. In this case, both voting age and block within block group are attributes of concern. The total $\rho$ of queries involving these attributes is $\approx 0.555$. 
This is a fairly low setting of $\rho$. The significance level vs. power trade-offs for significance levels $0.01$, $0.05$ and $0.10$ have the corresponding power values are $0.10$, $0.28$, and
$0.41$, respectively.
% total rho va_block: 0.5548829692906987
%At level=0.01 0.1015279029259174
% At level=0.05 0.27712644186977287
% At level=0.10 0.40978530856860423
The full significance level vs. power trade-off curve is shown in Figure \ref{fig:scenario_f_pl}, 
while the Bayesian $(\epsilon,\delta)$-curve is shown in Figure \ref{fig:scenario_f_pbdp}.

\subsubsection{Scenario G:} \target is concerned about protecting inferences about voting age and race, but also believes that knowledge about specific block within \target's block group would be too revealing about voting age and race. In this case, voting age, race, and block within block group are attributes of concern. The $\rho$ for this scenario is  $\approx 0.969$ and so the semantics are very similar to Scenario C.

%total rho va_race_block: 0.9685663156536115
%At level=0.01 0.1750130279488662
%At level=0.05 0.4001169109210638
%At level=0.10 0.5438974413818907

\subsubsection{Scenario H:} \target is concerned about protecting inferences about voting age and ethnicity, but also believes that knowledge about specific block within \target's block group would be too revealing about voting age and ethnicity. In this case, voting age, ethnicity, and block within block group are attributes of concern. The $\rho$ for this scenario is  $\approx 0.968$ and so the semantics are very similar to Scenario C.

% total rho va_eth_block: 0.9682525218360662
% At level=0.01 0.17495490924847856
% At level=0.05 0.400029795851353
% At level=0.10 0.5438080335539366

\subsection{Discussion}
In this section, we considered details of the allocation of the overall production privacy-loss budget to different queries by the TopDown Algorithm and appropriate accounting rules to use to compute the $\rho$ of a selected set of queries. These $\rho$ values can be used to provide semantics to different scenarios, such as those studied in this section.

These scenarios are by no means complete, and different people may have different concerns. They can evaluate the privacy impact of their concerns by selecting the set of queries that may be potentially revealing in their opinion (for example, they may feel that race queries below the county level may be revealing). Once the $\rho$ for the subset of queries is computed, the power at a given significance level $\level$ can be computed as:
\begin{align*}
1-\Phi_{\frac{1}{2\rho}}(\Phi_{\frac{1}{2\rho}}^{-1}(1-\level)-1),
\end{align*}
where $\Phi_{\frac{1}{2\rho}}$ is the CDF of the zero-mean Gaussian distribution with variance $\sigma^2=\frac{1}{2\rho}$. For the continuous Gaussian mechanism, this happens to mathematically correspond to the difficulty of distinguishing between a $N(0, \sigma^2)$ and $N(1, \sigma^2)$ random variable using a hypothesis test based on one sample.

Meanwhile, the $(\epsilon,\delta)$-curve of Theorem \ref{thm:bayes_pbdp} for the continuous Gaussian can be computed from the formula:
\begin{align*}
\epsilon = \log\left(\frac{\delta}{\Phi_1(-\Phi_1^{-1}(1-\delta) -\sqrt{2\rho})}\right).
\end{align*}

%% file: related.tex
Differential privacy has many variations, only some of which were discussed in this paper. See Desfontaines and Pejó \cite{sokdps} for a more comprehensive survey of variants of differential privacy.

There are several sources that explicitly treat the privacy guarantees as a causal property in which an intervention in the data (such as replacing a record) can occur before providing an input to $\mech$ \cite{kasiviswanathan2014semantics,dpcausal,BassilyGKS13}. These papers all contain Bayesian elements. The Bayesian variations of differential privacy (e.g., \cite{kasiviswanathan2014semantics,dpcausal,BassilyGKS13,bdp2,kifer14:tods,DandekarBasuBressan,DKM20}) also incorporate noise from the data into the constraints on a mechanisms output or privacy-loss random variable.

Among these, Kifer and Machanavajjhala \cite{kifer14:tods} study the semantics of a proposed definition called \emph{Pufferfish},
Bassily et al. \cite{BassilyGKS13} study the semantics of a proposed definition called \emph{Coupled-worlds Privacy}, and Kasiviswanathan and Smith \cite{kasiviswanathan2014semantics} study the semantics of approximate differential privacy under Bayesian inference. The Pufferfish definition comes with an absolute upper bound on the posterior-to-posterior ratio of a suitably chosen counterfactual world  \cite{kifer14:tods} under a restricted set of attacker priors. Bassily et al. \cite{BassilyGKS13} propose several variations of their privacy definition, one of which compares posteriors using inequalities similar to approximate differential privacy, also under a restricted set of priors. Similar studies were performed by Desfontaines et al. \cite{DKM20}, studying Bayesian semantics of privacy definitions similar to pure differential privacy, and proposing definitions that compare posteriors similarly to approximate differential privacy. Kasiviswanathan and Smith study the Bayesian guarantees of approximate differential privacy under arbitrary priors, hence, similar in spirit to the Bayesian guarantees in Section \ref{sec:bayes}. They compare posteriors using total variation distance and obtain a non-trivial bound (i.e., total variation distance less than 1) under a restricted set of parameters, such as $\epsilon<1/4$ and $\delta$ much smaller than the inverse of the number of people in the data.
 
For Bayesian privacy definitions that require a limited set of priors for modeling the data, it is also important to study how privacy degrades under  miss-specification of the priors, also known as \emph{close adversaries}. Song et al. \cite{wassmechanism} provided such an analysis for Pufferfish. 
 
An alternative approach, known as \emph{restricted sensitivity} \cite{restrictedsens} flips the idea typically taken by Bayesian approaches. Instead of providing privacy protections only when certain assumptions are met, one can instead provide privacy protections with no assumptions about the data, but only guarantee statistical validity of the results under certain conditions on the data. 

The idea of considering more fine-grained protection (i.e., protection of different parts of a record) has a very long history and is often achieved by redefining the choice of neighbors (e.g., \cite{dmns06,dwork:2006,elementlevel,BassilyGKS13,kifer14:tods,He14:sigmod,chatzimetrics,mcsherrynetwork,NissimRS07,hljm09}).

Finally, the use of the Gaussian distribution is often justified with central limit theorems for privacy loss random variables \cite{SommerMM19,gaussdp}.

%% file: conc.tex
In this paper, we surveyed the major differential privacy frameworks that are important for understanding the privacy semantics of the redistricting data from the 2020 Census -- namely, pure, approximate, R\'enyi, zero-concentrated, and $f$ differential privacy. The semantics considered interpretability of the privacy parameters, the frequentist hypothesis testing guarantees, and the Bayesian posterior-to-posterior guarantees.

We interpreted the guarantees in the context of the production settings for the disclosure avoidance system implemented for the 2020 Census major data products including the redistricting data. The main purposes of this paper were (1) to explain the privacy protections used in the redistricting data, (2) to serve as a guide for how other organizations can study the privacy semantics of their data releases, and (3) to provide a guide for what peer-reviewers can look for when asked to evaluate new privacy proposals.

The paper also generated some novel results such as a new motivation for $f$-DP via \pbdp and the Bayesian semantics for R\'enyi and zero-concentrated DP.

There are some topics relevant to the privacy guarantees of the 2020 Census redistricting data that are not covered. These include accounting for the policy-determined data \emph{invariants} -- properties of the confidential input data that were replicated without DP protection in the privacy-protected data. These include the exact as-enumerated state population totals, the number of housing units in each level of geography, and the number of occupied group quarters facilities in each major type for each level of geography. These more complex topics will be the subject of a separate paper.

%% file: appendix.tex
\section{Proofs from Section \ref{sec:privdefs}}\label{app:privdefs}
\printProofs[privdefs]

\section{Proofs from Section \ref{sec:freq}}\label{app:freq}
\printProofs[freq]

\section{Proofs from Section \ref{sec:bayes}}\label{app:bayes}
\printProofs[bayes]